\documentclass[12pt]{article}
\usepackage{a4wide}
\usepackage{bbold}
 \usepackage{pstricks}
\usepackage{amsmath,amssymb,bbm}
\usepackage{multirow}
 \usepackage{graphicx}
 \usepackage[verbose]{wrapfig}
\usepackage{comment}
\newsavebox{\uuunit}
\sbox{\uuunit}
    {\setlength{\unitlength}{0.825em}
     \begin{picture}(0.6,0.7)
        \thinlines
        \put(0,0){\line(1,0){0.5}}
        \put(0.15,0){\line(0,1){0.7}}
        \put(0.35,0){\line(0,1){0.8}}
       \multiput(0.3,0.8)(-0.04,-0.02){12}{\rule{0.5pt}{0.5pt}}
     \end {picture}}

\def\2{\frac12}
\def\4{\frac14}

\newcommand{\be}{\begin{equation}}
\newcommand{\ee}{\end{equation}}
\newcommand{\bea}{\begin{eqnarray}}
\newcommand{\eea}{\end{eqnarray}}

\usepackage{hyperref}
\hypersetup{
linkbordercolor={blue}
}

\hypersetup{
    bookmarks=false,         
    unicode=false,          
    pdftoolbar=true,        
    pdfmenubar=true,        
    pdffitwindow=false, 
    pdfstartview={FitH},    
    pdftitle={Towards a classification of branes in theories with eight supercharges},    
    pdfauthor={Eric A.Bergshoeff,Fabio Riccioni,Luca Romano},     
    pdfsubject={},   
    pdfnewwindow=true,      
    colorlinks=false,       
    linkcolor=blue,          
    citecolor=blue,        
    filecolor=blue,      
    urlcolor=blue           
    linkbordercolor={blue},
    citebordercolor={purple},
    urlbordercolor={gray}
}

\def\equationautorefname~#1\null{%
  eq.~(#1)\null
}

\def\figureautorefname~#1\null{%
  Fig.~#1\null
}

\begin{document}

\begin{titlepage}
\begin{center}


\vskip 1.5cm

{\Large \bf  Towards a classification of branes \\
\vskip .5cm
in theories with eight supercharges}

\vskip 2cm

{\bf Eric A.~Bergshoeff\,$^1$, Fabio Riccioni\,$^2$  and Luca Romano\,$^3$}

\vskip 40pt

{\em $^1$ \hskip -.1truecm Centre for Theoretical Physics,
University of Groningen, \\ Nijenborgh 4, 9747 AG Groningen, The
Netherlands \vskip 5pt }

{email: {\tt E.A.Bergshoeff@rug.nl}} \\

\vskip 15pt

{\em $^2$ \hskip -.1truecm
 INFN Sezione di Roma,   Dipartimento di Fisica, Universit\`a di Roma ``La Sapienza'',\\ Piazzale Aldo Moro 2, 00185 Roma, Italy
 \vskip 5pt }

{email: {\tt Fabio.Riccioni@roma1.infn.it}} \\

\vskip 15pt

{\em $^3$ \hskip -.1truecm  Dipartimento di Fisica and INFN Sezione di Roma, Universit\`a di Roma ``La Sapienza'',\\ Piazzale Aldo Moro 2, 00185 Roma, Italy
 \vskip 5pt }

{email: {\tt  Luca.Romano@roma1.infn.it}} \\

\end{center}

\vskip 0.5cm

\begin{center} {\bf ABSTRACT}\\[3ex]
\end{center}

We provide a classification of half-supersymmetric branes in quarter-maximal supergravity theories with scalars parametrising coset manifolds. Guided by the results previously obtained for the half-maximal theories, we are able to show that half-supersymmetric branes correspond to the real longest weights of the representations of the brane charges, where  the reality properties of the weights are  determined from the  Tits-Satake diagrams associated to the global symmetry groups.
We show that the resulting brane structure is universal for all theories that can be uplifted to six dimensions. We also show that when viewing these theories as low-energy theories for the suitably compactified heterotic string, the classification we obtain  is in perfect agreement with the wrapping rules derived in previous works for the same theory compactified on tori. Finally, we relate the branes to the R-symmetry representations of the central charges and we show that in general the degeneracies of the BPS conditions are twice those of the half-maximal theories and four times those of the maximal ones.

\end{titlepage}

\newpage
\setcounter{page}{1} \tableofcontents

\newpage

\setcounter{page}{1} \numberwithin{equation}{section}

\section{Introduction}

In general, BPS states are massive representations of extended supersymmetry algebras such that the inequality relating their mass and their central charge is saturated. This implies that the representations are short and thus the relation between the mass and the charge is quantum-mechanically exact.
Although one can explicitly determine the spectrum of string theories only at the perturbative level, non-perturbative BPS branes  manifest themselves
in the low-energy supergravity effective actions as solutions preserving a portion of the supersymmetry of the corresponding theory. It is for this reason that the study of these objects in supergravity has played a crucial role in understanding non-perturbative string dualities. Among all BPS branes, the ones that preserve the largest amount of supersymmetry, {\it i.e.}~1/2-supersymmetric branes, are special because one can think of the branes preserving less supersymmetry as bound states of them. In this sense, 1/2-supersymmetric branes are the basic building blocks of all the BPS branes of a theory.

The structure of BPS brane solutions of maximal supergravity theories was analysed originally in \cite{Ferrara:1997ci,Ferrara:1997uz,Lu:1997bg} for the solutions that are electrically or magnetically charged under the form fields of the theory that carry propagating degrees of freedom (with the exception of the scalar fields). Such BPS branes 
always have three or more transverse directions.
It turns out that one can also consider
branes with two, one or zero transverse directions, called respectively defect branes, domain walls and space-filling branes. Although these branes are not globally consistent as supergravity solutions they can nonetheless be described in terms of a world-volume effective action. One can consider as prototype examples the D7 and D9-branes of the IIB theory and the D8-brane of the IIA theory. It is well known that a single D7-brane does not have finite energy \cite{Greene:1989ya,Gibbons:1995vg}. To obtain finite-energy solutions one must construct multiple brane configurations which include orientifolds.
The IIA D8-brane can be viewed as a solution of the massive IIA theory \cite{Romans} whose consistency also requires orientifolds
\cite{Polchinski:1995df}. Finally, the D9-brane of the IIB theory plays a crucial role in the Type-I orientifold construction \cite{Polchinski:1995mt,Angelantonj:2002ct}.
Although none of these objects is a consistent single-brane configuration, one can construct for each of them a kappa-symmetric effective action, whose existence will be considered in this paper as a guiding principle for classifying branes.

In general, a $p$-brane is electrically charged under a $(p+1)$-form. Indeed, one can consider a $(D-p-4)$-brane, magnetically charged under a $(p+1)$-form, as being electrically charged with respect to its dual $(D-p-3)$-form. As already mentioned, this covers all the branes with at least three transverse directions. 
The branes with two, one or zero transverse directions are charged under $(D-2)$, $(D-1)$ and $D$-forms. The $(D-2)$-forms are dual to the scalars and they always belong to the adjoint representation of the global symmetry group of the theory. On the contrary, the $(D-1)$- and the $D$-forms are not associated to any degree of freedom and their representations can only be determined by requiring the closure of the supersymmetry and gauge algebras. In the case of the IIA and IIB theories, the full list of such fields was determined \cite{Bergshoeff:2005ac,Bergshoeff:2006qw,Bergshoeff:2010mv} and it includes the forms associated to the D8- and D9-branes together with additional forms. This was generalised to all maximal supergravity theories in any dimension in \cite{Riccioni:2007au,Bergshoeff:2007qi} using the $\text{E}_{11}$ Kac-Moody algebra \cite{West:2001as} and in \cite{deWit:2008ta} using the embedding-tensor formalism~\cite{Nicolai:2000sc}.
Based on this, recently a complete classification of 1/2-supersymmetric branes in maximal supergravity theories has been obtained by requiring that the worldvolume degrees of freedom of the brane fit in a multiplet with sixteen supersymmetries \cite{Bergshoeff:2011qk,Bergshoeff:2012ex}. The brane charges that are selected in this way correspond to particular components of the representations of the fields, and for branes with two or less transverse directions the number of these components is less than the dimension of the representation. The same result was obtained in
\cite{Kleinschmidt:2011vu} observing that the branes correspond to the particular subset of $\text{E}_{11}$ roots that are real.

A group-theoretic reformulation of the results of \cite{Bergshoeff:2011qk,Bergshoeff:2012ex}, as well as a natural explanation of why they coincide with the analysis of \cite{Kleinschmidt:2011vu}, was given in \cite{Bergshoeff:2013sxa}, where it was observed that the 1/2-supersymmetric $p$-branes correspond to the longest weights of the representations of the $(p+1)$-forms. The longest weights are all those weights that can be chosen as highest weights given a particular choice of simple roots, and thus counting the longest weights corresponds to counting all the components of the representation that satisfy the highest-weight constraint. Already in \cite{Ferrara:1997ci} it was observed that the 1/2-supersymmetric branes with more than two transverse directions satisfy the highest-weight constraint. Our analysis shows that actually this applies to all the branes in the theory, regardless of the dimension of the world-volume \cite{Bergshoeff:2012ex}.

A simple intuition for the highest-weight constraints is given by decomposing the global symmetry group with respect to its $\text{SO}(d,d)$ subgroup in $10-d$ dimensions. This symmetry does not act on the string dilaton and thus it is a  perturbative symmetry of the low-energy action, whose discrete counterpart is T-duality. For the fundamental 1/2-BPS 0-branes, with charges $Q_A$ belonging to the vector representation of $\text{SO}(d,d)$, the highest weight constraint is $Q^2=0$, and the longest weights correspond to the lightlike directions of $Q_A$. This was generalised in \cite{Bergshoeff:2011zk,Bergshoeff:2011ee}  to a set of rules,  named `light-cone rules', for the various $\text{SO}(d,d)$ representations of the fields in the theory. The branes correspond to the longest weights of $\text{SO}(d,d)$, and the light-cone rules select precisely the components associated to the longest weights.

The classification of 1/2-supersymmetic branes was extended to theories with sixteen supercharges in \cite{Bergshoeff:2012jb}.
Considering only ungauged theories, the global symmetry in $10-d$ dimensions for supergravity coupled to $d+n$ abelian vector multiplets is $\text{SO}(d,d+n)$, and the classification results in applying the light-cone rules of the maximal theory to this group. Clearly, the fact that in this case the group is not maximally non-compact (for $n \neq 0,1$) implies that the light-cone rules select fewer components for a given representation. Also, for a given representation the invariant constraint is the same as in the maximal theory. In the particular case of branes with more than two transverse directions, the constraints of single-brane states  were already discussed in \cite{Ferrara:1997ci}. Within the  classification of black hole orbits using the language of Jordan algebras  \cite{Borsten:2011ai}, these branes correspond to so-called rank-1 orbits.

The first goal of this paper is to give a more rigorous group-theoretic interpretation of the results obtained in the half-maximal case. We will show that in order to extend the longest-weight rule - valid for the maximally non-compact groups of the maximal theories - to the half-maximal theories, one has to introduce the Tits-Satake diagram describing the real form of the orthogonal Lie algebra corresponding to the group $\text{SO}(d,d+n)$. From the Tits-Satake diagram one reads the reality properties of the roots, and hence those of the weights, and it turns out that the light-cone rules  select the {\it real} longest weights of the representations of  $\text{SO}(d,d+n)$. This is consistent with the result of the maximal theories, because the Tits-Satake diagram for the maximally non-compact form of a given algebra results in roots, and therefore in weights, that are all real.

Guided by these findings, we then move to the main result of the paper. We consider matter-coupled supergravity theories with eight supercharges, with scalars parametrising coset manifolds. Each simple factor of the global symmetry  will be a particular group in a given real form, and we will consider its Tits-Satake diagram, which gives the reality properties of the roots and the weights of the corresponding algebra. From the representations of the $(p+1)$-form fields, we will then select those associated to 1/2-BPS branes by identifying the  real longest weights. We will consider first the particular set of theories that gives rise to symmetric manifolds upon reduction to three dimensions. Such theories do not contain hypermultiplets in dimension higher than three. The four- and five-dimensional black hole analysis for these theories was initiated in \cite{Ferrara:1997uz} and culminated in the general classification of \cite{Borsten:2011ai}. Again, the single-brane states we identify correspond to the rank-1 orbits of this last paper.
The main outcome of our analysis is that the brane classification that we obtain is universal for all the theories that can be uplifted to six dimensions. We then consider the inclusion of the hypermultiplets, and we manage to classify also the branes in the hyper-sector using a suitable truncation of the theory with sixteen supercharges.

The classification of branes in the maximal theories in terms of the T-duality $\text{SO}(d,d)$ symmetry revealed that the ten-dimensional branes satisfy
specific {\it wrapping rules} upon dimensional reduction \cite{Bergshoeff:2011mh,Bergshoeff:2011ee}. These wrapping rules were shown to apply also to the branes of the heterotic theory compactified on tori, as well as to the IIA theory compactified on K3 \cite{Bergshoeff:2012jb,Bergshoeff:2013spa}. Considering the six-dimensional models with eight supercharges as resulting from the heterotic theory compactified on K3, we will show that our classification in any dimension is consistent with the wrapping rules.

Finally, we will discuss the relation between the branes we obtain and the central charges of the corresponding supersymmetry algebra.
In the maximal theories there is a one-to-one correspondence between branes and central charges  for the branes with at least three transverse directions. This leads to the fact that there is no degeneracy for the BPS conditions, i.e.~each brane has its own BPS condition. Instead, the BPS conditions  are degenerate for the branes with two or fewer transverse directions. In particular, for the defect branes the degeneracy of the BPS conditions is always two \cite{Bergshoeff:2011se}. In general, bound states of degenerate branes give configurations that  preserve the same amount of supersymmetry as the constituent branes. This is actually a general feature of the branes of the theories with sixteen supercharges, for which one finds degeneracies that are always twice the degeneracies of the maximal theories. In this paper we will find that for the branes of the theories with eight supercharges the degeneracies of the BPS conditions are twice those of the half-maximal theories, and hence four times those of the maximal ones.

The paper is organized as follows. In section 2 we give a basic review of the classification of real forms of simple Lie algebras in terms of Tits-Satake diagrams. In section 3 we apply these techniques to show that the light-cone rules that classify the branes of the heterotic string on a torus correspond to identifying the real longest weights of the associated representations.  We will then apply the real-longest-weight rule to classify the branes of theories with eight supercharges in section 4. In section 5 we use these results to discuss the wrapping rules for the heterotic string on K3. Finally, in section 6 we compare the classification of the branes in theories with eight supercharges with the central charges present in the corresponding supersymmetry algebras and use this to derive the degeneracy of the BPS conditions. Our conclusions are given in section 7.

\section{Real forms and Tits-Satake diagrams}

The aim of this section is to give a pedagogical introduction to the classification of real forms of simple Lie algebras in terms of Tits-Satake diagrams. We do not want by any means to give an exhaustive account of the subject. The idea is just to give the information that the reader will need to understand the rest of the paper. For a more detailed and rigorous analysis we refer to e.g.~\cite{grouptheory}.

Given a complex Lie algebra $\mathfrak{g}^{\mathbb{C}}$, a real form $\mathfrak{g}$ exists if it admits a basis such that all the structure constants are real. When this happens, the complex algebra can be written in terms of its real form as
  \begin{equation}
  \mathfrak{g}^{\mathbb{C}}=\mathfrak{g}\oplus i\mathfrak{g} \quad .
  \end{equation}
As a prototype example we consider the real forms of the complex algebra $\mathfrak{sl}(2,\mathbb{C})$.
Taking as generators the real $2\times 2$ matrices
\begin{equation}
\rho_1 = \sigma_1 \qquad \rho_2 = i \sigma_2 \qquad \rho_3 = \sigma_3
\quad ,
\end{equation}
where the $\sigma_i$'s are the Pauli matrices, one obtains the commutation relations
  \begin{equation}
  [ \rho_1 , \rho_2 ] = -2 \rho_3 \qquad [ \rho_1 , \rho_3 ] = -2 \rho_2 \qquad [ \rho_2 , \rho_3 ] = -2 \rho_1
  \quad ,
  \end{equation}
and considering this algebra on the real numbers one generates the
real form $\mathfrak{sl}(2,\mathbb{R})$. Similarly, taking the
anti-hermitian generators $\tau_i = i \sigma_i$ one gets
 \begin{equation}
  [ \tau_i , \tau_j ] = -2 \epsilon_{ijk} \tau_k
  \quad ,
  \end{equation}
 which corresponds to the real form $\mathfrak{su}(2)$. Although in the case of $\mathfrak{su}(2)$ the
 generators are not real, in both cases the structure constants are real and thus each of the algebras
 defines  a different real form. Up to automorphisms, these are all the possible real forms of the complex algebra  $\mathfrak{sl}(2,\mathbb{C})$.  We now want to generalise this classification
 to any complex Lie algebra and determine all its possible different real forms.

A crucial ingredient in the classification of the various real forms
of a given complex algebra is the Killing form, which is defined as
  \begin{equation}
  B(X,Y) = {\rm Tr} ({\rm ad} X {\rm ad } Y ) \quad
  \end{equation}
for any elements $X, Y$ of the complex algebra. The eigenvectors of
the Killing form with positive eigenvalue are the non-compact
generators, while those with negative eigenvalue are the compact
generators. In particular, a real Lie algebra is compact when its
Killing metric is negative definite. For the two real forms of
$\mathfrak{sl}(2,\mathbb{C})$ defined above, one gets
\begin{equation}
  B_{\mathfrak{sl}(2,\mathbb{R})}=\begin{pmatrix}
 8&0&0\\
  0&-8&0\\
  0&0&8
\end{pmatrix}
  \qquad B_{\mathfrak{su}(2)}=\begin{pmatrix}
 -8&0&0\\
  0&-8&0\\
  0&0&-8
\end{pmatrix}\quad \,. \label{sl2killingforms}
\end{equation}
The fact that  the latter is negative definite implies that
${\mathfrak{su}(2)}$ is the compact real form as we know.

When dealing with a generic Lie algebra, it is convenient to introduce the Chevalley basis, which is defined by the commutation relations
\begin{subequations}
\begin{flalign}
 &[H_{\alpha},E_{\beta}]=A_{\beta\alpha}E_{\beta}\label{Chev1relat}\\
 &[H_{\alpha},E_{-\beta}]=-A_{\beta\alpha}E_{-\beta}\\
 &[E_{\alpha},E_{\beta}]=\left\{\begin{array}{ll}
 N_{\alpha,\beta}E_{\alpha+\beta}& \mbox{if } \alpha+\beta \ {\rm root}\\
 H_{\alpha}  & \mbox{if } \alpha+\beta=0 \\
 0&\mbox{if }\alpha+\beta\ {\rm not} \ {\rm root}
 \end{array}\,.
 \right.
\end{flalign}\label{rootvectorrelation1}
\end{subequations}
Here $H_\alpha$ are the Cartan generators, $E_\alpha$ and
$E_{-\alpha}$ are the positive root and negative root generators
respectively,  $A_{\alpha\beta}$ are the entries of the Cartan
matrix  and $N_{\alpha,\beta}$ are real structure constants. Given
that the structure constants are all real, this algebra on the real
numbers defines a particular real form. Computing the Killing form
one obtains
  \begin{equation}
  B (H_\alpha ,  H_\beta ) \propto \frac{4\langle \alpha , \beta \rangle}{\langle\alpha,\alpha\rangle \langle\beta,\beta \rangle}
  \qquad B (E_\alpha , E_\beta ) \propto \frac{2}{\langle\alpha,\alpha\rangle}\delta_{\alpha,-\beta} \quad
  ,
  \label{killingformgeneral}
  \end{equation}
where $\langle \alpha , \beta \rangle$ denotes the scalar product between two roots.
This implies that all the Cartan generators are non-compact because the eigenvalues of $ B (H_\alpha ,  H_\beta )$ are all positive, while the eigenvectors of $B (E_\alpha , E_\beta ) $ are $E_\alpha + E_{-\alpha}$ with positive eigenvalues and $E_\alpha - E_{-\alpha}$ with negative eigenvalues.
To summarise, the non-compact and compact generators  are
 \begin{equation}
 {\rm non-compact:} \ \ H_\alpha \ \ E_\alpha + E_{-\alpha} \ \ \ {\rm compact:} \ \ E_\alpha - E_{-\alpha} \quad .\label{generatorssplitform}
 \end{equation}
This real form is known as the {\it split} form or maximally
non-compact real form, and we denote it with $\mathfrak{s}$. It is
clear that the algebra generated by
\begin{equation}
{\rm compact:}
 \ \ i H_\alpha  \ \ i (E_\alpha + E_{-\alpha}) \ \ E_\alpha - E_{-\alpha} \label{generatorscompactform}
 \end{equation}
is also a real form. This is the compact form $\mathfrak{c}$ as it can be deduced from eq.~\eqref{killingformgeneral}.

In the particular case of $\mathfrak{sl}(2,\mathbb{C})$ one can represent the Chevalley generators as
 \begin{equation}
 E_+=\frac{1}{2}(\sigma_{1}+i\sigma_{2})=\begin{pmatrix}
 0&1\\
 0&0
\end{pmatrix}\ \
E_{-} =\frac{1}{2}(\sigma_{1}-i\sigma_{2})=\begin{pmatrix}
 0&0\\
 1&0
\end{pmatrix}\ \ H =
\sigma_{3}=\begin{pmatrix}
 1&0\\
 0&-1
\end{pmatrix}\ \ .
\end{equation}
It is straightforward to show that $\rho_1 = E_+ + E_-$, $\rho_2
= E_+ - E_-$ and $\rho_3 =H$ lead to the real form
$\mathfrak{sl}(2,\mathbb{R})$ with non-compact generators $\rho_1$
and $\rho_3$ and compact generator $\rho_2$, exactly as in the first
Killing form in eq.~\eqref{sl2killingforms}. Similarly, in the
compact case one gets $\tau_1 = i (E_+ + E_- )$, $\tau_2 = E_+ -
E_-$ and $\tau_3 = i H$, which are indeed all compact and lead to
the second Killing form in eq. \eqref{sl2killingforms}. Although in
the case of $\mathfrak{sl}(2,\mathbb{C})$ these are all the possible
real forms up to automorphisms, for larger Lie algebras there are
additional real forms.

Given a real form $\mathfrak{g}$, one defines the {\it Cartan involution} $\theta$ as an involution such that
\begin{equation}
 B_{\theta}(X,Y)= B(X,\theta Y)
\end{equation}
is negative definite. Clearly, in the case of the compact form
$\mathfrak{c}$ the Cartan involution is the identity, while in the
case of the split form $\mathfrak{s}$ the Cartan involution is such
that $\theta H_\alpha =-H_{\alpha}$ and $\theta E_{\alpha} = -
E_{-\alpha}$. In general, classifying the real forms of a given
complex Lie algebra corresponds to classifying all the possible
Cartan involutions. One can always define a basis of generators that
are eigenvectors of the Cartan involution $\theta$. Those with $+1$
eigenvalue are compact while those with $-1$ eigenvalue are
non-compact.  We call  $\mathfrak{t}$ the set of generators with
eigenvalue $+1$ and  $\mathfrak{p}$ the set of those with eigenvalue
$-1$, such that
  \begin{equation}
 \mathfrak{g}=\mathfrak{t}\oplus\mathfrak{p}\label{cartandec1} \quad .
\end{equation}
In particular, in the adjoint representation one has
 \begin{equation}
 {\rm ad } (\theta X ) = - ({\rm ad} X )^\dagger \quad ,
\end{equation}
so that for a compact generator, that is $X \in \mathfrak{t}$, one has that  ${\rm ad}X$ is anti-hermitian and thus has imaginary eigenvalues, while for a non-compact generator, that is  $X \in \mathfrak{p}$, one has that  ${\rm ad}X$  is hermitian and thus has real eigenvalues. In particular, the non-zero eigenvalues of the Cartan matrices in the adjoint ${\rm ad}H$ are the roots $\alpha (H)$, implying that one can classify the roots in the following way:
\begin{itemize}
 \item a root is a \textbf{\textit{real root}} if it takes real values, that is if it vanishes for $H \in \mathfrak{t}$;
 \item a root is an \textbf{\textit{imaginary root}} if it takes imaginary values, that is if it vanishes for $H \in \mathfrak{p}$;
 \item a root is a \textbf{\textit{complex root}} if it takes complex values and hence if it does not vanish  on either $\mathfrak{t}$ or $\mathfrak{p}$.
\end{itemize}
The $\theta$ involution on the Cartan generators induces a dual involution on the roots as
\begin{equation}
 \theta (\alpha(H))=\alpha(\theta (H)) \quad , \label{thetaalphaHalphathetaH}
\end{equation}
from which it follows that for a real root one has $\theta \alpha = -\alpha$ and for an imaginary root one has $\theta \alpha = \alpha$.  From eq.~\eqref{thetaalphaHalphathetaH} and the relation
$ \theta H_\alpha = H_{\theta \alpha}$,
 it also follows that
  \begin{equation}
    \langle\theta \alpha_{k},\alpha_{i}\rangle= \langle\alpha_{k},\theta \alpha_{i}\rangle
  \label{thetaalphaalpha}
\end{equation}
for any pair of roots.

The fact that the Cartan involution on the generators induces a dual
Cartan involution on the roots allows one to represent such an
involution on the Dynkin diagram of the Lie algebra. This can be
achieved in two different ways, leading to the so-called Vogan
diagrams and Tits-Satake diagrams. In this paper we will only be
interested in the latter because, as we will see in the next
section, there is a natural connection between real roots as
classified in terms of Tits-Satake diagrams and brane states in
supergravity theories. The difference between the two constructions
stems from the different choices that one can make for the Cartan
subalgebra. In the case of the Vogan diagrams one chooses the Cartan
subalgebra to be maximally compact, which leads to the absence of
real roots. In the case of Tits-Satake diagrams, instead, one makes
the opposite choice, that is one chooses the Cartan subalgebra to be
maximally non-compact.

A Tits-Satake diagram is a  ``decorated'' Dynkin diagram from which
one can deduce the whole structure of the real form, and in
particular the compact and non-compact generators. In order to do
this, an additional ingredient is required. We have already
mentioned that in order to construct a Tits-Satake diagram we have
to put as many Cartan generators as possible in the non-compact
part $\mathfrak{p}$. When this happens, that is when one chooses a Cartan
subalgebra that is maximally non-compact, there are no non-compact
generators associated to the imaginary roots. This means that if
$\theta \alpha =\alpha$, then this implies that $\theta E_{\alpha} =
E_{\alpha}$.

We now describe how to read all the relevant information from a Tits-Satake diagram. We divide all the simple roots in those that are fixed under $\theta$, that we denote with $\beta_n$, and the rest, that we denote with $\alpha_i$. The action of $\theta$ on the simple roots $\alpha_i$ is
\begin{equation}
\theta\alpha_{i}=-\alpha_{\pi(i)}+\sum_{n}a_{in}\beta_{n}\,,
\label{complexrootTitsSatake1}
\end{equation}
where $\pi(k)$ is an involutive ($\pi^{2}=1$) permutation of the
indices. The coefficients $a_{in}$ are determined by imposing
\begin{equation}
 \langle\alpha_{i}+\alpha_{\pi(i)},\beta_{m}\rangle=\sum_{n} a_{in}\langle\beta_{n},\beta_{m}\rangle \ , \label{thiseqdeterminesain}
\end{equation}
which follows from eq. \eqref{thetaalphaalpha} and the fact that the
simple roots $\beta_n$ are invariant under $\theta$. The Tits-Satake
diagram is then drawn from the corresponding Dynkin diagram with the
following additional rules:\\
\begin{wrapfigure}[7]{r,h!}{0.4\textwidth}
 \scalebox{0.5}{
\begin{pspicture}(0,-1.89)(10.41,2.85)
\definecolor{color79b}{RGB}{0,0,0}
\psline[linewidth=0.02cm](3.8,1.65)(6.2,1.65)
\pscircle[linewidth=0.02,dimen=outer,fillstyle=solid](5.0,1.65){0.4}
\psline[linewidth=0.02cm](3.8,4)(6.2,4)
\pscircle[linewidth=0.02,dimen=outer,fillstyle=solid,fillcolor=color79b](5.0,4){0.4}
\psline[linewidth=0.02cm,linestyle=dashed,dash=0.16cm 0.16cm](6.0,1.65)(7.0,1.65)
\psline[linewidth=0.02cm,linestyle=dashed,dash=0.16cm 0.16cm](3.0,1.65)(4.0,1.65)
\psline[linewidth=0.02cm,linestyle=dashed,dash=0.16cm 0.16cm](3.0,4)(4.0,4)
\psline[linewidth=0.02cm,linestyle=dashed,dash=0.16cm 0.16cm](6.0,4)(7.0,4)
\psline[linewidth=0.02cm](0.8,-2.55)(3.2,-2.55)
\pscircle[linewidth=0.02,dimen=outer,fillstyle=solid](2.0,-2.55){0.4}
\psline[linewidth=0.02cm,linestyle=dashed,dash=0.16cm 0.16cm](3.0,-2.55)(4.0,-2.55)
\psline[linewidth=0.02cm,linestyle=dashed,dash=0.16cm 0.16cm](0.0,-2.55)(1.0,-2.55)
\psline[linewidth=0.02cm](7.2,-2.55)(9.6,-2.55)
\pscircle[linewidth=0.02,dimen=outer,fillstyle=solid](8.4,-2.55){0.4}
\psline[linewidth=0.02cm,linestyle=dashed,dash=0.16cm 0.16cm](9.4,-2.55)(10.4,-2.55)
\psline[linewidth=0.02cm,linestyle=dashed,dash=0.16cm 0.16cm](6.4,-2.55)(7.4,-2.55)
\psbezier[linewidth=0.06,arrowsize=0.05291667cm 2.0,arrowlength=1.4,arrowinset=0.4]{<->}(2.1036007,-1.9986596)(2.1036007,-0.94999975)(4.113303,-0.45268014)(5.1178346,-0.42567003)(6.1223655,-0.39865994)(8.503601,-0.62723136)(8.533966,-1.99866)
\rput(5.0848436,0.795){\Large \textbf{\textit{complex}}  1-cycle ($\rm{mod} \ \beta \ {\rm roots}$)}
\rput(4.924844,3.145){\Large \textbf{\textit{imaginary}} (compact generator)}
\rput(5.1215625,-3.605){\Large \textbf{\textit{complex}} 2-cycle ($\rm{mod} \ \beta \ {\rm roots}$)}
\end{pspicture}
}
\end{wrapfigure}
\vspace{-1cm}
\begin{enumerate}
 \item to each root $\beta$ (imaginary simple root) one associates a black painted node
 \item to each simple root $\alpha_i$ such that $\pi (i) =i$ one associates an unpainted node

 \item for each two complex simple roots $\alpha_i$ and $\alpha_j$ such that $\pi (i) = j$ one draws an arrow joining the  two corresponding unpainted nodes.
\end{enumerate}
The behaviour of all the other roots under $\theta$ clearly follows
from the behaviour of the simple roots. This
means that from the Tits-Satake diagram one knows how the Cartan
involution acts on all the roots. In the case in which $\pi (i) =i$
and the node associated to the simple root $\alpha_i$ is not
connected to any painted node in the Tits-Satake diagram, then
clearly from eq.~\eqref{thiseqdeterminesain} it follows that $a_{in}
=0$ and thus $\theta \alpha_i =-\alpha_i$, which means that the root
is real.

\begin{figure}[h]
\centering
\scalebox{.5} 
{
\begin{pspicture}(0,-0.7601563)(4.8,0.80015624)
\definecolor{color383b}{rgb}{0.996078431372549,0.996078431372549,0.996078431372549}
\psline[linewidth=0.02cm](0.6000001,-0.34984365)(4.4,-0.34984365)
\pscircle[linewidth=0.02,dimen=outer,fillstyle=solid](4.4,-0.36015624){0.4}
\pscircle[linewidth=0.02,dimen=outer,fillstyle=solid,fillcolor=color383b](0.4,-0.36015624){0.4}
\rput(0.351875,0.45046875){\huge 1}
\rput(4.3620315,0.45046875){\huge 2}
\end{pspicture}
}

\caption{\label{sl3split} \sl The Tits-Satake diagram of $\mathfrak{sl}(3,\mathbb{R})$.}
\end{figure}
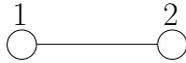

In order to better understand how the construction works explicitly,
and how one can easily translate the action of the Cartan involution
on the roots that one reads from the Tits-Satake diagram to the
Cartan involution on the generators, we consider the example of the
real forms of $\mathfrak{sl}(3,\mathbb{C})$. The Chevalley
generators of  $\mathfrak{sl}(3,\mathbb{C})$ in the $3 \times 3$
matrix representation are the Cartan generators
   \begin{equation}
 H_{\alpha_1}=\begin{pmatrix}
 1&0& 0\\
 0&-1& 0\\
 0 & 0&0 \\
\end{pmatrix}\quad
H_{\alpha_2 } =\begin{pmatrix}
 0&0& 0\\
 0&1 & 0\\
 0& 0&-1\\
\end{pmatrix}
\end{equation}
and the positive root generators
   \begin{equation}
 E_{\alpha_1}=\begin{pmatrix}
 0&1& 0\\
 0&0& 0\\
 0 & 0&0 \\
\end{pmatrix}\quad
E_{\alpha_2 } =\begin{pmatrix}
 0&0& 0\\
 0&0 & 1\\
 0& 0&0\\
\end{pmatrix}\quad E_{\alpha_1 + \alpha_2}  =
\begin{pmatrix}
 0&0& 1\\
 0&0 & 0\\
 0 & 0 &0\\
\end{pmatrix}
\end{equation}
while the negative root generators can be chosen as $E_{-\alpha} = (
E_{\alpha} )^\dagger$. The Tits-Satake diagram of the split form
 $\mathfrak{sl}(3,\mathbb{R})$ is just the $A_2$ Dynkin diagram in
Fig.~\ref{sl3split}:\,\footnote{The Tits-Satake diagram of the split
form by definition always coincides with the corresponding Dynkin
diagram.}

\begin{figure}[h]
\centering
\scalebox{.5} 
{
\begin{pspicture}(0,-0.7601563)(4.8,0.80015624)
\definecolor{color376b}{rgb}{0.00392156862745098,0.00392156862745098,0.00392156862745098}
\psline[linewidth=0.02cm](0.6000001,-0.34984365)(4.4,-0.34984365)
\pscircle[linewidth=0.02,dimen=outer,fillstyle=solid,fillcolor=color376b](4.4,-0.36015624){0.4}
\pscircle[linewidth=0.02,dimen=outer,fillstyle=solid,fillcolor=black](0.4,-0.36015624){0.4}
\rput(0.351875,0.45046875){\huge 1}
\rput(4.3620315,0.45046875){\huge 2}
\end{pspicture}
}

\caption{\label{su3compact} \sl The Tits-Satake diagram of $\mathfrak{su}(3)$.}
\end{figure}
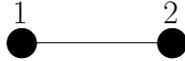

\noindent From this diagram it follows that
\begin{equation}
\theta \alpha_1 = -\alpha_1 \qquad \theta \alpha_2 = -\alpha_2 \quad ,
\end{equation}
implying that $\theta H_{\alpha_i} = -H_{\alpha_i}$ and $\theta
E_{\alpha_i} = -E_{-\alpha_i}$. This naturally leads to the compact
and non-compact generators of the split form as in
eq.~\eqref{generatorssplitform}. Similarly, the Tits-Satake diagram
for the compact $\mathfrak{su}(3)$ case is given in Fig.~\ref{su3compact}.
This leads to the following action of the Cartan
involution on the simple roots:
\begin{equation}
\theta \beta_1 = \beta_1 \qquad \theta \beta_2 = \beta_2 \quad .
\end{equation}

\noindent In terms of the generators, this means that $\theta H_{\beta_i} =
H_{\beta_i}$ and $\theta E_{\beta_i} = E_{\beta_i}$. The  last
relation is due to the fact that if a root is imaginary the
corresponding root generator must be compact. All the generators are
thus fixed under $\theta$, which means that they are all
anti-hermitian, i.e.~compact, and therefore the right basis of
$\theta$-fixed generators is as in
eq.~\eqref{generatorscompactform}.

The most interesting case is the third real form
$\mathfrak{su}(2,1)$, whose Tits-Satake diagram is given in Fig.~\ref{su21titssatakediagram}:
\begin{figure}[h]
\centering
\scalebox{.5} 
{
\begin{pspicture}(0,-1.4146875)(4.7999997,1.4346875)
\psline[linewidth=0.02cm](0.38999996,0.3013125)(4.59,0.3013125)
\pscircle[linewidth=0.02,dimen=outer,fillstyle=solid](4.4,0.300625){0.4}
\pscircle[linewidth=0.02,dimen=outer,fillstyle=solid](0.4,0.3006251){0.4}
\rput(0.2821875,1.085){\huge 1}
\rput(4.2923436,1.085){\huge 2}
\psbezier[linewidth=0.04,arrowsize=0.05291667cm 2.0,arrowlength=1.4,arrowinset=0.4]{<->}(0.38999996,-0.3746875)(0.38999996,-1.1746875)(1.65,-1.3946875)(2.31,-1.3746876)(2.97,-1.3546875)(4.35,-1.3146875)(4.39,-0.3746875)
\end{pspicture}
}

\caption{\label{su21titssatakediagram} \sl The Tits-Satake diagram of $\mathfrak{su}(2,1)$.}
\end{figure}
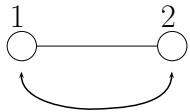

\noindent From this diagram one reads that
 \begin{equation}
 \theta \alpha_1 = - \alpha_2 \quad ,
 \end{equation}
which in terms of the Chevalley generators means
  \begin{equation}
  \theta H_{\alpha_1} = - H_{\alpha_2} \qquad \theta E_{\alpha_1} =- E_{-\alpha_2} \qquad \theta E_{\alpha_2} = - E_{-\alpha_1} \quad .
  \end{equation}
From the relation $[E_{\alpha_1} , E_{\alpha_2} ] = E_{\alpha_1 + \alpha_2}$ one also derives
 \begin{equation}
 \theta E_{\alpha_1 + \alpha_2 } = E_{-\alpha_1 -\alpha_2}\,,
 \end{equation}
 which has opposite sign with respect to the split case because in this case the Cartan involution exchanges the simple roots. From these rules and the fact that the generators in $\mathfrak{t}$ are anti-hermitian and those in $\mathfrak{p}$ are hermitian one can derive the whole set of compact and non-compact generators. The complete list of compact and non-compact generators that one gets from the Tits-Satake diagrams of $\mathfrak{sl}(3,\mathbb{C})$ is given in Table \ref{generatorsofsl3allrealforms} for all the real forms.

\begin{table}[t!]
\begin{center}
\begin{tabular}{|c|c|c|c|}
\hline \rule[-1mm]{0mm}{1mm} real form & generators & $\mathfrak{t}$ & $\mathfrak{p}$\\
\hline \hline \rule[-1mm]{0mm}{1mm}  $\mathfrak{sl}(3,\mathbb{R})$ & Cartan &  & $H_{\alpha_1}$\\
 \rule[-1mm]{0mm}{1mm} & &   & $H_{\alpha_2} $\\
\cline{2-4} \rule[-1mm]{0mm}{1mm} &  root  &  $E_{\alpha_1} - E_{-\alpha_1} $ & $ E_{\alpha_1} + E_{-\alpha_1}  $\\
 \rule[-1mm]{0mm}{1mm} & generators &  $E_{\alpha_2} - E_{-\alpha_2} $ & $ E_{\alpha_2} + E_{-\alpha_2}  $\\
  \rule[-1mm]{0mm}{1mm} &  &  $E_{\alpha_1+ \alpha_2} - E_{-\alpha_1 -\alpha_2} $ & $ E_{\alpha_1 + \alpha_2} + E_{-\alpha_1 -\alpha_2}  $\\
\hline   \hline \rule[-1mm]{0mm}{1mm}  $\mathfrak{su}(3)$ & Cartan &  $iH_{\alpha_1}$& \\
   \rule[-1mm]{0mm}{1mm}  &  &  $iH_{\alpha_2}$& \\
   \cline{2-4} \rule[-1mm]{0mm}{1mm} &  root  &  $E_{\alpha_1} - E_{-\alpha_1} $ & \\
   \rule[-1mm]{0mm}{1mm} & generators  &  $i (E_{\alpha_1} + E_{-\alpha_1} ) $ & \\
 \rule[-1mm]{0mm}{1mm} &   &  $E_{\alpha_2} - E_{-\alpha_2} $ & \\
   \rule[-1mm]{0mm}{1mm} &  &  $i (E_{\alpha_2} + E_{-\alpha_2} ) $ & \\
     \rule[-1mm]{0mm}{1mm} &  &  $E_{\alpha_1+ \alpha_2} - E_{-\alpha_1 -\alpha_2} $ & \\
       \rule[-1mm]{0mm}{1mm} &  &  $i (E_{\alpha_1+ \alpha_2} + E_{-\alpha_1 -\alpha_2} ) $ & \\
  \hline  \hline \rule[-1mm]{0mm}{1mm}  $\mathfrak{su}(2,1)$ & Cartan &  $i(H_{\alpha_1} - H_{\alpha_2})$& $H_{\alpha_1} + H_{\alpha_2}$\\
   \cline{2-4} \rule[-1mm]{0mm}{1mm} &  root  &  $E_{\alpha_1} + E_{\alpha_2} - E_{-\alpha_1} - E_{-\alpha_2} $ & $E_{\alpha_1} + E_{\alpha_2} + E_{-\alpha_1} + E_{-\alpha_2} $\\
   \rule[-1mm]{0mm}{1mm} &  generators &  $i (E_{\alpha_1} - E_{\alpha_2} + E_{-\alpha_1} - E_{-\alpha_2} ) $ & $i (E_{\alpha_1} - E_{\alpha_2} - E_{-\alpha_1} + E_{-\alpha_2} ) $\\
       \rule[-1mm]{0mm}{1mm} &  &  $i (E_{\alpha_1+ \alpha_2} + E_{-\alpha_1 -\alpha_2} ) $ & $i (E_{\alpha_1+ \alpha_2} - E_{-\alpha_1 -\alpha_2} ) $\\
   \hline
\end{tabular}
\caption{ \label{generatorsofsl3allrealforms} \sl
The generators of the three different real forms of  $\mathfrak{sl}(3,\mathbb{C})$ as obtained from the corresponding Tits-Satake diagrams.
}
\end{center}
\end{table}

In the next section we will show how the classification of single
1/2-BPS branes in the heterotic string compactified on a torus can be
naturally rephrased in terms of reality properties of the roots and
the weights of the different real forms of $\mathfrak{so}(p)$ as
derived from the corresponding Tits-Satake diagrams. In section 4 we
will  generalise this to any real form. In particular, we will show
how this  naturally leads to a classification of single 1/2-BPS branes
in theories with 8 supercharges in which the scalars parametrise
coset manifolds.

\section{Branes of the heterotic string on a torus}

The heterotic theory compactified on a torus $T^d$, with generic Wilson lines so that the gauge group is
$\text{U}(1)^{2d+16}$, has a low-energy action possessing 16 supersymmetries and describing a supergravity multiplet coupled to $d+16$ vector multiplets. The global symmetry of this low-energy action is $\text{SO}(d,d+16)$. In \cite{Bergshoeff:2012jb} the half-supersymmetric single branes of this theory have been classified by analysing their Wess-Zumino terms. The result of that analysis is that, given a $p$-brane charge in a specific $\text{SO}(d,d+16)$ representation,  not all its components  correspond to single 1/2-BPS branes. More specifically, given a $p$-brane charge
in a tensor representation  of the duality group
$\text{SO}(d,d+n)$, we split the $2d+n$ duality indices into $2d$ `lightlike' indices  $i  \pm\ (i=1,...,d)$
and the remaining $n$ `spacelike' indices.
A given component of the charge corresponds to a
half-supersymmetric $p$-brane if one of the following situations apply:

\begin{description}

\item{1.} {\sl antisymmetric tensor representations}\,:\
the antisymmetric indices
are of the form $i\pm j\pm k \pm \dots $ with $i\,,j\,,k\,,\dots  $ all different.

\item{2.} {\sl mixed-symmetry tensor representations}\,: For a charge $T_{ A_1\dots A_m,B_1\dots B_n}\ (m>n)$ in a representation corresponding to a 2-column Young ta\-bleaux of heights $m$ and $n$, on top of the previous rule the following  additional rule applies:
each of the antisymmetric $B$  indices of $T_{A_1\dots A_m,B_1\dots B_n}$
has to be parallel to one of the antisymmetric $A$ indices.

\end{description}

The `light-cone' rules listed above are a natural extension of the light-cone rules obtained in \cite{Bergshoeff:2011zk,Bergshoeff:2011ee,Bergshoeff:2012ex} in the context of the classification of branes of the maximally supersymmetric theories in any dimension $10-d$ with respect to the T-duality group $\text{SO}(d,d)$. The difference with the half-maximal case is that clearly for  $\text{SO}(d,d)$ {\it all} the $2d$ vector indices can be written as lightlike indices  $i\pm\ (i=1,...,d)$. In group-theoretical terms, the difference between the maximal and half-maximal case is that the T-duality symmetry of the maximal theory is maximally non-compact, while in the half-maximal case it is in a different real form of the orthogonal group. The general analysis of \cite{Bergshoeff:2013sxa}, which applies to both U-duality and T-duality representations of the maximal theories, gives a simple group-theoretic explanation for the light-cone rule: all the branes correspond to the longest weights of the
representation, and in the T-duality case these longest weights correspond to the components selected by the light-cone rule. This simple result is based on the fact that the symmetry groups of the maximal theories are always maximally non-compact. In the half-maximal case, the T-duality group is not maximally non-compact and the correspondence between branes and weights has to be refined. This is the aim of this section.\,\footnote{Some of the results in this section were already contained in the first appendix of \cite{Bergshoeff:2012jb}.}
More precisely, our strategy consists in identifying the Chevalley generators for the orthogonal groups, so that, by looking at the Cartan involution on the simple roots as derived from the Tits-Satake diagram, we can show that the real roots, that is those such that $\theta \alpha = -  \alpha$,
precisely correspond to the generators whose components satisfy the light-cone rules.  We will then
show how this can be generalised to any representation.

\begin{figure}[h]
 \centering
\scalebox{0.5} 
{
\begin{pspicture}(0,-2.935)(18.720625,2.895)
\psline[linewidth=0.02cm](0.2,0.505)(4.2,0.505)
\psline[linewidth=0.02cm](14.4,0.505)(18.0,2.505)
\psline[linewidth=0.02cm](14.4,0.505)(18.0,-1.495)
\psline[linewidth=0.02cm](13.2,0.505)(14.2,0.505)
\psline[linewidth=0.02cm](8.6,0.505)(9.6,0.505)
\psline[linewidth=0.02cm](4.6,0.505)(8.4,0.505)
\pscircle[linewidth=0.02,dimen=outer,fillstyle=solid](14.4,0.4946875){0.4}
\pscircle[linewidth=0.02,dimen=outer,fillstyle=solid](4.4,0.4946875){0.4}
\pscircle[linewidth=0.02,dimen=outer,fillstyle=solid](8.4,0.4946875){0.4}
\pscircle[linewidth=0.02,dimen=outer,fillstyle=solid](17.9,2.495){0.4}
\psline[linewidth=0.02cm,linestyle=dashed,dash=0.16cm 0.16cm](9.4,0.505)(13.2,0.505)
\pscircle[linewidth=0.02,dimen=outer,fillstyle=solid](17.9,-1.505){0.4}
\pscircle[linewidth=0.02,dimen=outer,fillstyle=solid](0.4,0.4946875){0.4}
\rput(17.960312,1.315){\huge $n$}
\rput(14.505938,-0.685){\huge $n$-2}
\rput(18.108593,-2.685){\huge $n$-1}
\rput(0.351875,-0.685){\huge 1}
\rput(4.3620315,-0.685){\huge 2}
\rput(8.367969,-0.685){\huge 3}
\end{pspicture}
}

\caption{\sl The $\mathfrak{so}(n,n)$ Tits-Satake diagram.\label{sonntitssatake}}
\end{figure}
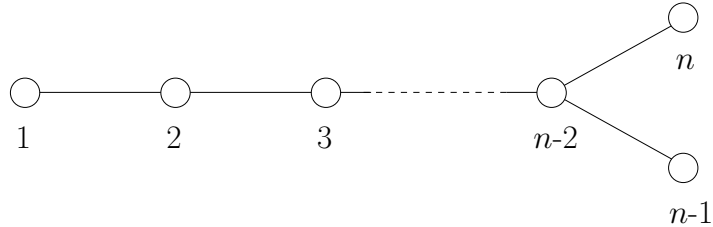

We first apply the results of \cite{Bergshoeff:2013sxa} to the T-duality group $\text{SO}(n,n)$ of the maximally supersymmetric theory in $10-n$ dimensions. We denote the nodes of the Tits-Satake diagram of $\text{SO}(n,n)$ as in
Fig.~\ref{sonntitssatake}.
 Working in light-cone coordinates  $I = i\pm$, $i=1,...,n$, with invariant metric
  \begin{equation}
  {\eta}_{i+\  j-} =   {\eta}_{i-\  j+} = \delta_{ij} \quad , \quad   {\eta}_{i+\  j+} =   {\eta}_{i- \ j-}  =0 \quad ,\label{metriclight-cone}
\end{equation}
the Chevalley generators acting on the vector representation  can be chosen to be
  \begin{eqnarray}
 & &   H_{\alpha_i} =  {\bf e}_{i+}{}^{i+} - {\bf e}_{i-}{}^{i-}  -  {\bf e}_{(i+1)+}{}^{(i+1)+} + {\bf e}_{(i+1)-}{}^{(i+1)-} \qquad  \quad  \ \ \ \quad  i= 1,...,n-1\nonumber \\
  & &   H_{\alpha_n} =  {\bf e}_{(n-1)+}{}^{(n-1)+} - {\bf e}_{(n-1)-}{}^{(n-1)-}  +  {\bf e}_{n+}{}^{n+} - {\bf e}_{n-}{}^{n-} \nonumber \\
   & & E_{\alpha_i} = {\bf e}_{i+}{}^{(i+1)+} - {\bf e}_{(i+1)-}{}^{i-} \qquad \qquad \quad \	\qquad \qquad \qquad \quad  \qquad \ \  i= 1,...,n-1\nonumber \\
   & & E_{\alpha_n} =  {\bf e}_{(n-1)+}{}^{n-} - {\bf e}_{n+}{}^{(n-1)-} \nonumber \\
  & & E_{- \alpha_i} =  {\bf e}_{(i+1)+}{}^{i+}- {\bf e}_{i-}{}^{(i+1)-} \  \qquad \qquad \qquad \qquad\qquad \quad \qquad \quad i= 1,...,n-1\nonumber \\
   & & E_{-\alpha_n} =   {\bf e}_{n-}{}^{(n-1)+} -{\bf e}_{(n-1)-}{}^{n+} \quad .\label{chevalleygeneratorsdn}
\end{eqnarray}
Here  we denote with ${\bf e}_{I}{}^{J}$ the $2 n \times 2n $ matrix that has entry 1 on the $I$th row and $J$th column and 0 otherwise.
From eq.~\eqref{chevalleygeneratorsdn} we see that lowering the column index of the generators by means of the metric
of eq.~\eqref{metriclight-cone}, the root generators precisely correspond to the charges $T_{i\pm \ j\pm}$, $i \neq j$ in the antisymmetric tensor representation,  as resulting from the light-cone rules. In the case of the split real form
$\text{SO}(n,n)$ the Cartan involution acts on all the roots as $\theta \alpha = -  \alpha$, so that all the roots are real. In particular, the simple roots are associated to the charges
\begin{equation}
  \alpha_1 \rightarrow T_{1+ \ 2-} \qquad \alpha_2 \rightarrow T_{2+\  3-} \qquad ...\quad \alpha_{n-1} \rightarrow T_{(n-1)+\  n-} \qquad \alpha_n \rightarrow T_{(n-1) +\  n+ } \quad  . \label{rootsandchargessonn}
\end{equation}
Similarly, all the weights of any representation are real, and in particular for any representation associated to brane states in the maximal supergravity theories we can identify the 1/2-BPS branes as resulting from the light-cone rule to the longest weights of the representation. This is exactly the application of the results of \cite{Bergshoeff:2013sxa} to the representations of the T-duality group. For instance, for the ${2d \choose 3}$-dimensional  representation with three antisymmetric indices, the half-supersymmmetric charges are $T_{i\pm \ j\pm \ k\pm}$, which makes a total of ${d \choose 3} \times 2^3$ components, associated to all the longest weights of the representation.

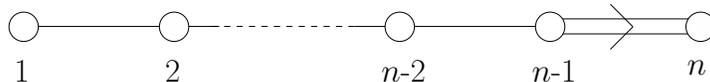
\begin{figure}[h]
\centering
\scalebox{0.5} 
{
\begin{pspicture}(0,-1.04)(18.8,1.01)
\definecolor{color346b}{rgb}{0.996078431372549,0.996078431372549,0.996078431372549}
\psline[linewidth=0.02cm](10.6,0.410313)(14.4,0.410313)
\psline[linewidth=0.02cm](14.6,0.210313)(18.4,0.210313)
\psline[linewidth=0.02cm](9.2,0.410313)(10.2,0.410313)
\psline[linewidth=0.02cm](4.6,0.410313)(5.6,0.410313)
\psline[linewidth=0.02cm](0.6,0.410313)(4.4,0.410313)
\pscircle[linewidth=0.02,dimen=outer,fillstyle=solid,fillcolor=color346b](4.4,0.4){0.4}
\psline[linewidth=0.02cm,linestyle=dashed,dash=0.16cm 0.16cm](5.4,0.410313)(9.2,0.410313)
\rput(0.351875,-0.79){\huge 1}
\rput(4.3620315,-0.79){\huge 2}
\psline[linewidth=0.02cm](14.4,0.610313)(18.2,0.610313)
\pscircle[linewidth=0.02,dimen=outer,fillstyle=solid,fillcolor=color346b](18.4,0.4){0.4}
\psline[linewidth=0.02cm](16.6,0.4)(16.0,1.0)
\psline[linewidth=0.02cm](16.6,0.4)(16.0,-0.2)
\rput(18.300312,-0.69){\huge $n$}
\pscircle[linewidth=0.02,dimen=outer,fillstyle=solid](14.4,0.4){0.4}
\rput(14.508594,-0.79){\huge $n$-1}
\pscircle[linewidth=0.02,dimen=outer,fillstyle=solid](0.4,0.4){0.4}
\pscircle[linewidth=0.02,dimen=outer,fillstyle=solid](10.4,0.4){0.4}
\rput(10.505938,-0.79){\huge $n$-2}
\end{pspicture}
}

\caption{\sl The $\mathfrak{so}(n,n+1)$ Tits-Satake diagram.\label{sonn+1titssatake}}
\end{figure}

One can  perform the same analysis for the maximally non-compact group $\text{SO}(n,n+1)$, whose Tits-Satake diagram is given in Fig.~\ref{sonn+1titssatake}.
In this case, we split the $2n+1$ coordinates in $2n$ lightlike coordinates and one space coordinate.  Denoting with 1 the single space index and again with  $i\pm$, $i=1,...,n$ the lightlike directions, the correspondence between roots and charges is
\begin{equation}
  \alpha_1 \rightarrow T_{1+ \ 2-} \qquad \alpha_2 \rightarrow T_{2+ \ 3-} \qquad ...\quad \alpha_{n-1} \rightarrow T_{(n-1)+\ n-} \qquad \alpha_n \rightarrow T_{ n + \ 1 } \quad , \label{rootgeneratorSOnn+1}
\end{equation}
where the last root $\alpha_n$  is the short simple root. In this case the symmetric  invariant metric is as before with the addition of ${\eta}_{ 1 \ 1 } = 1$ that lowers the index in the spacelike direction.
For any $n$, this algebra contains $n$ short positive roots, which are associated to the charges $T_{i+ \ 1}$, and $n$ short negative roots, associated to the charges $T_{i- \ 1}$.  All the other roots are long and are associated to the 1/2-supersymmetric charges $T_{i\pm\  j\pm}$, $i \neq j$, as selected by the light-cone rule. As in the $\text{SO}(d,d)$ case, this real form is maximally non-compact and thus all roots are real, and  the light-cone rule selects the real longest roots. Similarly, for other representations the light-cone rule selects the real longest weights exactly as before.

We now want to consider different real forms of the orthogonal group.  What we want to show is that for any real form, the 1/2-supersymmetric branes that one obtains from the light-cone rules are exactly the real longest roots and real longest weights that result from the Cartan involution acting on the simple roots as dictated by the corresponding Tits-Satake diagram.  A crucial ingredient in the construction is the identification of  the restricted-root subalgebra as the maximally non-compact algebra which has as simple roots the so-called `restricted' simple roots,
\begin{equation}
\alpha_R = \frac{1}{2} (\alpha - \theta \alpha )
 \quad .
 \end{equation}
Clearly, the restricted simple root coincides with the simple root if the latter is real.

\begin{figure}[h]
 \centering
\scalebox{0.5} 
{
\begin{pspicture}(0,-2.935)(20.62,2.895)
\psline[linewidth=0.02cm](0.2,0.505)(4.2,0.505)
\psline[linewidth=0.02cm](14.4,0.505)(18.0,2.505)
\psline[linewidth=0.02cm](14.4,0.505)(18.0,-1.495)
\psline[linewidth=0.02cm](13.2,0.505)(14.2,0.505)
\psline[linewidth=0.02cm](8.6,0.505)(9.6,0.505)
\psline[linewidth=0.02cm](4.6,0.505)(8.4,0.505)
\pscircle[linewidth=0.02,dimen=outer,fillstyle=solid](14.4,0.4946875){0.4}
\pscircle[linewidth=0.02,dimen=outer,fillstyle=solid](4.4,0.4946875){0.4}
\pscircle[linewidth=0.02,dimen=outer,fillstyle=solid](8.4,0.4946875){0.4}
\pscircle[linewidth=0.02,dimen=outer,fillstyle=solid](17.9,2.495){0.4}
\psline[linewidth=0.02cm,linestyle=dashed,dash=0.16cm 0.16cm](9.4,0.505)(13.2,0.505)
\pscircle[linewidth=0.02,dimen=outer,fillstyle=solid](17.9,-1.505){0.4}
\pscircle[linewidth=0.02,dimen=outer,fillstyle=solid](0.4,0.4946875){0.4}
\rput{-90.0}(18.095,19.105){\psarc[linewidth=0.04,arrowsize=0.05291667cm 4.0,arrowlength=1.4,arrowinset=0.2]{<->}(18.6,0.505){2.0}{0.0}{180.0}}
\rput(17.760313,1.315){\huge $n$}
\rput(17.908594,-2.685){\huge $n$-1}
\rput(14.505938,-0.685){\huge $n$-2}
\rput(0.351875,-0.685){\huge 1}
\rput(4.3620315,-0.685){\huge 2}
\rput(8.367969,-0.685){\huge 3}
\end{pspicture}
}

\caption{\sl The $\mathfrak{so}(n-1,n+1)$ Tits-Satake diagram. \label{son-1n+1titssatake}}
\end{figure}
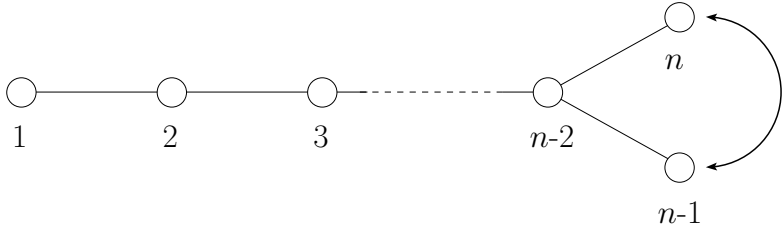

The real form $\text{SO}(n-1,n+1)$ of $\text{D}_n$ corresponds to the Tits-Satake diagram in
Fig.~\ref{son-1n+1titssatake}, from which we read the Cartan involution
  \begin{equation}
  \theta \alpha_i = -\alpha_i \qquad   i = 1, ..., n-2 \qquad \quad \theta \alpha_{n-1} =-\alpha_n \quad .
  \end{equation}
The restricted roots are
  \begin{equation}
  (\alpha_i )_R = \alpha_i \quad (i =1,...,n-2) \qquad (\alpha_{n-1})_R = (\alpha_{n})_R  = \frac{1}{2} [ \alpha_{n-1} + \alpha_n ] \quad ,\label{restrictedrootsson-1n+1}
  \end{equation}
which are the simple roots of $\text{B}_{n-1}$, which means that the restricted-root subalgebra of
$\mathfrak{so}(n-1,n+1)$  is  $\mathfrak{so}(n-1,n)$.  In particular, the last root in \eqref{restrictedrootsson-1n+1}
is the short root of $\text{B}_{n-1}$ and it has multiplicity two because $\alpha_{n-1}$ and $\alpha_n$ are exchanged under $\theta$.
If we now identify the last restricted root with the light-cone charges using eq.~\eqref{rootsandchargessonn}, we get
\begin{equation}
(\alpha_{n-1} )_R \rightarrow \frac{1}{2} [ T_{(n-1)+ \ n-} + T_{(n-1)+ \ n+ } ]\quad .
\end{equation}
We recognise this as the charge $T_{(n-1)+ \ 1}$ which is the generator associated to the short simple root of  $\text{SO}(n-1,n)$ (see eq.~\eqref{rootgeneratorSOnn+1}~identifying the roots with the charges for $\text{SO}(n,n+1)$). This shows that the light-cone rules reproduce exactly the group theory analysis in this case. The roots associated to the 1/2-supersymmetric branes, which are the real roots, can also be seen as the longest roots of  the restricted-root subalgebra $\mathfrak{so}(n-1,n)$. The same applies to all the other representations: the weights that correspond to 1/2-supersymmetric branes are real longest weights. These can be obtained from the restricted roots as follows. One takes the highest weight of the representation written as a linear combination of simple roots and projects it onto the restricted roots using
eq.~\eqref{restrictedrootsson-1n+1}. This projected weight is the highest weight of the corresponding representation of $\text{SO}(n-1,n)$. If the restricted highest weight coincides with the highest weight (i.e.~the highest weight is real), that weight is associated to a brane and so are all the restricted weights of the same length. Otherwise there are no branes.

\begin{figure}[h]
 \centering
\scalebox{0.5} 
{
\begin{pspicture}(0,-2.935)(24.520624,2.895)
\definecolor{color889b}{rgb}{0.996078431372549,0.996078431372549,0.996078431372549}
\psline[linewidth=0.02cm](9.2,0.505)(10.2,0.505)
\psline[linewidth=0.02cm](20.4,0.505)(24.0,2.505)
\psline[linewidth=0.02cm](20.4,0.505)(24.0,-1.495)
\psline[linewidth=0.02cm](19.2,0.505)(20.2,0.505)
\psline[linewidth=0.02cm](14.6,0.505)(15.6,0.505)
\psline[linewidth=0.02cm](10.6,0.505)(14.4,0.505)
\pscircle[linewidth=0.02,dimen=outer,fillstyle=solid,fillcolor=black](20.4,0.4946875){0.4}
\pscircle[linewidth=0.02,dimen=outer,fillstyle=solid](10.4,0.4946875){0.4}
\pscircle[linewidth=0.02,dimen=outer,fillstyle=solid,fillcolor=black](14.4,0.4946875){0.4}
\pscircle[linewidth=0.02,dimen=outer,fillstyle=solid,fillcolor=black](23.9,2.495){0.4}
\psline[linewidth=0.02cm,linestyle=dashed,dash=0.16cm 0.16cm](15.4,0.505)(19.2,0.505)
\pscircle[linewidth=0.02,dimen=outer,fillstyle=solid,fillcolor=black](23.9,-1.505){0.4}
\psline[linewidth=0.02cm](4.6,0.505)(5.6,0.505)
\psline[linewidth=0.02cm](0.6,0.505)(4.4,0.505)
\pscircle[linewidth=0.02,dimen=outer,fillstyle=solid](0.4,0.4946875){0.4}
\pscircle[linewidth=0.02,dimen=outer,fillstyle=solid,fillcolor=color889b](4.4,0.4946875){0.4}
\psline[linewidth=0.02cm,linestyle=dashed,dash=0.16cm 0.16cm](5.4,0.505)(9.2,0.505)
\rput(0.351875,-0.885){\huge 1}
\rput(4.3620315,-0.885){\huge 2}
\rput(10.37125,-0.885){\huge $p$}
\rput(14.468594,-0.885){\huge $p$+1}
\rput(20.305937,-0.885){\huge $n$-2}
\rput(23.908594,-2.685){\huge $n$-1}
\rput(23.960312,1.315){\huge $n$}
\end{pspicture}
}

\caption{\sl The $\mathfrak{so}(p,2n-p)$ Tits-Satake diagram for $p<n-1$. The nodes between 2 and $p$ are unpainted, while the nodes between $p+1$ and $n-2$ are painted. \label{sop2n-ptitssatake}}
\end{figure}
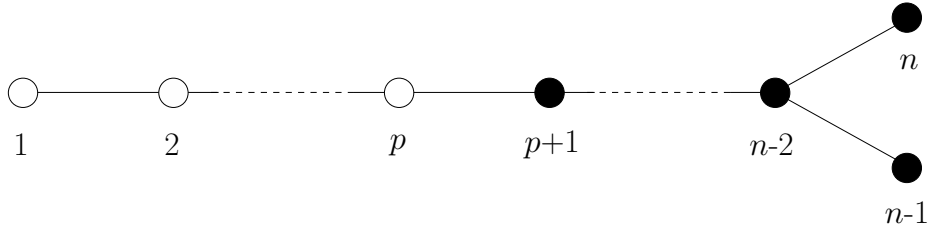

\begin{figure}[h]
\centering
\scalebox{0.5} 
{
\begin{pspicture}(0,-0.99140626)(25.247187,0.96140623)
\psline[linewidth=0.02cm](9.2,0.36171925)(10.2,0.36171925)
\psline[linewidth=0.02cm](20.6,0.16171925)(24.4,0.16171925)
\psline[linewidth=0.02cm](19.2,0.36171925)(20.2,0.36171925)
\psline[linewidth=0.02cm](14.6,0.36171925)(15.6,0.36171925)
\psline[linewidth=0.02cm](10.6,0.36171925)(14.4,0.36171925)
\pscircle[linewidth=0.02,dimen=outer,fillstyle=solid](10.4,0.35140625){0.4}
\pscircle[linewidth=0.02,dimen=outer,fillstyle=solid,fillcolor=black](14.4,0.35140625){0.4}
\psline[linewidth=0.02cm,linestyle=dashed,dash=0.16cm 0.16cm](15.4,0.36171925)(19.2,0.36171925)
\rput(10.97125,-0.43859375){\huge $p$}
\rput(15.468594,-0.43859375){\huge $p$+1}
\psline[linewidth=0.02cm](20.4,0.56171924)(24.2,0.56171924)
\pscircle[linewidth=0.02,dimen=outer,fillstyle=solid,fillcolor=black](20.4,0.35140625){0.4}
\pscircle[linewidth=0.02,dimen=outer,fillstyle=solid,fillcolor=black](24.4,0.35140625){0.4}
\psline[linewidth=0.02cm](22.6,0.35140625)(22.0,0.95140624)
\psline[linewidth=0.02cm](22.6,0.35140625)(22.0,-0.24859375)
\rput(20.554531,-0.63859373){\huge $n$-1}
\rput(24.54453,-0.53859377){\huge $n$}
\psline[linewidth=0.02cm](4.6,0.36171925)(5.6,0.36171925)
\psline[linewidth=0.02cm](0.6,0.36171925)(4.4,0.36171925)
\pscircle[linewidth=0.02,dimen=outer,fillstyle=solid](0.4,0.35140625){0.4}
\pscircle[linewidth=0.02,dimen=outer,fillstyle=solid](4.4,0.35140625){0.4}
\psline[linewidth=0.02cm,linestyle=dashed,dash=0.16cm 0.16cm](5.4,0.36171925)(9.2,0.36171925)
\rput(0.951875,-0.43859375){\huge 1}
\rput(4.9620314,-0.43859375){\huge 2}
\end{pspicture}
}
\caption{\sl The $\mathfrak{so}(p,2n-p+1)$ Tits-Satake diagram. The nodes between 2 and $p$ are unpainted, while the nodes between $p+1$ and $n-1$ are painted.\label{sop2n-p+1titssatake}}
\end{figure}
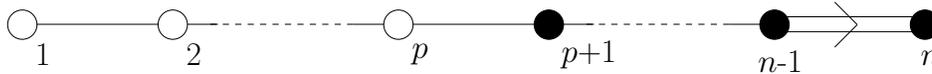

This can be generalised to all the other real forms. If the $\text{D}_n$ Tits-Satake diagram has $n-p$ painted nodes as in the  diagram of Fig.~\ref{sop2n-ptitssatake}, this corresponds to the real form  $\text{SO}(p, 2n-p)$. The painted roots, that are those fixed under $\theta$, are $\beta_{p+1},...,\beta_n$.
The Cartan involution acts on the other roots as
 \begin{eqnarray}
 & & \theta \alpha_i  = -\alpha_i \qquad \quad \quad (i = 1,...,p-1)\,, \nonumber \\
 & & \theta\alpha_{p} =-\alpha_{p} - 2\beta_{p+1}- ...-2\beta_{n-2} -\beta_{n-1} -\beta_n  \quad .
 \end{eqnarray}
From this we derive the restricted roots, which are
\begin{eqnarray}
& &  (\alpha_i )_R = \alpha_i \quad \qquad \quad (i =1,...,p-1)\,, \nonumber \\
& & (\alpha_{p} )_R = \alpha_{p} + \beta_{p+1} +...+\beta_{n-2} + \frac{1}{2}  [ \beta_{n-1} +\beta_n ] \quad . \label{restrictedrootsgenericp}
\end{eqnarray}
These are the simple roots of $\text{B}_{p}$, and thus the restricted-root subalgebra is $\mathfrak{so}(p,p+1)$. The last restricted root is the short root, and the corresponding charge can by identified using eq.~\eqref{rootsandchargessonn}. The result is
  \begin{equation}
  (\alpha_{p} )_R \rightarrow \frac{1}{2} [ T_{p+ \ (p+1) - } +  T_{p+ \ (p+1) + } ]  \label{shortrootofBp}
  \quad .
  \end{equation}
  Indeed, the combination of  simple roots that occurs in the second of
eqs.~\eqref{restrictedrootsgenericp} is half the sum of two roots. One is
the simple root $\alpha_{p}$, corresponding to the charge $T_{p+ \ (p+1) - }$ which is the first charge on the right-hand side of eq.~\eqref{shortrootofBp}, while  the other is the sum $\alpha_{p} + 2 \beta_{p+1} +...+2\beta_{n-2} +  \beta_{n-1} +\beta_n$,  which is a simple root of $\text{D}_n$ and in the split form it is associated to the charge  $T_{p+ \ (p+1) + }$. We recognise in eq.~\eqref{shortrootofBp} the charge $T_{p+ \ 1 }$ which is not a 1/2-supersymmetric charge because of the light-cone rule. The roots that are associated to the supersymmetric branes as given by the light-cone rule are the real ones, that are the longest roots of the maximally non-compact algebra. The analysis for the other representations works exactly as in the previous case. Moreover, the same result is obtained if one considers any real form of the algebra $\text{B}_n$, whose Tits-Satake diagram is given in Fig.~\ref{sop2n-p+1titssatake}.

In the next section we will apply these results to the theories that possess eight supercharges and whose scalars parametrise a symmetric manifold. These theories, that exist in dimension six and lower, have global symmetries that are groups in various real forms. We will conjecture that the 1/2-supersymmetric branes of these theories can be obtained by requiring that the charges are associated to the real longest weights of the representation, as deduced from the Tits-Satake diagram of the real form.

\section{Branes in theories with eight supercharges}

In the previous sections we have seen how the half-supersymmetric branes of maximal and half-maximal theories are associated to the components of the representation of the internal symmetry corresponding to the real longest  weights, with the reality properties of the roots as resulting from the Tits-Satake diagram of the symmetry group.
This can be naturally extended to classify branes in theories with lower supersymmetry that have scalars that describe coset manifolds. In this section we will perform this classification for theories with eight supercharges, obtaining the single 1/2-BPS branes of these theories. These branes preserve four supercharges on their worldvolume which implies that the worldvolume can be at most four-dimensional.

Theories with eight supercharges exist in six dimensions and below. The bosonic fields of the supersymmetric multiplets in six dimensions are the metric and a selfdual 2-form in the gravity multiplet,  an anti-selfdual 2-form and a scalar in the tensor multiplet, a gauge vector in  the vector multiplet and, finally, four scalars in the hypermultiplet. The latter multiplet has the same field content in any dimension below six. In five dimensions the gravity multiplet contains the metric and a vector and the vector multiplet describes a vector and a scalar. In four dimensions the gravity multiplet describes the metric and a vector as in five dimensions, while the vector multiplet contains a vector and two scalars. Finally, in three dimensions only scalars propagate. Therefore, as far as propagating degrees of freedom are concerned, one only has hypermultiplets.

In the first subsection we will consider the branes of the six-dimensional theory. In subsection 4.2 we will
extend the analysis to  any dimension for those theories that give rise to symmetric theories upon reduction to three dimensions. Such theories do not contain hypermultiplets in dimensions higher than three. Finally, in subsection 4.3 we will show how the branes in the hypermultiplet sector can be included by considering the theories with eight supercharges as resulting from truncations of the theories with sixteen supercharges.

\subsection{Branes in 6D minimal matter-coupled supergravity}

 We  first consider the six-dimensional case. If gravity couples to $n_T$ tensor multiplets, the scalars in the tensor multiplets parametrise the symmetric manifold $\text{SO}(1,n_T)/\text{SO}(n_T)$, while the $1+n_T$ 2-forms of the gravity multiplet and of the tensor multiplets transform in the vector representation of $\text{SO}(1,n_T)$ \cite{Romans6D}. Given that the symmetry is orthogonal, we can apply the light-cone rules to declare that there are two 1/2-supersymmetric 1-branes with charges along the two lightlike directions in  $\text{SO}(1,n_T)$. Moreover, the fact that there are no scalars associated to the vectors, which is the same as saying that the group rotating $n_V$ abelian vectors is the compact form  $\text{SO}(n_V)$, implies that there are no 1/2-supersymmetric 0-branes.  Given that the highest worldvolume dimension for a 1/2-BPS state is four, in the six-dimensional case we can have at most defect branes. There cannot be any defect branes charged under the duals of the scalars in the tensor multiplets.
The reason for this is that such fields are 4-forms $A_{4, A_1 A_2}$ belonging to the adjoint of $\text{SO}(1,n_T)$, and thus the corresponding charge cannot lead to a 1/2-BPS 3-brane according to the light-cone rule. Using the language of the previous section, this means that the highest weight of the representation is not real. So the only 1/2-BPS defect branes can be those that are charged under the duals of the scalars in the hypermultiplets.

In any dimension, the four scalars in the hypermultiplets parametrise a quaternionic manifold. For our method to be applied, we consider the special case in which the quaternionic manifold is a symmetric manifold $G/H$.
This leads to the following manifolds:
\begin{eqnarray}
  & &
  \text{SO} (4,n_H) /[ \text{SO}(4) \times \text{SO}(n_H) ] \nonumber \\
    & & \text{F}_{4(4)} / [ \text{USp}(6) \times \text{SU}(2) ] \qquad \quad (n_H  = 7 )  \nonumber \\
  & &  \text{E}_{6(2)} / [\text{ SU}(6) \times \text{SU}(2) ]   \qquad \quad (n_H  = 10 ) \nonumber \\
  & &
  \text{E}_{7(-5)}/[\text{SO}(12) \times\text{ SU}(2)  ] \qquad \quad (n_H  = 16 ) \nonumber \\
  & &
  \text{E}_{8(-24)} / [\text{E}_7 \times \text{SU}(2) ] \qquad \quad (n_H  = 28 ) \nonumber \\
  & &
  \text{G}_{2(2)} / \text{SO}(4) \qquad \quad  (n_H  = 2 )  \nonumber \\
   & &
  \text{SU}(n_H,2) / [ \text{SU}(n_H)\times \text{SU}(2) \times \text{U}(1) ]   \nonumber \\
    & &    \text{USp}(2 n_H,2) / [ \text{USp}(2 n_H)\times \text{USp}(2) ]   \quad , \label{quatsymmspaces}
  \end{eqnarray}
where $n_H$ is the number of hypermultiplets. Given that the charge of a defect brane is always in the adjoint of $G$, the number of 1/2-supersymmetric defect branes  in the hyper-sector in any dimension (and in particular the number of 3-branes in six dimensions)  is given by the number of real longest roots of $G$. Clearly, while for the first manifold in
eq.~\eqref{quatsymmspaces} this counting amounts to applying the light-cone rules for a charge with two antisymmetric indices of $\text{SO} (4,n_H)$, in general we must count the real longest roots using the Cartan involution as obtained from the corresponding Tits-Satake diagram.
The Tits-Satake diagrams and the associated Cartan involutions of the real forms listed in eq.~\eqref{quatsymmspaces} are given in Table \ref{quaternionictitssatake}.

\begin{table}[h!]
\renewcommand{\arraystretch}{1.25}
\begin{center}
\resizebox{\textwidth}{!}{
 \begin{tabular}{|c|c|c|}
 \hline
 G/H&Tits-Satake diagram of $G$ &Cartan involution\tabularnewline
 \hline\hline

 &&\tabularnewline
 &&$\theta(\alpha_{1})=-\alpha_{1}$\tabularnewline
  &&$\theta(\alpha_{2})=-\alpha_{2}$\tabularnewline
 &\multirow{-4}{*}{
 \centering
\scalebox{0.4} 
{
\begin{pspicture}(0,-3.3059375)(26.600624,3.3059375)
\psline[linewidth=0.02cm](4.6,0.3340625)(8.4,0.3340625)
\psline[linewidth=0.02cm](0.6,0.3340625)(4.4,0.3340625)
\psline[linewidth=0.02cm](12.6,0.3340625)(16.4,0.3340625)
\psline[linewidth=0.02cm](12.6,0.3340625)(13.6,0.3340625)
\psline[linewidth=0.02cm](11.2,0.3340625)(12.2,0.3340625)
\psline[linewidth=0.02cm](8.6,0.3340625)(12.4,0.3340625)
\psline[linewidth=0.02cm](22.4,0.3340625)(26.0,2.3340626)
\psline[linewidth=0.02cm](22.4,0.3340625)(26.0,-1.6659375)
\psline[linewidth=0.02cm](21.2,0.3340625)(22.2,0.3340625)
\psline[linewidth=0.02cm](16.6,0.3340625)(17.6,0.3340625)
\pscircle[linewidth=0.02,dimen=outer,fillstyle=solid,fillcolor=black](22.4,0.32375){0.4}
\pscircle[linewidth=0.02,dimen=outer,fillstyle=solid](12.4,0.32375){0.4}
\pscircle[linewidth=0.02,dimen=outer,fillstyle=solid,fillcolor=black](16.4,0.32375){0.4}
\pscircle[linewidth=0.02,dimen=outer,fillstyle=solid,fillcolor=black](25.9,2.3240626){0.4}
\psline[linewidth=0.02cm,linestyle=dashed,dash=0.16cm 0.16cm](17.4,0.3340625)(21.2,0.3340625)
\pscircle[linewidth=0.02,dimen=outer,fillstyle=solid,fillcolor=black](25.9,-1.6759375){0.4}
\pscircle[linewidth=0.02,dimen=outer,fillstyle=solid](8.4,0.32375){0.4}
\rput(8.367969,-1.0559375){\huge 3}
\rput(12.385938,-1.0559375){\huge 4}
\rput(16.365,-1.0559375){\huge 5}
\rput(22.37,-1.0559375){\huge k}
\rput(25.848595,-3.0559375){\huge k+1}
\rput(25.845938,0.9440625){\huge k+2}
\pscircle[linewidth=0.02,dimen=outer,fillstyle=solid](4.4,0.32375){0.4}
\rput(4.3620315,-1.0559375){\huge 2}
\pscircle[linewidth=0.02,dimen=outer,fillstyle=solid](0.4,0.32375){0.4}
\rput(0.351875,-1.0559375){\huge 1}
\rput(12.3,2.5440625){\Huge n=2k}
\end{pspicture}
} }&$\theta(\alpha_{3})=-\alpha_{3}$\tabularnewline
 &&$\theta(\alpha_{4})=-\alpha_{4}-2\sum_{i=5}^{k}\beta_{i}$\tabularnewline
 &&$-2^{n\mod 2}(\beta_{k+1}+\beta_{k+2})$\tabularnewline
 &&$\theta(\beta_{i})=\beta_{i}\qquad i=5,...,k+2$\tabularnewline
 \multirow{-8}{*}{$\dfrac{\text{SO}(4,n)}{\text{SO}(4)\times \text{SO}(n)}$}&\multirow{-4}{*}{
 \centering
\scalebox{0.4} 
{
\begin{pspicture}(0,-2.0573437)(27.627188,2.0573437)
\psline[linewidth=0.02cm](0.6,-0.5042182)(4.4,-0.5042182)
\psline[linewidth=0.02cm](22.6,-0.70421827)(26.4,-0.70421827)
\psline[linewidth=0.02cm](21.2,-0.5042182)(22.2,-0.5042182)
\psline[linewidth=0.02cm](16.6,-0.5042182)(17.6,-0.5042182)
\psline[linewidth=0.02cm](12.6,-0.5042182)(16.4,-0.5042182)
\pscircle[linewidth=0.02,dimen=outer,fillstyle=solid,fillcolor=black](16.4,-0.51453125){0.4}
\psline[linewidth=0.02cm,linestyle=dashed,dash=0.16cm 0.16cm](17.4,-0.5042182)(21.2,-0.5042182)
\rput(12.385938,-1.7045312){\huge 4}
\rput(16.365,-1.7045312){\huge 5}
\psline[linewidth=0.02cm](22.4,-0.30421826)(26.2,-0.30421826)
\pscircle[linewidth=0.02,dimen=outer,fillstyle=solid,fillcolor=black](22.4,-0.51453125){0.4}
\pscircle[linewidth=0.02,dimen=outer,fillstyle=solid,fillcolor=black](26.4,-0.51453125){0.4}
\psline[linewidth=0.02cm](24.6,-0.51453125)(24.0,0.08546875)
\psline[linewidth=0.02cm](24.6,-0.51453125)(24.0,-1.1145313)
\rput(22.29453,-1.7045312){\huge $k+1$}
\rput(26.434532,-1.6045313){\huge $k+2$}
\psline[linewidth=0.02cm](8.6,-0.5042182)(12.4,-0.5042182)
\psline[linewidth=0.02cm](4.6,-0.5042182)(8.4,-0.5042182)
\pscircle[linewidth=0.02,dimen=outer,fillstyle=solid](8.4,-0.51453125){0.4}
\rput(8.367969,-1.7045312){\huge 3}
\pscircle[linewidth=0.02,dimen=outer,fillstyle=solid](4.4,-0.51453125){0.4}
\rput(4.3620315,-1.7045312){\huge 2}
\pscircle[linewidth=0.02,dimen=outer,fillstyle=solid](12.4,-0.51453125){0.4}
\pscircle[linewidth=0.02,dimen=outer,fillstyle=solid](0.4,-0.51453125){0.4}
\rput(0.351875,-1.7045312){\huge 1}
\rput(13.824532,1.6954688){\Huge $n=2k+1$}
\end{pspicture}
}
}&\tabularnewline\hline

&&\tabularnewline
 \multirow{-2}{*}{$\dfrac{\text{F}_{4(4)}}{\text{USp}(6)\times \text{SU}(2)}$}&\multirow{-2}{*}{
 \centering
\scalebox{0.4} 
{
\begin{pspicture}(0,-1.145)(12.8,1.115)
\definecolor{color57b}{rgb}{0.996078431372549,0.996078431372549,0.996078431372549}
\psline[linewidth=0.02cm](4.2,0.505)(0.6,0.505)
\psline[linewidth=0.02cm](8.4,0.705)(4.4,0.705)
\psline[linewidth=0.02cm](8.2,0.305)(4.6,0.305)
\psline[linewidth=0.02cm](12.4,0.505)(8.6,0.505)
\rput{180.0}(24.8,1.01){\pscircle[linewidth=0.02,dimen=outer,fillstyle=solid](12.4,0.505){0.4}}
\rput{180.0}(0.8,1.01){\pscircle[linewidth=0.02,dimen=outer,fillstyle=solid,fillcolor=color57b](0.4,0.505){0.4}}
\rput{180.0}(8.8,1.01){\pscircle[linewidth=0.02,dimen=outer,fillstyle=solid](4.4,0.505){0.4}}
\rput{180.0}(16.8,1.01){\pscircle[linewidth=0.02,dimen=outer,fillstyle=solid](8.4,0.505){0.4}}
\psline[linewidth=0.02cm](6.8,0.505)(6.2,-0.095)
\psline[linewidth=0.02cm](6.8,0.505)(6.2,1.105)
\rput(0.351875,-0.885){\huge 1}
\rput(4.3620315,-0.885){\huge 2}
\rput(8.367969,-0.885){\huge 3}
\rput(12.385938,-0.885){\huge 4}
\end{pspicture}
}
}&\multirow{-2}{*}{$\theta(\alpha_{i})=-\alpha_{i}\qquad i=1,2,3,4$}\tabularnewline\hline

 &&$\theta(\alpha_{1})=-\alpha_{5}$\tabularnewline
 &&$\theta(\alpha_{2})=-\alpha_{4}$\tabularnewline
 &&$\theta(\alpha_{3})=-\alpha_{3}$\tabularnewline
 &&$\theta(\alpha_{4})=-\alpha_{2}$\tabularnewline
 &&$\theta(\alpha_{5})=-\alpha_{1}$\tabularnewline

 \multirow{-6}{*}{$\dfrac{\text{E}_{6(2)}}{\text{SU}(6)\times \text{SU}(2)}$}&\multirow{-6}{*}{
 \centering
\scalebox{0.4} 
{
\begin{pspicture}(0,-4.674844)(16.8,4.7148438)
\psline[linewidth=0.02cm](0.6,-0.25796875)(16.2,-0.25796875)
\psline[linewidth=0.02cm](8.4,3.9420311)(8.4,-0.25796875)
\pscircle[linewidth=0.02,dimen=outer,fillstyle=solid](0.4,-0.25796875){0.4}
\pscircle[linewidth=0.02,dimen=outer,fillstyle=solid](8.4,3.7420313){0.4}
\pscircle[linewidth=0.02,dimen=outer,fillstyle=solid](16.4,-0.25796875){0.4}
\pscircle[linewidth=0.02,dimen=outer,fillstyle=solid](8.4,-0.25796875){0.4}
\pscircle[linewidth=0.02,dimen=outer,fillstyle=solid](12.4,-0.25796875){0.4}
\pscircle[linewidth=0.02,dimen=outer,fillstyle=solid](4.4,-0.25796875){0.4}
\psbezier[linewidth=0.06,arrowsize=0.05291667cm 4.0,arrowlength=1.4,arrowinset=0.2]{<->}(0.32453123,-0.83484375)(0.32453123,-2.0071514)(1.5156925,-3.1833103)(2.7545307,-3.765613)(3.993369,-4.3479156)(7.075781,-4.6448436)(8.307333,-4.6448436)(9.538885,-4.6448436)(12.474531,-4.6448436)(14.094531,-3.765613)(15.714531,-2.886382)(16.52453,-2.3002284)(16.52453,-0.83484375)
\psbezier[linewidth=0.06,arrowsize=0.05291667cm 4.0,arrowlength=1.4,arrowinset=0.2]{<->}(4.558437,-0.81576675)(4.558437,-1.440098)(5.1441727,-2.0664806)(5.7533517,-2.3765953)(6.3625307,-2.68671)(7.878258,-2.8448439)(8.541485,-2.8448439)(9.204712,-2.8448439)(10.524531,-2.8448439)(11.3296175,-2.3765953)(12.134704,-1.9083468)(12.524531,-1.596181)(12.524531,-0.8157669)
\rput(0.87640625,0.5651562){\huge 1}
\rput(8.8925,0.5651562){\huge 3}
\rput(12.710468,0.5651562){\huge 4}
\rput(16.48953,0.5651562){\huge 5}
\rput(8.900782,4.365156){\huge 6}
\rput(4.6865625,0.5651562){\huge 2}
\end{pspicture}
} }&$\theta(\alpha_{6})=-\alpha_{6}$\tabularnewline\hline

 &&$\theta(\alpha_{2})=-\alpha_{2} - \beta_1 - \beta_3$\tabularnewline
 &&$\theta(\alpha_{4})=-\alpha_{4} -\beta_3 -\beta_7$\tabularnewline
 &&$\theta(\alpha_{5})=-\alpha_{5}$\tabularnewline
 &&$\theta(\alpha_{6})=-\alpha_{6}$\tabularnewline

\multirow{-5}{*}{$\dfrac{\text{E}_{7(-5)}}{\text{SO}(12)\times \text{SU}(2)}$}&\multirow{-7}{*}{
\centering
\scalebox{0.4} 
{
\begin{pspicture}(0,-0.845)(20.8,2.805)
\definecolor{color168b}{rgb}{0.00392156862745098,0.00392156862745098,0.00392156862745098}
\definecolor{color172b}{rgb}{0.996078431372549,0.996078431372549,0.996078431372549}
\psline[linewidth=0.02cm](12.4,2.405)(12.4,-1.595)
\psline[linewidth=0.02cm](0.2,-1.595)(20.2,-1.595)
\pscircle[linewidth=0.02,dimen=outer,fillstyle=solid](20.4,-1.595){0.4}
\pscircle[linewidth=0.02,dimen=outer,fillstyle=solid,fillcolor=color168b](0.4,-1.595){0.4}
\pscircle[linewidth=0.02,dimen=outer,fillstyle=solid,fillcolor=color172b](4.4,-1.595){0.4}
\pscircle[linewidth=0.02,dimen=outer,fillstyle=solid,fillcolor=color168b](8.4,-1.595){0.4}
\pscircle[linewidth=0.02,dimen=outer,fillstyle=solid,fillcolor=color168b](12.4,2.405){0.4}
\pscircle[linewidth=0.02,dimen=outer,fillstyle=solid](16.4,-1.595){0.4}
\pscircle[linewidth=0.02,dimen=outer,fillstyle=solid](12.4,-1.595){0.4}
\rput(0.351875,-2.585){\huge 1}
\rput(20.37625,-2.585){\huge 6}
\rput(16.365,-2.585){\huge 5}
\rput(12.385938,-2.585){\huge 4}
\rput(4.3620315,-2.585){\huge 2}
\rput(8.367969,-2.585){\huge 3}
\rput(13.1640625,2.415){\huge 7}
\end{pspicture}
}
}&
$\theta(\beta_{i})=\beta_{i} \ \ \ \ \ \  i=1,3,7$
\tabularnewline\hline

 &&$\theta(\alpha_{1})=-\alpha_{1}$\tabularnewline
 &&$\theta(\alpha_{2})=-\alpha_{2}$\tabularnewline
 &&$\theta(\alpha_{3})=-\alpha_{3}-2\beta_{4}-2\beta_{5}-\beta_{6}-\beta_{8}$\tabularnewline
 &&$\theta(\alpha_{7})=-\alpha_{7}-\beta_{4}-2\beta_{5}-2\beta_{6}-\beta_{8}$\tabularnewline
  \multirow{-5}{*}{$\dfrac{\text{E}_{8(-24)}}{\text{E}_{7}\times \text{SU}(2)}$}&\multirow{-5}{*}{
 \centering
\scalebox{0.4} 
{
\begin{pspicture}(0,-2.845)(24.8,2.805)
\definecolor{color148b}{rgb}{0.00392156862745098,0.00392156862745098,0.00392156862745098}
\psline[linewidth=0.02cm](16.4,2.405)(16.4,-1.595)
\psline[linewidth=0.02cm](0.2,-1.595)(24.4,-1.595)
\pscircle[linewidth=0.02,dimen=outer,fillstyle=solid](0.4,-1.595){0.4}
\pscircle[linewidth=0.02,dimen=outer,fillstyle=solid](4.4,-1.595){0.4}
\pscircle[linewidth=0.02,dimen=outer,fillstyle=solid](8.4,-1.595){0.4}
\pscircle[linewidth=0.02,dimen=outer,fillstyle=solid,fillcolor=color148b](12.4,-1.595){0.4}
\pscircle[linewidth=0.02,dimen=outer,fillstyle=solid,fillcolor=color148b](16.4,-1.595){0.4}
\pscircle[linewidth=0.02,dimen=outer,fillstyle=solid,fillcolor=color148b](16.4,2.405){0.4}
\pscircle[linewidth=0.02,dimen=outer,fillstyle=solid](24.4,-1.595){0.4}
\pscircle[linewidth=0.02,dimen=outer,fillstyle=solid,fillcolor=color148b](20.4,-1.595){0.4}
\rput(0.351875,-2.585){\huge 1}
\rput(17.175,2.415){\huge 8}
\rput(24.364063,-2.585){\huge 7}
\rput(20.37625,-2.585){\huge 6}
\rput(16.365,-2.585){\huge 5}
\rput(12.385938,-2.585){\huge 4}
\rput(8.367969,-2.585){\huge 3}
\rput(4.3620315,-2.585){\huge 2}
\end{pspicture}
}
}&$\theta(\beta_{i})=\beta_{i}\qquad i=4,5,6,8$\tabularnewline\hline

&&\tabularnewline
\multirow{-2}{*}{$\dfrac{\text{G}_{2(2)}}{\text{SO}(4)}$}&\multirow{-2}{*}{
\centering
\scalebox{0.4} 
{
\begin{pspicture}(0,-1.0275)(8.649062,0.9875)
\psline[linewidth=0.02cm](1.941875,0.3775)(5.541875,0.3775)
\psline[linewidth=0.02cm](1.741875,0.1775)(5.741875,0.187813)
\psline[linewidth=0.02cm](1.741875,0.587813)(5.741875,0.5775)
\pscircle[linewidth=0.02,dimen=outer,fillstyle=solid](1.741875,0.3775){0.4}
\pscircle[linewidth=0.02,dimen=outer,fillstyle=solid](5.741875,0.3775){0.4}
\psline[linewidth=0.02cm](3.941875,0.3775)(3.341875,0.9775)
\psline[linewidth=0.02cm](3.941875,0.3775)(3.341875,-0.2225)
\rput(1.8464063,-0.6125){\huge 1}
\rput(5.846406,-0.6125){\huge 2}
\end{pspicture}
}
}&\multirow{-2}{*}{$\theta(\alpha_{i})=-\alpha_{i}\qquad i=1,2$}\tabularnewline
\hline

&&$\theta(\alpha_{1})=-\alpha_{3}$\tabularnewline
&&$\theta(\alpha_{2})=-\alpha_{2}$\tabularnewline
&\multirow{-3}{*}{
 \centering
\scalebox{0.4} 
{
\begin{pspicture}(0,-1.8104135)(13.022344,1.8304136)
\psline[linewidth=0.02cm](4.746875,0.67072606)(12.946875,0.67072606)
\pscircle[linewidth=0.02,dimen=outer,fillstyle=solid](8.622344,0.65760106){0.4}
\pscircle[linewidth=0.02,dimen=outer,fillstyle=solid](12.622344,0.65760106){0.4}
\pscircle[linewidth=0.02,dimen=outer,fillstyle=solid](4.6223435,0.65760106){0.4}
\rput(4.49875,1.4807261){\huge 1}
\rput(12.714844,1.4807261){\huge 3}
\rput(8.508906,1.4807261){\huge 2}
\psbezier[linewidth=0.04,arrowsize=0.05291667cm 5.0,arrowlength=1.4,arrowinset=0.4]{<->}(4.686875,0.030726083)(4.686875,-0.76927394)(5.6270986,-1.7904136)(8.686875,-1.7292739)(11.746652,-1.6681342)(12.686875,-0.9692739)(12.686875,0.030726083)
\rput(1.34375,0.99072605){\Huge $n=2$}
\end{pspicture}
}
}&$\theta(\alpha_{3})=-\alpha_{1}$\tabularnewline \cline{2-3}

 &&$\theta(\alpha_{1})=-\alpha_{n+1}$\tabularnewline
 &&$\theta(\alpha_{2})=-\alpha_{n}-\sum_{i=3}^{n-1}\beta_{i}$\tabularnewline
 &&$\theta(\beta_{i})=\beta_{i}\qquad i=3,..,n-1$\tabularnewline
 &&$\theta(\alpha_{n})=-\alpha_{2}-\sum_{i=3}^{n-1}\beta_{i}$\tabularnewline

\multirow{-8}{*}{$\dfrac{\text{SU}(n,2)}{\text{SU}(n)\times \text{SU}(2)\times \text{U}(1)}$}&\multirow{-5}{*}{
\centering
\scalebox{0.4} 
{
\begin{pspicture}(0,-3.3838048)(23.165155,3.4038053)
\definecolor{color300b}{rgb}{0.00392156862745098,0.00392156862745098,0.00392156862745098}
\psline[linewidth=0.02cm](0.52453125,0.49505517)(9.524531,0.49505517)
\psline[linewidth=0.02cm](13.324532,0.49505517)(22.724531,0.49505517)
\pscircle[linewidth=0.02,dimen=outer,fillstyle=solid](4.4,0.48193017){0.4}
\rput(4.2865624,1.3050551){\huge 2}
\rput(8.2925,1.3050551){\huge 3}
\psbezier[linewidth=0.04,arrowsize=0.05291667cm 5.0,arrowlength=1.4,arrowinset=0.4]{<->}(4.4445314,-0.22494483)(4.4445314,-1.0249448)(6.179919,-2.4460845)(11.384531,-2.4249449)(16.589144,-2.4038053)(18.324532,-1.2249448)(18.324532,-0.22494483)
\pscircle[linewidth=0.02,dimen=outer,fillstyle=solid](4.4,0.48193017){0.4}
\pscircle[linewidth=0.02,dimen=outer,fillstyle=solid](0.4,0.48193017){0.4}
\rput(0.27640623,1.3050551){\huge 1}
\pscircle[linewidth=0.02,dimen=outer,fillstyle=solid,fillcolor=color300b](8.4,0.48193017){0.4}
\pscircle[linewidth=0.02,dimen=outer,fillstyle=solid](18.4,0.48193017){0.4}
\rput(18.284843,1.3050551){\huge n}
\rput(22.393126,1.3050551){\huge n+1}
\pscircle[linewidth=0.02,dimen=outer,fillstyle=solid](18.4,0.48193017){0.4}
\pscircle[linewidth=0.02,dimen=outer,fillstyle=solid,fillcolor=color300b](14.4,0.48193017){0.4}
\rput(14.433125,1.3050551){\huge n-1}
\pscircle[linewidth=0.02,dimen=outer,fillstyle=solid](22.4,0.48193017){0.4}
\psline[linewidth=0.02cm,linestyle=dashed,dash=0.16cm 0.16cm](9.324532,0.49505517)(13.724531,0.49505517)
\psbezier[linewidth=0.04,arrowsize=0.05291667cm 5.0,arrowlength=1.4,arrowinset=0.4]{<->}(0.44453123,-0.22494483)(0.44453123,-1.0249448)(3.005148,-3.2460845)(11.324532,-3.3049448)(19.643915,-3.363805)(22.52453,-1.2249448)(22.52453,-0.22494483)
\rput(11.341406,2.9550552){\Huge $n>2$}
\end{pspicture}
}
}&$\theta(\alpha_{n+1})=-\alpha_{1}$\tabularnewline\hline

&&$\theta(\beta_{1})=\beta_{1}$\tabularnewline
&&$\theta(\alpha_{2})=-\alpha_{2}-2\beta_{1}$\tabularnewline
&\multirow{-3}{*}{
\centering
\scalebox{0.4} 
{
\begin{pspicture}(0,0)(4.8,0)
\definecolor{color45b}{rgb}{0.00392156862745098,0.00392156862745098,0.00392156862745098}
\psline[linewidth=0.02cm](0.6,-0.724062)(4.4,-0.724062)
\psline[linewidth=0.02cm](0.4,-0.324062)(4.2,-0.324062)
\pscircle[linewidth=0.02,dimen=outer,fillstyle=solid,fillcolor=color45b](0.4,-0.534375){0.4}
\pscircle[linewidth=0.02,dimen=outer,fillstyle=solid](4.4,-0.534375){0.4}
\psline[linewidth=0.02cm](2.2,-0.534375)(2.8,0.065625)
\psline[linewidth=0.02cm](2.2,-0.534375)(2.8,-1.134375)
\rput(4.3620315,-1.524375){\huge 2}
\rput(0.351875,-1.524375){\huge 1}
\rput(2.416875,1.325625){\Huge $n=1$}
\end{pspicture}
}
}&\tabularnewline\cline{2-3}

&&$\theta(\beta_{i})=\beta_{i}\qquad i\neq 2$\tabularnewline
&&$\theta(\alpha_{2})=-\alpha_{2}-\beta_{1}$\tabularnewline
\multirow{-6}{*}{$\dfrac{\text{USp}(2n,2)}{\text{USp}(2n)\times \text{USp}(2)}$}&\multirow{-3}{*}{
\centering
\scalebox{0.4} 
{
\begin{pspicture}(0,-1.779375)(19.640625,1.779375)
\definecolor{color53b}{rgb}{0.00392156862745098,0.00392156862745098,0.00392156862745098}
\psline[linewidth=0.02cm](0.2,-0.529375)(8.4,-0.529375)
\psline[linewidth=0.02cm](14.2,-0.529375)(14.4,-0.529375)
\psline[linewidth=0.02cm](14.6,-0.719062)(18.4,-0.719062)
\psline[linewidth=0.02cm](13.2,-0.519062)(14.2,-0.519062)
\psline[linewidth=0.02cm](8.6,-0.519062)(9.6,-0.519062)
\psline[linewidth=0.02cm](14.4,-0.319062)(18.2,-0.319062)
\pscircle[linewidth=0.02,dimen=outer,fillstyle=solid,fillcolor=color53b](14.4,-0.529375){0.4}
\pscircle[linewidth=0.02,dimen=outer,fillstyle=solid,fillcolor=color53b](8.4,-0.529375){0.4}
\pscircle[linewidth=0.02,dimen=outer,fillstyle=solid,fillcolor=color53b](18.4,-0.529375){0.4}
\psline[linewidth=0.02cm,linestyle=dashed,dash=0.16cm 0.16cm](9.4,-0.519062)(13.2,-0.519062)
\psline[linewidth=0.02cm](16.2,-0.529375)(16.8,0.070625)
\psline[linewidth=0.02cm](16.2,-0.529375)(16.8,-1.129375)
\rput(18.868593,-1.519375){\huge n+1}
\rput(8.367969,-1.519375){\huge 3}
\pscircle[linewidth=0.02,dimen=outer,fillstyle=solid](4.4,-0.529375){0.4}
\pscircle[linewidth=0.02,dimen=outer,fillstyle=solid,fillcolor=color53b](0.4,-0.529375){0.4}
\rput(4.3620315,-1.519375){\huge 2}
\rput(0.351875,-1.519375){\huge 1}
\rput(14.360312,-1.519375){\huge n}
\rput(10.216875,1.330625){\Huge $n>1$}
\end{pspicture}
}
}&$-2\sum_{i=3}^{n}\beta_{i}-\beta_{n+1}$\tabularnewline\hline

 \end{tabular}
}
\end{center}\caption{\label{quaternionictitssatake}\sl The Tits-Satake diagrams of the groups corresponding to the quaternionic symmetric spaces given in eq.~\eqref{quatsymmspaces}. The last column gives the Cartan involution.}
\end{table}

We can analyse each case in more detail. As already mentioned, for the symmetry  $\text{SO} (4,n_H)$ we can use the  analysis of the previous section. The restricted simple roots generate the maximally non-compact algebra $\text{SO} (4,5)$ for $n_H>4$, while for $n_H =4$ the algebra $\text{SO} (4,4)$ is already maximally non-compact, and for $n_H<4$ the restricted-root algebra is  $\text{SO} (n_H +1,n_H)$. If $n_H >1$, the highest weight of the adjoint representation is always real, and the number of real longest roots is $2n_H (n_H-1)$ for $n_H<4$ and 24 otherwise.

In the $ \text{F}_{4(4)}$ case, all the roots are real because the group is maximally non-compact. The number of longest roots is 24.
The  $\text{E}_{6(2)}$, $\text{E}_{7(-5)}$ and
$\text{E}_{8(-24)}$ cases all give the same result as the $ \text{F}_{4(4)}$ case because the restricted-root algebra is in all cases $ \text{ F}_{4(4)}$, and the highest weight of the adjoint representation is in all cases real and coincides with the highest weight of the adjoint of $ \text{F}_{4(4)}$.
This can be verified using the Cartan involutions in Table \ref{quaternionictitssatake}. As an example, we consider explicitly the $\text{E}_{6(2)}$ case. From the Cartan involution given in the third row, last column of Table \ref{quaternionictitssatake} we get
\begin{equation}
(\alpha_1 )_R = \tfrac{1}{2} [ \alpha_1 + \alpha_5 ]  \qquad ( \alpha_{2} )_R =  \tfrac{1}{2} [ \alpha_2 + \alpha_4 ]  \qquad (\alpha_3 )_R = \alpha_3 \qquad  (\alpha_6 )_R =\alpha_6 \quad .
\end{equation}
 We recognise in $(\alpha_6 )_R$ and in $(\alpha_3 )_R$ the first and the second long simple roots of $ \text{ F}_{4(4)}$, while $( \alpha_{2} )_R$ and
$(\alpha_1 )_R$ are the third and fourth short simple roots of $ \text{F}_{4(4)}$. Denoting with $\tilde{\alpha}_i$\ $(i=1,...,4)$ the nodes of $\text{ F}_{4(4)}$, as labeled in Table \ref{quaternionictitssatake}, we summarise this as
\begin{eqnarray}
& & \tilde{\alpha}_1 = (\alpha_6 )_R = \alpha_6 \nonumber \\
& & \tilde{\alpha}_2 = (\alpha_3 )_R = \alpha_3 \nonumber \\
& & \tilde{\alpha}_3 = (\alpha_2 )_R = \tfrac{1}{2} [ \alpha_2 + \alpha_4 ] \nonumber \\
& & \tilde{\alpha}_4 = (\alpha_1 )_R = \tfrac{1}{2} [ \alpha_1 + \alpha_5 ] \quad . \label{rootsoff4rootsofe6}
\end{eqnarray}
The highest weight of the adjoint of $\text{E}_{6(2)}$ is
\begin{equation}
\Lambda_{\text{E}_6}  =\alpha_1 + 2 \alpha_2 + 3 \alpha_3 + 2\alpha_4 + \alpha_5 + 2 \alpha_6 \quad ,
\end{equation}
and acting with the Cartan involution one finds $(\Lambda_{\text{E}_6} )_R = \Lambda_{\text{E}_6}$. Moreover, using
eq.~\eqref{rootsoff4rootsofe6} one obtains
\begin{equation}
\Lambda_{\text{E}_6}  = 2 \tilde{\alpha}_1 + 3 \tilde{\alpha}_2 + 4 \tilde{\alpha}_3 + 2 \tilde{\alpha}_4 \quad , \label{highestweightofadjofF4}
\end{equation}
which is the highest weight of the adjoint of $ \text{F}_{4(4)}$. This implies that the number of real longest roots of $\text{E}_{6(2)}$ is equal to the number of longest roots of $ \text{ F}_{4(4)}$, which is 24.
Exactly the same result holds for  $\text{E}_{7(-5)}$ and
$\text{E}_{8(-24)}$. In the first (second) case  the first two long simple roots of the restricted root algebra $ \text{ F}_{4(4)}$ are associated  to nodes 6 and 5 (nodes 1 and 2) of the corresponding Tits-Satake diagrams given in Table \ref{quaternionictitssatake}.
The highest weights of the adjoint representation of these two real forms are
  \begin{eqnarray}
  & & \Lambda_{\text{E}_7} = \beta_1 + 2 \alpha_2 + 3 \beta_3 + 4 \alpha_4 + 3\alpha_5 + 2 \alpha_6 + 2 \beta_7 \nonumber \\
  & & \Lambda_{\text{E}_8} = 2 \alpha_1 + 3 \alpha_2 + 4 \alpha_3 + 5 \beta_4 + 6\beta_5 + 4 \beta_6 + 2 \alpha_7+ 3 \beta_8 \quad ,
  \end{eqnarray}
respectively. Using the Cartan involutions given in Table \ref{quaternionictitssatake} one can verify that in both cases these highest weights satisfy $\Lambda = \Lambda_R$ and that their expressions are
given by the same  eq.~\eqref{highestweightofadjofF4} when expressed in terms of the simple roots of $ \text{ F}_{4(4)}$.
This means that for all the $\text{E}$ groups in eq.~\eqref{quatsymmspaces} the number of 3-branes in the hypermultiplet sector of the six-dimensional theory is 24 precisely as for the  $ \text{F}_{4(4)}$ case.

One can perform the same analysis for the last three groups in eq.~\eqref{quatsymmspaces}. In particular,
$\text{G}_{2(2)}$ is maximally non-compact, and thus the number of defect branes is given in this case by the number of longest roots, which is 6.
We show in Table \ref{hyperbranessixdim} the outcome of the analysis for all the groups in eq.~\eqref{quatsymmspaces}. In the second column of the table we have listed the restricted-root algebras and in the last column the number of real longest roots, giving the number of 3-branes in the six-dimensional theory. Taking $n_H \geq 4$ in the first row, we see that the result is universal in the first five cases. These cases have the common feature that the algebra has four non-compact Cartan generators, while in the last three cases their number is equal to 2, 2 and 1 respectively. The outcome of this  analysis is that all the branes of the six-dimensional theory can be determined using our methods for the special case  that the hypermultiplet scalars parametrise a  symmetric manifold.

\begin{table}
\begin{center}
\begin{tabular}{|c|c|c|}
\hline \rule[-1mm]{0mm}{6mm} $G$ & restricted-root algebra & $\#$ of real longest roots\\
\hline
\hline  \rule[-1mm]{0mm}{6mm}
 &  $\mathfrak{so} (n_H +1,n_H)$ \ $n_H<4$ & $2n_H (n_H-1)$  \\
  $\text{SO} (4,n_H)$ &  $\mathfrak{so} (4,4)$ \ $n_H=4$ &  24 \\
   &  $\mathfrak{so} (4,5)$\  $n_H>4$ &  24 \\
\hline \rule[-1mm]{0mm}{6mm}
$ \text{ F}_{4(4)}$ & $ \mathfrak{f}_{4(4)}$  & 24 \\
\hline \rule[-1mm]{0mm}{6mm}
$\text{E}_{6(2)}$ & $ \mathfrak{f}_{4(4)}$  & 24 \\
\hline \rule[-1mm]{0mm}{6mm}
$\text{E}_{7(-5)}$ & $ \mathfrak{f}_{4(4)}$  & 24  \\
\hline \rule[-1mm]{0mm}{6mm}
$\text{E}_{8(-24)}$ & $ \mathfrak{f}_{4(4)}$  &  24  \\
\hline  \rule[-1mm]{0mm}{6mm}
$\text{G}_{2(2)}$ & $\mathfrak{g}_{2(2)}$  & 6 \\
\hline
\rule[-1mm]{0mm}{6mm}
 $\text{SU}(n_H,2)$ & $\mathfrak{sp} (4,\mathbb{R})$  & 2\\
\hline \rule[-1mm]{0mm}{6mm}
 $\text{USp}(2 n_H,2)$ & $\mathfrak{so}(1,2)$  & 0 \\
\hline
\end{tabular}
\end{center}
  \caption{\sl The number of real longest roots for the groups associated to the symmetric quaternionic spaces
  given in eq.~\eqref{quatsymmspaces}. These numbers correspond to the number of  1/2-supersymmetric BPS defect branes in the hyper-sector. \label{hyperbranessixdim}}
\end{table}

This completes our discussion of the six-dimensional case. In the remaining part of this section we will  consider the theories in lower dimensions. In the next subsection we will first consider a special class of theories,
 namely those  that give rise to the symmetric quaternionic manifolds of eq.~\eqref{quatsymmspaces} upon reduction to three dimensions.
The full list of coset manifolds and number of multiplets resulting from the uplift of the three-dimensional symmetric theories is given in Table \ref{upliftchain} \cite{gunaydinsierratownsend}. We can see from the table that these theories do not contain hypermultiplets in any dimension higher than three. Moreover,
only the three-dimensional theories with symmetry group containing four non-compact Cartan generators can be uplifted all the way to six dimensions. What is special about these theories is that the full spectrum, including the $(D-1)$- and $D$-forms, can be obtained by considering the very-extended Kac-Moody algebra $G^{+++}$, where $G$ is the symmetry of the three-dimensional theory. In subsection 4.2 we will use the results coming from the Kac-Moody analysis to obtain the representations of all the fields in any dimension. We will next  obtain the number of branes by considering the Cartan involution on the highest weights of these representations. The remarkable outcome of this analysis will be that there is a universal structure of branes for all the theories that can be uplifted to six dimensions.

\begin{table}\small
\begin{center}
\begin{tabular}{|c|c|c|c|}
\hline \rule[-1mm]{0mm}{6mm} $D=3$ & $D=4$ & $D=5$ & $D=6$ \\
\hline
\hline  \rule[-1mm]{0mm}{8mm}
& & & $\text{SO}(1,1)$ \\
  $\dfrac{\text{SO}(4,n)}{\text{SO}(4)\times \text{SO}(n)}$ &  $\dfrac{\text{SO}(2,n-2)}{\text{SO}(2)\times  \text{SO}(n-2)} \times \dfrac{\text{SU}(1,1)}{\text{U}(1)}$ & $\dfrac{\text{SO}(1,n-3)}{\text{SO}(n-3)} \times \mathbb{R}^+$  & $n_T =1 ,  n_V = n-4$ \\
\cline{4-4} \rule[-1mm]{0mm}{8mm}
$n_H =n$ & $ n_V = n-1$  & $n_V=n-2$ & $\dfrac{\text{SO}(1,n-3)}{\text{SO}(n-3)}$ \\
& & & $n_T=n-3 , n_V=0$ \\
\hline \rule[-1mm]{0mm}{8mm}
$\dfrac{\text{F}_{4(4)}}{\text{USp}(6)\times \text{SU}(2)}$ & $\dfrac{\text{Sp}(6, \mathbb{R})}{\text{U}(3)}$ & $\dfrac{\text{SL}(3,\mathbb{R})}{\text{SO}(3)}$ & $\dfrac{\text{SO}(1,2)}{\text{SO}(2)}$ \\
\rule[-1mm]{0mm}{8mm}
$n_H =7$ & $ n_V = 6$  & $n_V =5$& $n_T = 2, n_V=2$\\
\hline \rule[-1mm]{0mm}{8mm}
$\dfrac{\text{E}_{6(2)}}{\text{SU}(6)\times \text{SU}(2)}$ & $\dfrac{\text{SU}(3,3)}{\text{SU}(3)\times \text{SU}(3)}$ & $\dfrac{\text{SL}(3,\mathbb{C})}{\text{SU}(3)}$ & $\dfrac{\text{SO}(1,3)}{\text{SO}(3)}$ \\
\rule[-1mm]{0mm}{8mm}
$n_H =10$ & $ n_V = 9$  & $n_V =8$& $n_T = 3, n_V=4$\\
\hline \rule[-1mm]{0mm}{8mm}
$\dfrac{\text{E}_{7(-5)}}{\text{SO}(12)\times \text{SU}(2)}$ & $\dfrac{\text{SO}^*(12)}{\text{U}(6)}$ & $\dfrac{\text{SU}^*(6)}{\text{USp}(6)}$ & $\dfrac{\text{SO}(1,5)}{\text{SO}(5)}$ \\
\rule[-1mm]{0mm}{8mm}
$n_H =16$ & $ n_V = 15$  & $n_V =14$& $n_T = 5, n_V=8$\\
\hline \rule[-1mm]{0mm}{8mm}
$\dfrac{\text{E}_{8(-24)}}{\text{E}_7\times \text{SU}(2)}$ & $\dfrac{\text{E}_{7(-25)}}{\text{E}_6 \times\text{SO}(2)}$ & $\dfrac{\text{E}_{6(-26)}}{\text{F}_4}$ & $\dfrac{\text{SO}(1,9)}{\text{SO}(9)}$ \\
\rule[-1mm]{0mm}{8mm}
$n_H =28$ & $ n_V = 27$  & $n_V =26$& $n_T = 9, n_V=16$\\
\hline  \rule[-1mm]{0mm}{8mm}
$\dfrac{\text{G}_{2(2)}}{\text{SO}(4)}$ & $\dfrac{\text{SU}(1, 1)}{\text{U}(1)}$ & $1$ & - \\
\rule[-1mm]{0mm}{8mm}
$n_H =2$ & $ n_V = 1$  & $n_V =0$& \\
\hline  \rule[-1mm]{0mm}{8mm}
  $\dfrac{\text{SU}(n,2)}{\text{SU}(n)\times \text{SU}(2)\times \text{U}(1)}$ &  $\dfrac{\text{SU}(n-1,1)}{\text{SU}(n-1)\times  \text{U}(1)}$ &  - & - \\
 \rule[-1mm]{0mm}{6mm}
$n_H =n$ & $ n_V = n-1$  &  & \\
\hline
\rule[-1mm]{0mm}{8mm}
$\dfrac{\text{USp}(2n,2)}{\text{USp}(2n)\times \text{USp}(2)}$ & -   &  - & - \\
\rule[-1mm]{0mm}{8mm}
 $n_H = n$ &  &  & \\
\hline
\end{tabular}
\end{center}
  \caption{\sl The coset manifolds resulting from the  uplift of the three-dimensional symmetric manifolds listed in the first column. The table gives in  each case the resulting number of multiplets. The first chain of theories (first row) can be uplifted in two different ways to six dimensions, giving in one case a theory with one tensor multiplet, and in the other case a theory with only tensor multiplets and no vector multiplets. \label{upliftchain}}
\end{table}

In subsection 4.3 we will discuss the general case, in which also hypermultiplets in dimensions higher than three are present. This case  is more complicated to analyse. Indeed, the dimensional reduction to three dimensions now leads to two different hypermultiplet sectors.
 In the special case of symmetric manifolds the global symmetry is then given by the product of two groups
 $G_1 \times G_2$ occurring in
 eq.~\eqref{quatsymmspaces}. A possible way to proceed using generalised Kac-Moody constructions was conjectured in
 \cite{Kleinschmidt:2008jj}.
Here we will make a different proposal. We will show that,  for the particular case that the three-dimensional symmetry is $\text{SO}(4,n)\times \text{SO}(4,m)$, the brane structure in any dimension can be obtained by requiring that the theory is a truncation of the theory with sixteen supercharges whose three-dimensional symmetry is $\text{SO}(8,n+m)$. We will  argue that this result is universal for all the theories that admit an uplift to six dimensions.

\subsection{Symmetric theories in three dimensions and their oxidation}

In this subsection we want to determine the branes for all the chains of theories in Table \ref{upliftchain}.
In \cite{Riccioni:2008jz} it was shown that for these theories one can use the Kac-Moody analysis to obtain all the fields of the theory. This includes the $(D-1)$- and $D$-forms, that are not associated to propagating degrees of freedom. One considers the very-extended Kac-Moody algebra $G^{+++}$, whose Dynkin diagram is obtained starting from the affine extension of the three-dimensional symmetry group $G$ and attaching two more simply-laced nodes to the affine node.
The spectrum of the three-dimensional theory is then obtained by decomposing the adjoint representation of $G^{+++}$ in terms of $\text{GL}(3,\mathbb{R}) \times G$. In particular, the antisymmetric representations of $\text{GL}(3,\mathbb{R})$ are associated to the form fields in the theory, and this method thus gives their representations under $G$. Similarly, decomposing the same algebra in terms of representations of $\text{GL}(4,\mathbb{R})$ gives the spectrum of the four-dimensional theory, whose internal symmetry is given by the nodes of $G^{+++}$ which are not connected to the simple $\text{SL}(4,\mathbb{R})$ part of $\text{GL}(4,\mathbb{R})$. The same applies to higher dimensions.

The Kac-Moody method gives a  simple way of understanding the chains of symmetry groups in Table \ref{upliftchain} and why these theories can only be uplifted at most to six dimensions \cite{Riccioni:2008jz} (see also \cite{Keurentjes:2002xc}). In the case of  $\text{SO}(4,n)$, the affine node is attached to node 2 of the Tits-Satake diagram in the first row of Table \ref{quaternionictitssatake}. Deleting node 2 one obtains the symmetry $\text{SO}(2,n-2) \times \text{SU}(1,1)$ of the
four-dimensional theory, where the Cartan generator associated to the deleted node is the $\mathbb{R}^+$ factor of $\text{GL}(4,\mathbb{R})$.
Deleting nodes 1 and 3 (with node 2 being part of the $\text{SL}(5,\mathbb{R})$ in the Kac-Moody algebra) gives the symmetry $\text{SO}(1,n-3) \times \mathbb{R}^{+}$ of the five-dimensional theory. The extra internal  $\mathbb{R}^{+}$ symmetry is due to the fact that two non-compact nodes are deleted. There are two possible six-dimensional theories. The first one is obtained by deleting nodes 1 and 4 (with nodes 2 and 3 being part of the $\text{SL}(6,\mathbb{R})$ in the Kac-Moody algebra) giving a symmetry $\text{SO}(1,1) \simeq \mathbb{R}^+$, associated to the dilaton,  times the compact symmetry $\text{SO}(n-4)$ which is associated to the vector multiplets and does not correspond to a scalar manifold. The second one is obtained by deleting node 3 (with nodes 1 and 2 being part of the $\text{SL}(6,\mathbb{R})$ in the Kac-Moody algebra) which results in the symmetry $\text{SO}(1,n-3)$. This reproduces all the symmetries in the first row of Table \ref{upliftchain}. Note that a further uplift to seven
dimensions would correspond to deleting node 5, but this is impossible because node 5 is a compact node of the Tits-Satake diagram, and thus cannot be the scaling symmetry of a seven-dimensional theory.

This construction can be repeated for all the groups in Table \ref{quaternionictitssatake}. The uplift of the $\text{F}_{4(4)}$ theory, where the affine node is attached to node 1 of the Tits-Satake diagram in the second row of  Table \ref{quaternionictitssatake}, corresponds to deleting nodes 1 (four dimensions), 2 (five dimensions) and 3 (six dimensions). This precisely gives rise to the chain of groups in the second row of Table \ref{upliftchain}.  The $\text{E}_{6(2)}$ theory, where the affine node is attached to node 6 of the Tits-Satake diagram,
 is uplifted by deleting node 6 (four dimensions), 3 (five dimensions) and nodes 2 and 4 (six dimensions)\,\footnote{Note that the deletion of nodes 2 and 4 gives only one non-compact Cartan generator because the two nodes are connected by an arrow, resulting in one compact and one non-compact Cartan generator.}. The resulting real forms are given in the third row of Table \ref{upliftchain}. Similarly, the $\text{E}_{7(-5)}$ theory,
where the affine node is attached to node 6,
  is uplifted by deleting nodes 6, 5 and 4, and the $\text{E}_{8(-24)}$ theory,
  where the affine node is attached to node 1,
  is uplifted by deleting nodes 1, 2 and 3. The last three cases cannot be uplifted to six dimensions. The $\text{G}_{2(2)}$ theory is uplifted to four dimension by deleting node 1 and to pure five dimensional supergravity by deleting node 2. The $\text{SU}(n,2)$ theory can only be uplifted to four dimensions by deleting nodes 1 and $n+1$. Finally, the $\text{USp}(2n,2)$ theory cannot be uplifted at all.

The fact that in the uplifting process one always deletes either an unpainted node or a pair of unpainted nodes that are connected by an arrow implies that one can read the Cartan involution acting on the roots of the diagram of the symmetry group of the uplifted theory by simply looking at the Cartan involution in Table \ref{quaternionictitssatake} and ignoring the roots that have been deleted. This has the consequence  that the restricted root algebras of the symmetry groups resulting from uplifting the  $\text{E}_{6(2)}$, $\text{E}_{7(-5)}$ and $\text{E}_{8(-24)}$ theories are the same in any dimension and are equal to the algebras that result from uplifting the $\text{F}_{4(4)}$ theory. As we will see, this will imply that the whole structure of 1/2-BPS branes of these theories coincides.

Given the fields of the theory as obtained using the Kac-Moody method, we want to select those components that are associated to 1/2-supersymmetric branes. In the case of the maximal theory, this was achieved in \cite{Kleinschmidt:2011vu}, where it was shown that the branes correspond to the roots of positive squared length of the Kac-Moody algebra $\text{E}_{8(8)}^{+++}$. This gives exactly the same classification of \cite{Bergshoeff:2011zk,Bergshoeff:2011ee,Bergshoeff:2011qk,Bergshoeff:2012ex}, based on supergravity methods. Indeed, it was shown in
\cite{Bergshoeff:2013sxa} that what one is actually counting in both cases are the longest weights of each representation of the fields. The theories with sixteen supercharges also admit a Kac-Moody description \cite{Schnakenburg:2004vd}, which was used in \cite{Bergshoeff:2007vb} to determine the full spectrum of the theory. The real form of the algebra is in this case not maximally non-compact, and thus the weights of the representations of the fields are not necessarily real. In order to obtain the (single) 1/2-supersymmetric branes, one selects the representations whose highest weights correspond to roots of maximum positive squared length of the Kac-Moody algebra, and then counts the real longest weights of those representations. This is exactly what we did in the previous section, which reproduces the supergravity analysis of \cite{Bergshoeff:2012jb}.

In this subsection we want to perform the same analysis for the Kac-Moody algebras associated to the theories with eight supercharges. We will first list all the representations of the fields whose highest weights correspond to roots of maximum positive squared length of the Kac-Moody algebra\,\footnote{This was determined  using the software Simplie \cite{Bergshoeff:2007qi}.}. We will then count the number of real longest weights in those representations. This number is the number of single 1/2-supersymmetric branes.
 We will find that for all  theories only 0-branes and 1-branes occur. As we have seen in the previous subsection, in six dimensions there are no 0-branes, and the 1-branes are selfdual. In five dimensions the 0-branes are dual to the 1-branes. In four dimensions the 0-branes are selfdual while the 1-branes are defect branes. Finally, in three dimensions the 0-branes are defect branes and the 1-branes are domain walls. Only the latter branes are not associated to propagating degrees of freedom. This implies that, actually, the Kac-Moody method is only essential to count the three-dimensional domain walls.

\begin{table}[t]
\begin{center}
\begin{tabular}{|c|c|c|c|}
\hline \rule[-1mm]{0mm}{6mm} dim. & type of brane & field & $\#$ of branes \\
\hline
\hline  \rule[-1mm]{0mm}{6mm}
$D=3$ & 0-brane & $A_{1, A_1 A_2}$ & 24\\
\cline{2-4} \rule[-1mm]{0mm}{6mm} & 1-brane & $A_{2, AB}$ &8\\
\cline{3-4} \rule[-1mm]{0mm}{6mm} &  & $A_{2, A_1...A_4}$ &16\\
\hline \hline \rule[-1mm]{0mm}{6mm}
$D=4$ & 0-brane & $A_{1, Aa}$ & 8\\
\cline{2-4} \rule[-1mm]{0mm}{6mm} & 1-brane & $A_{2, ab}$ & 2\\
\cline{3-4} \rule[-1mm]{0mm}{6mm} &  & $A_{2, A_1 A_2}$ &4\\
\hline \hline \rule[-1mm]{0mm}{6mm}
$D=5$ & 0-brane  & $A_{1,A}$ $(\alpha=0)$ & 2 \\
\cline{3-4} \rule[-1mm]{0mm}{6mm} &  & $A_1$ $(\alpha=-2)$ & 1\\
\cline{2-4} \rule[-1mm]{0mm}{6mm} & 1-brane & $A_2$ $(\alpha=0)$ & 1\\
\cline{3-4} \rule[-1mm]{0mm}{6mm} &  & $A_{2,A}$ $(\alpha=-2)$ & 2\\
\hline \hline \rule[-1mm]{0mm}{6mm}
$D=6$ & 1-brane & $A_{2,A}$ & 2\\
\hline
\end{tabular}
\end{center}
  \caption{\sl The branes of the theories resulting from the uplift of the $\text{SO}(4,n)$ theory (with $n \geq4$). In six dimensions the $A$ index is either an $\text{SO}(1,1)$ index or an $\text{SO}(1,n-3)$ index, according to the two possible oxidations.
   \label{branesofSOtheories}}
\end{table}

We start by considering the $\text{SO}(4,n)^{+++}$ case, giving the branes for the chain of theories listed in the first row of Table \ref{upliftchain}. In this case we do not actually need the Tits-Satake machinery that we have developed in this paper, since we can actually use the light-cone rules\,\footnote{We will only consider the case in which $n \geq 4$, leaving the  $n<4$ case to the reader.}.
In three dimensions, the global symmetry is $\text{SO}(4,n)$ and the fields associated to the roots of maximum positive squared length of the Kac-Moody algebra are
\begin{equation}
 A_{1, A_1 A_2}   \qquad A_{2, AB} \qquad A_{2, A_1 ...A_4} \qquad A_{3, A B_1 ...B_5} \quad , \label{threedimSOSimplie}
\end{equation}
where our notation means that the fields belong to the irreducible representation with Young tableaux made of different columns, each of length $n$ equal to the number of repeated indices $A_1 A_2 ...A_n$.  Using the light-cone rule, we find that the 1-forms lead to ${4 \choose 2} \times 2^2 = 24$ 0-branes, while the 2-forms $A_{2, AB}$ give 8 1-branes and $A_{2, A_1 ...A_4}$ give ${4 \choose 4} \times 2^4 = 16$ 1-branes. The 3-forms do not give any brane.
In four dimensions, the global symmetry is $\text{SO}(2,n-2)\times \text{SL}(2,\mathbb{R})$, and the relevant fields are
 \begin{equation}
 A_{1, A a}   \quad A_{2, ab} \quad A_{2, A_1 A_2} \quad A_{3, A_1 A_2 A_3 a} \quad A_{4, ab A_1 ...A_4} \quad A_{4, A B_1 ...B_3} \quad  ,\label{fourdimSOSimplie}
\end{equation}
where $a$ labels the doublet of $\text{SL}(2,\mathbb{R})$ and the pair $ab$ is symmetrised. Applying the light-cone rule, together with the longest weight rule for the split form $\text{SL}(2,\mathbb{R})$, one finds that the 1-forms give $4 \times 2 = 8$ 0-branes, while the 2-forms $A_{2, ab}$ give 2 1-branes and the 2-forms $A_{2, A_1 A_2}$ give ${2 \choose 2} \times 2^2 = 4$ 1-branes. All the other fields give no branes.
In five dimensions the global symmetry is $\text{SO}(1,n-3) \times \mathbb{R}^+$.
Like in the cases of maximal and half-maximal supergravity \cite{Bergshoeff:2010xc,Bergshoeff:2012jb}, it is convenient, for $D>4$, to classify the potentials according to a number $\alpha$ that specifies how the tension of the brane that couples to the potential
scales with respect to the the $\mathbb{R}^+$ dilaton\,\footnote{As we will see in the next section, the $\mathbb{R}^+$ dilaton is the heterotic string dilaton.}. This scaling $\alpha$ follows from the $\mathbb{R}^+$ weight and the rank of the potential.
According to the Kac-Moody analysis we have the following fields:
  \begin{eqnarray}
  & & \alpha=0: \ \ \,\qquad A_{1,{A}} \quad A_2 \,, \nonumber \\
& & \alpha=-2: \qquad A_{1} \quad A_{2,{A}} \quad A_{3, {A}_1 {A}_2} \quad A_{4, {A}_1 {A}_2 {A}_3} \quad A_{5 , {A}_1 ...{A}_4} \,,\nonumber \\
  & & \alpha=-4: \qquad A_{4,{A}_1 {A}_2} \quad A_{5, {A}, {B}_1 {B}_2} \quad .
  \label{fivedimSOSimplie}
  \end{eqnarray}
 The 1-form $A_1$ gives one 0-brane, while the 1-form $A_{1,A}$ gives 2 0-branes. Similarly, one obtains $1+2$ 1-branes. All the other fields give no branes. Finally, there are two possible six-dimensional theories. The first is the one with $n-4$ vector multiplets and one tensor multiplet. We list the fields as representations of the compact global symmetry $\text{SO}(n-4)$\,\footnote{There  is also  a non-compact symmetry $\text{SO}(1,1)$.}, obtaining
  \begin{equation}
  A_{1,A} \ \  2\times A_2 \  \ A_{3,A}\  \ A_{4,A_1 A_2}\  \ A_{5, A}\  \ A_{5, A_1 A_2 A_3}\  \ A_{6 , A_1 ...A_4}\  \ A_{6, AB} \quad .
  \end{equation}
Only the fields with no internal indices can give branes. This means that there are only two 1-branes, one with $\alpha=0$ and one with $\alpha=-2$. Like above,  $\alpha$ is related to the $\text{SO}(1,1) \simeq \mathbb{R}^+$ weight and the
rank of the potential.
 The second six-dimensional theory has $n-3$ tensor multiplets and no vector multiplets. The global symmetry is   $\text{SO}(1,n-3)$ and the relevant fields are
\begin{equation}
A_{2, A} \qquad A_{4, A_1 A_2} \quad .
\end{equation}
Using the light-cone rule, one finds again only two 1-branes. We have summarised the final result in Table \ref{branesofSOtheories}.

\begin{table}
\begin{center}
\begin{tabular}{|c|c|c|c|c|}
\hline \rule[-1mm]{0mm}{6mm} dim. & type of brane & repr. & highest weight & $\#$ of branes \\
\hline
\hline  \rule[-1mm]{0mm}{6mm}
$D=3$ & 0-brane & ${\bf 52} \ (1\ 0\ 0\ 0 )$ & $2\alpha_1 + 3 \alpha_2 +4 \alpha_3+ 2 \alpha_4$ & 24\\
\cline{2-5} \rule[-1mm]{0mm}{6mm} & 1-brane & ${\bf 324} \ (0 \ 0 \ 0\ 2)$ & $2\alpha_1 + 4\alpha_2 + 6 \alpha_3 + 4 \alpha_4$ & 24\\
\hline \hline \rule[-1mm]{0mm}{6mm}
$D=4$ & 0-brane & ${\bf 14} \ (1\ 0\ 0 )$ & $\frac{3}{2} \alpha_2 + 2 \alpha_3+ \alpha_4$ &  8\\
\cline{2-5} \rule[-1mm]{0mm}{6mm} & 1-brane & ${\bf 21} \ (0 \ 0 \ 2)$ & $\alpha_2 + 2 \alpha_3 + 2 \alpha_4$ & 6\\
\hline \hline \rule[-1mm]{0mm}{6mm}
$D=5$ & 0-brane & ${\bf 6}\  (2\ 0 )$ & $ \tfrac{4}{3} \alpha_3 + \tfrac{2}{3} \alpha_4$ &3\\
\cline{2-5} \rule[-1mm]{0mm}{6mm} & 1-brane & ${\bf \overline{6}} \ (0 \ 2)$ & $ \tfrac{2}{3} \alpha_3 + \tfrac{4}{3} \alpha_4$ & 3\\
\hline \hline \rule[-1mm]{0mm}{6mm}
$D=6$ & 1-brane & ${\bf 3}\  (2)$ & $\alpha_4$ & 2\\
\hline
\end{tabular}
\end{center}
  \caption{\sl The branes of the theories resulting from the uplift of the $\text{F}_{4(4)}$ theory. The symmetries in dimensions 4, 5 and 6 can be read from the second row of Table \ref{upliftchain}. They correspond to deleting  nodes 1, 2 and 3, respectively,  of the  $\text{F}_{4(4)}$ Tits-Satake diagram in Table \ref{quaternionictitssatake}. The number of branes corresponds to the number of longest weights in the representation.
   \label{F4chain}}
\end{table}

We now move on to consider the other theories. The $\text{F}_{4(4)}$ chain of theories in the second row of Table \ref{upliftchain} is simple to analyse because the symmetry groups are all maximally non-compact. This implies that the weights of the representations are all real, and thus once one has identified the relevant representations, one only has to count the number of longest weights of these representations. The outcome of this analysis is summarised in Table \ref{F4chain}. Remarkably, the number of branes that one obtains in any dimension coincides with that of the previous chain of theories.

We next consider the $\text{E}_{6(2)}$,  $\text{E}_{7(-5)}$ and  $\text{E}_{8(-24)}$ cases. What one finds is that the global symmetry, in all  cases and in any dimension, is  such that the restricted root algebra is the $\text{F}_{4(4)}$ chain of symmetries. Moreover the real highest weights, when written in terms of the restricted simple roots, are exactly the highest weights of the representations listed in Table \ref{F4chain}. This means that the number of 1/2-BPS branes in each dimension is the same for all these four theories. The final result is summarised in Tables \ref{E6chain}, \ref{E7chain} and \ref{E8chain}.

\begin{table}\footnotesize
\begin{center}
\begin{tabular}{|c|c|c|c|c|c|}
\hline \rule[-1mm]{0mm}{6mm} dim. & brane & repr. & highest weight & restricted repr. & $\#$ \\
\hline
\hline  \rule[-1mm]{0mm}{6mm}
$D=3$ & 0-brane & ${\bf 78} \ (0\ 0\ 0\ 0\ 0\ 1)$ & $\alpha_1 + 2 \alpha_2 + 3 \alpha_3 + 2 \alpha_4 + \alpha_5 + 2 \alpha_6$ &  ${\bf 52} \ (1\ 0\ 0\ 0 )$ & 24\\
\cline{2-6} \rule[-1mm]{0mm}{6mm} & 1-brane & ${\bf 650} \ (1 \ 0 \ 0 \ 0\ 1 \ 0 )$ & $2 \alpha_1 + 3 \alpha_2 + 4 \alpha_3 + 3 \alpha_4 + 2 \alpha_5 + 2 \alpha_6$ & ${\bf 324} \ (0 \ 0 \ 0\ 2)$ & 24\\
\cline{2-6}
\rule[-1mm]{0mm}{6mm} & 2-brane & ${\bf 5824} \ (1 \ 1 \ 0 \ 0\ 0 \ 0 )$ & $3 \alpha_1 + 5 \alpha_2 + 6 \alpha_3 + 4 \alpha_4 + 2 \alpha_5 + 3 \alpha_6$ & - & -\\
\cline{3-6}
\rule[-1mm]{0mm}{6mm}
&  & ${\bf \overline{5824}} \ (0 \ 0 \ 0 \ 1\ 1 \ 0 )$ & $2 \alpha_1 + 4 \alpha_2 + 6 \alpha_3 + 5 \alpha_4 + 3 \alpha_5 + 3 \alpha_6$ & - & -\\
\hline \hline
\rule[-1mm]{0mm}{6mm}
$D=4$ & 0-brane & ${\bf 20} \ (0 \ 0\ 1\ 0\ 0 )$ & $\tfrac{1}{2} \alpha_1 + \alpha_2 + \tfrac{3}{2} \alpha_3 + \alpha_4 +\tfrac{1}{2} \alpha_5$ &  ${\bf 14} \ (1\ 0\ 0 )$ & 8\\
\cline{2-6} \rule[-1mm]{0mm}{6mm} & 1-brane & ${\bf 35} \ (1 \ 0 \ 0 \ 0\  1 )$ & $ \alpha_1+\alpha_2 + \alpha_3 +\alpha_4 +\alpha_5$ & ${\bf 21} \ (0 \ 0 \ 2)$ & 6\\
\cline{2-6} \rule[-1mm]{0mm}{6mm} & 2-brane & ${\bf 70} \ (1\ 1 \ 0 \ 0\ 0)$ & $ \tfrac{3}{2} \alpha_1+ 2 \alpha_2 + \tfrac{3}{2} \alpha_3 + \alpha_4+ \tfrac{1}{2} \alpha_5 $ & - & -\\
\cline{3-6} \rule[-1mm]{0mm}{6mm} &  & ${\bf \overline{70}} \ (0\ 0 \ 0 \ 1\ 1)$ & $ \tfrac{1}{2} \alpha_1+  \alpha_2 + \tfrac{3}{2} \alpha_3 + 2\alpha_4+ \tfrac{3}{2} \alpha_5 $ & - & -\\
\cline{2-6} \rule[-1mm]{0mm}{6mm} & 3-brane & ${\bf {280}} \ (2\ 0 \ 0\ 1\  0)$ & $   2\alpha_1 + 2 \alpha_2 + 2 \alpha_3 + 2 \alpha_4 + \alpha_5
$ & - & -\\
\cline{3-6} \rule[-1mm]{0mm}{6mm} &  & ${\bf \overline{280}} \ (0 \ 1 \ 0\ 0\  2)$ & $   \alpha_1 + 2 \alpha_2 + 2 \alpha_3 + 2 \alpha_4 + 2\alpha_5
$ & - & -\\
\hline
\hline  \rule[-1mm]{0mm}{6mm}
$D=5$ & 0-brane & ${\bf (\overline{3},3)}\  (0 \ 1 \ 1 \ 0 )$ & $\tfrac{1}{3} \alpha_1 + \tfrac{2}{3} \alpha_2 + \tfrac{2}{3} \alpha_4 + \tfrac{1}{3} \alpha_5 $ &  ${\bf 6}\  (2\ 0 )$ & 3\\
\cline{2-6} \rule[-1mm]{0mm}{6mm} & 1-brane & ${\bf ({3},\overline{3})}\  (1 \ 0 \ 0 \ 1 )$ & $\tfrac{2}{3} \alpha_1 + \tfrac{1}{3} \alpha_2 + \tfrac{1}{3} \alpha_4 + \tfrac{2}{3} \alpha_5 $ &  ${\bf \overline{6}}\  (0\ 2 )$ & 3\\
\cline{2-6} \rule[-1mm]{0mm}{6mm} & 2-brane & ${\bf ({8},{1})}\  (1 \ 1 \ 0 \ 0 )$ & $ \alpha_1 +  \alpha_2  $ &  - & -\\
\cline{3-6} \rule[-1mm]{0mm}{6mm} &  & ${\bf ({1},{8})}\  (0 \ 0 \ 1 \ 1 )$ & $ \alpha_4 +  \alpha_5  $ &  - & -\\
\cline{2-6} \rule[-1mm]{0mm}{6mm} & 3-brane & ${\bf ({6},{3})}\  (2 \ 0 \ 1 \ 0 )$ & $ \tfrac{4}{3}\alpha_1 + \tfrac{2}{3} \alpha_2 + \tfrac{2}{3} \alpha_4 + \tfrac{1}{3} \alpha_5 $ &  - & -\\
\cline{3-6} \rule[-1mm]{0mm}{6mm} &  & ${\bf (\overline{3},\overline{6})}\  (0 \ 1 \ 0 \ 2 )$ &  $ \tfrac{1}{3}\alpha_1 + \tfrac{2}{3} \alpha_2 + \tfrac{2}{3} \alpha_4 + \tfrac{4}{3} \alpha_5 $ &  - & -\\
\cline{2-6} \rule[-1mm]{0mm}{6mm} & 4-brane & ${\bf ({15},\overline{3})}\  (2 \ 1 \ 0 \ 1 )$ & $ \tfrac{5}{3}\alpha_1 + \tfrac{4}{3} \alpha_2 + \tfrac{1}{3} \alpha_4 + \tfrac{2}{3} \alpha_5 $ &  - & -\\
\cline{3-6} \rule[-1mm]{0mm}{6mm} &  & ${\bf ({3},\overline{15})}\  (1 \ 0 \ 1 \ 2 )$ &  $ \tfrac{2}{3}\alpha_1 + \tfrac{1}{3} \alpha_2 + \tfrac{4}{3} \alpha_4 + \tfrac{5}{3} \alpha_5 $ & - & -\\
\hline
\hline  \rule[-1mm]{0mm}{6mm}
$D=6$ & 0-brane & ${\bf ({2},1)}\  (1 \ 0  )$ & $\tfrac{1}{2} \alpha_1 $ &  - & -\\
\cline{3-6} \rule[-1mm]{0mm}{6mm} & & ${\bf ({1},2)}\  (0 \ 1  )$ & $\tfrac{1}{2} \alpha_5 $ &  - & -\\
\cline{2-6} \rule[-1mm]{0mm}{6mm} & 1-brane & ${\bf ({2},{2})}\  (1  \ 1 )$ & $\tfrac{1}{2} \alpha_1  + \tfrac{1}{2} \alpha_5 $ &  ${\bf {3}}\  ( 2 )$ & 2\\
\cline{2-6} \rule[-1mm]{0mm}{6mm}
 & 2-brane & ${\bf ({2},1)}\  (1 \ 0  )$ & $\tfrac{1}{2} \alpha_1 $ &  - & -\\
\cline{3-6} \rule[-1mm]{0mm}{6mm} & & ${\bf ({1},2)}\  (0 \ 1  )$ & $\tfrac{1}{2} \alpha_5 $ &  - & -\\
\cline{2-6} \rule[-1mm]{0mm}{6mm}
 & 3-brane & ${\bf ({1},3)}\  (0 \ 2  )$ & $ \alpha_5 $ &  - & -\\
\cline{3-6} \rule[-1mm]{0mm}{6mm} & & ${\bf ({3},1)}\  (2 \ 0  )$ & $ \alpha_1 $ &  - & -\\
\cline{2-6} \rule[-1mm]{0mm}{6mm}
 & 4-brane & ${\bf ({2},3)}\  (1 \ 2  )$ & $ \tfrac{1}{2} \alpha_1 +\alpha_5 $ &  - & -\\
\cline{3-6} \rule[-1mm]{0mm}{6mm} & & ${\bf ({3},2)}\  (2 \ 1  )$ & $ \alpha_1 + \tfrac{1}{2} \alpha_5 $ &  - & -\\
\cline{2-6} \rule[-1mm]{0mm}{6mm}
 & 5-brane & ${\bf ({4},2)}\  (3 \ 1  )$ & $ \tfrac{3}{2} \alpha_1 +\frac{1}{2} \alpha_5 $ &  - & -\\
\cline{3-6} \rule[-1mm]{0mm}{6mm} & & ${\bf ({2},4)}\  (1 \ 3  )$ & $ \tfrac{1}{2} \alpha_1 + \tfrac{3}{2} \alpha_5 $ & - & -\\
\hline
\end{tabular}
\end{center}
  \caption{\sl The branes resulting from the uplift of the $\text{E}_{6(2)}$ theory. The symmetries in dimensions 4, 5 and 6 can be read from the third row of Table \ref{upliftchain}. They correspond to deleting  nodes 6, 3 and 2 and 4 of the  $\text{E}_{6(2)}$ Tits-Satake diagram in Table \ref{quaternionictitssatake}, respectively. Only for the real longest weights we have listed in the fifth column the corresponding representations of the restricted root algebras, which coincide with the  $\text{F}_{4(4)}$ chain. The last column gives the
number of branes. \label{E6chain}}
\end{table}

\begin{table}\footnotesize
\begin{center}
\begin{tabular}{|c|c|c|c|c|c|}
\hline \rule[-1mm]{0mm}{6mm} dim. &  brane & repr. & highest weight & restricted repr. & $\#$  \\
\hline
\hline  \rule[-1mm]{0mm}{6mm}
$D=3$ & 0-brane & ${\bf 133} \ (0\ 0\ 0\ 0\ 0\ 1 \ 0)$ & $\beta_1 + 2 \alpha_2 + 3 \beta_3 + 4 \alpha_4 + 3\alpha_5 + 2 \alpha_6 + 2 \beta_7$ &  ${\bf 52} \ (1\ 0\ 0\ 0 )$ & 24\\
\cline{2-6} \rule[-1mm]{0mm}{6mm} & 1-brane & ${\bf 1539} \ (0 \ 1 \ 0 \ 0 \ 0\ 0 \ 0 )$ & $2 \beta_1 + 4 \alpha_2 + 5 \beta_3 + 6 \alpha_4 + 4 \alpha_5 + 2 \alpha_6 + 3 \beta_7$ & ${\bf 324} \ (0 \ 0 \ 0\ 2)$ & 24\\
\cline{2-6}
\rule[-1mm]{0mm}{6mm} & 2-brane & ${\bf 40755} \ (1 \ 0 \ 0 \ 0\ 0 \ 0 \ 1)$ & $3 \beta_1 + 5 \alpha_2 + 7 \beta_3 + 9 \alpha_4 + 6 \alpha_5 + 3 \alpha_6 + 5 \beta_7$ & - & -\\
\hline \hline
\rule[-1mm]{0mm}{6mm}
$D=4$ & 0-brane & ${\bf 32} \ (0\ 0\ 0\ 0\ 1 \ 0 )$ & $\tfrac{1}{2} \beta_1 + \alpha_2 + \tfrac{3}{2} \beta_3 + 2\alpha_4 + \tfrac{3}{2} \alpha_5+ \beta_7$ &  ${\bf 14} \ (1\ 0\ 0 )$ & 8\\
\cline{2-6} \rule[-1mm]{0mm}{6mm} & 1-brane & ${\bf 66} \ (0 \ 1\ 0 \ 0 \ 0\ 0)$ & $\beta_1 +2 \alpha_2 + 2\beta_3 + 2\alpha_4 + \alpha_5+ \beta_7$ & ${\bf 21} \ (0 \ 0 \ 2)$ & 6\\
\cline{2-6} \rule[-1mm]{0mm}{6mm} & 2-brane & ${\bf 352} \ (1\ 0 \ 0 \ 0\ 0\ 1 )$ & $\tfrac{3}{2}\beta_1 + 2\alpha_2 + \tfrac{5}{2}\beta_3 + 3\alpha_4 + \tfrac{3}{2}\alpha_5 + 2 \beta_7$ & - & -\\
\cline{2-6} \rule[-1mm]{0mm}{6mm} & 3-brane & ${\bf 462} \ (0\ 0 \ 0 \ 0\ 0 \ 2)$ & $\beta_1 + 2\alpha_2 + 3\beta_3 + 4\alpha_4 + 2\alpha_5+ 3 \beta_7$ & - & -\\
\cline{3-6} \rule[-1mm]{0mm}{6mm} &  & ${\bf {2079}} \ (1\ 0 \ 1 \ 0\  0\ 0)$ & $2\beta_1 + 3\alpha_2 + 4\beta_3 + 4\alpha_4 + 2\alpha_5+ 2 \beta_7$ & - & -\\
\hline
\hline  \rule[-1mm]{0mm}{6mm}
$D=5$ & 0-brane & ${\bf \overline{15}}\  (0 \ 0 \ 0\ 1 \ 0 )$ & $\tfrac{1}{3} \beta_1 + \tfrac{2}{3} \alpha_2 +\beta_3 + \tfrac{4}{3} \alpha_4 + \tfrac{2}{3} \beta_7 $ &  ${\bf 6}\  (2\ 0 )$ & 3\\
\cline{2-6} \rule[-1mm]{0mm}{6mm} & 1-brane & ${\bf 15}\  (0\ 1 \ 0 \ 0 \ 0 )$ & $\tfrac{2}{3} \beta_1 + \tfrac{4}{3} \alpha_2 +\beta_3 + \tfrac{2}{3} \alpha_4 + \tfrac{1}{3} \beta_7 $ &  ${\bf \overline{6}}\  (0\ 2 )$ & 3\\
\cline{2-6} \rule[-1mm]{0mm}{6mm} & 2-brane & ${\bf 35}\  (1 \ 0 \ 0 \ 0\ 1 )$ & $ \beta_1 +  \alpha_2 +\beta_3 + \alpha_4 + \beta_7 $ &  - & -\\
\cline{2-6} \rule[-1mm]{0mm}{6mm} & 3-brane & ${\bf \overline{21}}\  (0 \ 0 \ 0 \ 0\ 2 )$ & $ \tfrac{1}{3}\beta_1 + \tfrac{2}{3} \alpha_2 + \beta_3+ \tfrac{4}{3} \alpha_4 + \tfrac{5}{3} \beta_7 $ &  - & -\\
\cline{3-6} \rule[-1mm]{0mm}{6mm} &  & ${\bf 105}\  (1 \ 0 \ 1 \ 0 \ 0 )$ &  $ \tfrac{4}{3}\beta_1 + \tfrac{5}{3} \alpha_2 +2 \beta_3 + \tfrac{4}{3} \alpha_4 + \tfrac{2}{3} \beta_7 $ & - & -\\
\cline{2-6} \rule[-1mm]{0mm}{6mm} & 4-brane & ${\bf 384}\  (1 \ 1 \ 0 \ 0\ 1 )$ & $ \tfrac{5}{3}\beta_1 + \tfrac{7}{3} \alpha_2 +2\beta_3 + \tfrac{5}{3} \alpha_4 + \tfrac{4}{3} \beta_7 $ & - & -\\
\hline
\hline  \rule[-1mm]{0mm}{6mm}
$D=6$ & 0-brane & ${\bf (\overline{4},2)}\  (0 \ 0\ 1 \ 1  )$ & $\tfrac{1}{4} \beta_1 + \tfrac{1}{2} \alpha_2 +\tfrac{3}{4} \beta_3 + \tfrac{1}{2} \beta_7 $ &  - & -\\
\cline{2-6} \rule[-1mm]{0mm}{6mm} & 1-brane & ${\bf ({6},{1})}\  (0\ 1  \ 0\ 0 )$ & $\tfrac{1}{2} \beta_1  +\alpha_2 + \tfrac{1}{2} \beta_3 $ &  ${\bf {3}}\  ( 2 )$ & 2\\
\cline{2-6} \rule[-1mm]{0mm}{6mm}
 & 2-brane & ${\bf ({4},2)}\  (1 \ 0 \ 0\ 1 )$ & $\tfrac{3}{4} \beta_1 + \tfrac{1}{2} \alpha_2 + \tfrac{1}{4} \beta_3 + \tfrac{1}{2} \beta_7 $ & - & -\\
\cline{2-6} \rule[-1mm]{0mm}{6mm}
 & 3-brane & ${\bf ({1},3)}\  (0\ 0\ 0 \ 2  )$ & $ \beta_7 $ &  - & -\\
\cline{3-6} \rule[-1mm]{0mm}{6mm} & & ${\bf ({15},1)}\  (1\ 0\ 1 \ 0  )$ & $ \beta_1 + \alpha_2 + \beta_3$ & - & -\\
\cline{2-6} \rule[-1mm]{0mm}{6mm}
 & 4-brane & ${\bf ({20},2)}\  (1 \ 1 \ 0\  1  )$ & $ \tfrac{5}{4} \beta_1 +\tfrac{3}{2} \alpha_2 + \tfrac{3}{4} \beta_3 +\frac{1}{2}\beta_7 $ &  - & -\\
\cline{2-6} \rule[-1mm]{0mm}{6mm}
 & 5-brane & ${\bf ({10},3)}\  (2 \ 0\ 0  \ 2  )$ & $ \tfrac{3}{2} \beta_1+\alpha_2 + \tfrac{1}{2} \beta_3  + \beta_7 $ &  - & -\\
\cline{3-6} \rule[-1mm]{0mm}{6mm} & & ${\bf ({64},1)}\  (1 \ 1\ 1 \ 0  )$ & $ \tfrac{3}{2} \beta_1 + 2\alpha_2  + \tfrac{3}{2} \beta_3 $ & - & -\\
\hline
\end{tabular}
\end{center}
  \caption{\sl The branes of the theories resulting from the uplift of the $\text{E}_{7(-5)}$ theory. The symmetries in dimensions 4, 5 and 6 can be read from the fourth row of Table \ref{upliftchain}. They  correspond to deleting  nodes 6, 5 and 4 of the  $\text{E}_{7(-5)}$ Tits-Satake diagram in Table \ref{quaternionictitssatake}, respectively. \label{E7chain}}
\end{table}

\begin{table}\scriptsize
\begin{center}
\begin{tabular}{|c|c|c|c|c|c|}
\hline \rule[-1mm]{0mm}{5mm} dim. & brane & repr. & highest weight & restricted repr. & $\#$  \\
\hline
\hline  \rule[-1mm]{0mm}{5mm}
$D=3$ & 0-brane & ${\bf 248} \ (1 \ 0\ 0\ 0\ 0\ 0\ 0 \ 0)$ & $2 \alpha_1 + 3 \alpha_2 + 4 \alpha_3 + 5 \beta_4 + 6\beta_5 + 4 \beta_6 + 2 \alpha_7+ 3 \beta_8$ &  ${\bf 52} \ (1\ 0\ 0\ 0 )$ & 24\\
\cline{2-6} \rule[-1mm]{0mm}{5mm} & 1-brane & ${\bf 3875} \ (0 \ 0 \ 0 \ 0 \ 0\ 0 \ 1 \ 0 )$ & $2 \alpha_1 + 4 \alpha_2 + 6 \alpha_3 + 8 \beta_4 + 10 \beta_5 + 7 \beta_6 + 4 \alpha_7+ 5 \beta_8$ & ${\bf 324} \ (0 \ 0 \ 0\ 2)$ & 24\\
\cline{2-6}
\rule[-1mm]{0mm}{5mm} & 2-brane & ${\bf 147250} \ (0 \ 0 \ 0 \ 0\ 0 \ 0 \ 0\ 1)$ & $3 \alpha_1 + 6 \alpha_2 + 9 \alpha_3 + 12 \beta_4 + 15 \beta_5 + 10 \beta_6 + 5 \alpha_7+ 8\beta_8$ & - & -\\
\hline \hline
\rule[-1mm]{0mm}{5mm}
$D=4$ & 0-brane & ${\bf 56} \ (1 \ 0\ 0\ 0\ 0\ 0 \ 0 )$ & $\tfrac{3}{2} \alpha_2 +2 \alpha_3 + \tfrac{5}{2} \beta_4 + 3 \beta_5 + 2 \beta_6 + \alpha_7 + \tfrac{3}{2} \beta_8$ &  ${\bf 14} \ (1\ 0\ 0 )$ & 8\\
\cline{2-6} \rule[-1mm]{0mm}{5mm} & 1-brane & ${\bf 133} \ (0 \ 0 \ 0 \ 0\ 0\ 1 \ 0)$ & $ \alpha_2 + 2 \alpha_3 + 3 \beta_4 + 4 \beta_5 + 3 \beta_6 + 2 \alpha_7 + 2 \beta_8$ & ${\bf 21} \ (0 \ 0 \ 2)$ & 6\\
\cline{2-6} \rule[-1mm]{0mm}{5mm} & 2-brane & ${\bf 912} \ (0\ 0 \ 0 \ 0\ 0\ 0\ 1)$ & $ \tfrac{3}{2} \alpha_2 + 3 \alpha_3 + \tfrac{9}{2} \beta_4 + 6 \beta_5 + 4 \beta_6 + 2 \alpha_7 + \tfrac{7}{2} \beta_8$ & - & -\\
\cline{2-6} \rule[-1mm]{0mm}{5mm} & 3-brane & ${\bf {8645}} \ (0\ 0 \ 0\ 0  \ 1\ 0\ 0)$ & $    2 \alpha_2 + 4 \alpha_3 + 6 \beta_4 + 8 \beta_5 + 6 \beta_6 + 3 \alpha_7 + 4 \beta_8
$ & - & -\\
\hline
\hline  \rule[-1mm]{0mm}{5mm}
$D=5$ & 0-brane & ${\bf {27}}\  (1  \ 0 \ 0\ 0 \ 0\ 0 )$ & $\tfrac{4}{3}\alpha_3 + \tfrac{5}{3} \beta_4 +2 \beta_5 + \tfrac{4}{3}\beta_6 +\tfrac{2}{3}\alpha_7 + \beta_8$
 &  ${\bf 6}\  (2\ 0 )$ & 3\\
\cline{2-6} \rule[-1mm]{0mm}{5mm} & 1-brane & ${\bf \overline{27}}\  (0\ 0 \ 0 \ 0 \ 1\ 0 )$ & $\tfrac{2}{3} \alpha_3 + \tfrac{4}{3} \beta_4 +2 \beta_5 + \tfrac{5}{3} \beta_6 + \tfrac{4}{3} \alpha_7+ \beta_8 $ &  ${\bf \overline{6}}\  (0\ 2 )$ & 3\\
\cline{2-6} \rule[-1mm]{0mm}{5mm} & 2-brane & ${\bf 78}\  (0 \ 0 \ 0 \ 0\ 0 \ 1)$ & $\alpha_3 + 2 \beta_4 + 3 \beta_5 + 2 \beta_6 +\alpha_7 + 2 \beta_8$ &  - & -\\
\cline{2-6} \rule[-1mm]{0mm}{5mm} & 3-brane & ${\bf {351}}\  (0 \ 0 \ 0 \ 1\ 0\ 0 )$ & $\tfrac{4}{3} \alpha_3 +\tfrac{8}{3} \beta_4 + 4 \beta_5 + \tfrac{10}{3} \beta_6 + \tfrac{5}{3} \alpha_7 + 2 \beta_8$ &  - & -\\
\cline{2-6} \rule[-1mm]{0mm}{5mm} & 4-brane & ${\bf 1728}\  (0 \ 0 \ 0 \ 0\ 1\ 1 )$ & $\tfrac{5}{3} \alpha_3 + \tfrac{10}{3} \beta_4 + 5 \beta_5 + \tfrac{11}{3} \beta_6 + \tfrac{7}{3} \alpha_7 + 3 \beta_8$ & - & -\\
\hline
\hline  \rule[-1mm]{0mm}{5mm}
$D=6$ & 0-brane & ${\bf 16}\  (1 \ 0 \ 0\ 0 \ 0  )$ & $\tfrac{5}{4} \beta_4 + \tfrac{3}{2} \beta_5 + \beta_6 + \tfrac{1}{2} \alpha_7 + \tfrac{3}{4} \beta_8$ &  - & -\\
\cline{2-6} \rule[-1mm]{0mm}{5mm} & 1-brane & ${\bf 10}\  (0\ 0  \ 0\ 1\ 0 )$ & $\tfrac{1}{2} \beta_4 + \beta_5 + \beta_6 +\alpha_7 +\tfrac{1}{2} \beta_8$ &  ${\bf {3}}\  ( 2 )$ & 2\\
\cline{2-6} \rule[-1mm]{0mm}{5mm}
 & 2-brane & ${\bf \overline{16}}\  (0 \ 0 \ 0\ 0\ 1 )$ & $ \tfrac{3}{4} \beta_4 + \tfrac{3}{2}\beta_5 + \beta_6 + \tfrac{1}{2} \alpha_7 + \tfrac{5}{4} \beta_8$ & - & -\\
\cline{2-6} \rule[-1mm]{0mm}{5mm}
 & 3-brane & ${\bf 45}\  (0\ 0\ 1 \ 0\ 0  )$ & $ \beta_4 + 2 \beta_5 + 2 \beta_6 + \alpha_7 + \beta_8 $ &  - & -\\
\cline{2-6} \rule[-1mm]{0mm}{5mm}
 & 4-brane & ${\bf 144}\  (0 \ 0 \ 0\  1\ 1  )$ & $ \tfrac{5}{4} \beta_4 +\tfrac{5}{2} \beta_5 + 2 \beta_6 +\tfrac{3}{2} \alpha_7  +\tfrac{7}{4} \beta_8$ &  - & -\\
\cline{2-6} \rule[-1mm]{0mm}{5mm}
 & 5-brane & ${\bf 320}\  (0 \ 0\ 1\ 1 \ 0  )$ & $ \tfrac{3}{2} \beta_4+ 3 \beta_5 + 3  \beta_6 +2 \alpha_7 +\frac{3}{2} \beta_8 $ &  - & -\\
\hline
\end{tabular}
\end{center}
  \caption{\sl The branes of the theories resulting from the uplift of the $\text{E}_{8(-24)}$ theory. The symmetries in dimensions 4, 5 and 6 can be read from the fifth row of Table \ref{upliftchain}. They correspond to deleting  nodes 1, 2 and 3 of the  $\text{E}_{8(-24)}$ Tits-Satake diagram in Table \ref{quaternionictitssatake}, respectively. \label{E8chain}}
\end{table}

The fact that the first five chains of theories in Table \ref{upliftchain} all give the same results as far as the 1/2-supersymmetric branes are concerned can be understood as follows. First of all, one
observes that the groups in the $\text{F}_{4(4)}$, $\text{E}_{6(2)}$, $\text{E}_{7(-5)}$ and $\text{E}_{8(-24)}$ chains all admit maximal subgroups that are of the form of those in the first chain, modulo simple compact factors. Let us consider the three-dimensional case first.  We notice that the following set of maximal embeddings holds,
  \begin{eqnarray}
  && \text{F}_{4(4)} \supset \text{SO}(4,5)\,, \nonumber \\
  & & \text{E}_{6(2)} \supset \text{SO}(4,6) \times \text{U}(1)\,, \nonumber \\
  & & \text{E}_{7(-5)} \supset \text{SO}(4,8) \times \text{SU}(2)\,, \nonumber \\
  & & \text{E}_{8(-24)} \supset \text{SO}(4,12) \quad ,  \label{FEEEunderSOD=3}
  \end{eqnarray}
 where the subgroups are all $\text{SO}(4,n)$ apart from compact factors. We have already shown (see Table \ref{F4chain}) that the 0-branes and the 1-branes of the $\text{F}_{4(4)}$ theory are the longest weights of the  ${\bf 52}$ and the ${\bf 324}$ respectively.
   If one decomposes these representations under $\text{SO}(4,5)$ one obtains
   \begin{eqnarray}
   && {\bf 52 } = {\bf 16} + {\bf 36}\,, \nonumber \\
   & & {\bf 324} = {\bf 1} + {\bf 9} + {\bf 16} + {\bf 44} +{\bf 126} + {\bf 128} \quad .
   \end{eqnarray}
The longest weights belong to the ${\bf 36}$ in the first case and to the ${\bf 44}$ and ${\bf 126}$ in the second case. This is exactly what one would obtain in the $\text{SO}(4,5) $ theory, because the ${\bf 36}$ is the representation with two antisymmetric indices (giving 24 0-branes), the ${\bf 44}$ the one with two symmetric indices (giving 8 1-branes) and finally the ${\bf 126}$ is the one with four antisymmetric indices (giving 16 1-branes). In other words, as far as the branes are concerned the $\text{F}_{4(4)}$ theory is the same as the $\text{SO}(4,5)$ theory, and the fields that are responsible for the symmetry enhancement are not associated to branes.

The same occurs for the other three theories we are considering. The ${\bf 78}$ and the ${\bf 650}$ of $\text{E}_{6(2)}$, see Table \ref{E6chain}, decompose under  $\text{SO}(4,6) \times \text{U}(1)$ as
     \begin{eqnarray}
   && {\bf 78 }\ \, = {\bf 1}(0) + {\bf 16} (-3) + {\bf \overline{16}}(-3) + {\bf 45}(0)\,, \nonumber \\
   & & {\bf 650} = {\bf 1}(0) + {\bf 10}(6) + {\bf 10}(-6) + {\bf 16}(3) +{\bf \overline{16}} (-3)+ {\bf 45} (0) + {\bf 54} (0) \nonumber\,, \\
   & & \qquad \quad + {\bf 144} (-3) + {\bf\overline{144}}(3) +{\bf 210} (0) \quad .
   \end{eqnarray}
The numbers in brackets  are the $\text{U}(1)$ weights.
The real longest weights belong to the ${\bf 45}$ in the first case and to the ${\bf 54}$ and ${\bf 210}$ in the second case. Like in the previous case, these representations  correspond to two antisymmetric, two symmetric and four antisymmetric vector indices   of $\text{SO}(4,6)$.
Similarly, decomposing the ${\bf 133}$ and the ${\bf 1539}$ of $\text{E}_{7(-5)}$, see Table \ref{E7chain}, under $ \text{SO}(4,8) \times \text{SU}(2)$ one gets
       \begin{eqnarray}
   && {\bf 133 }\ \, = {\bf (1,3)} + {\bf (66,1)}  + {\bf (32,2)}\,,\nonumber \\
   & & {\bf 1539} = {\bf (1,1)} + {\bf (32,2)} + {\bf (77,1)} + {\bf (66,3)} +{\bf (495,1)}+ {\bf (352,2)} \quad .
   \end{eqnarray}
The real longest weights belong to the ${\bf (66,1)}$ in the first case and to the ${\bf (77,1)}$ and  ${\bf (495,1)}$ in the second case. Again, the ${\bf 66}$, ${\bf 77}$ and ${\bf 495}$ are  the representations with two antisymmetric, two symmetric and four antisymmetric vector indices of  $\text{SO}(4,6)$, respectively.  Finally, the ${\bf 248}$ and the ${\bf 3875}$ of $\text{E}_{8(-24)}$, see Table \ref{E8chain}, decompose under  $\text{SO}(4,12)$ as
  \begin{eqnarray}
   & & {\bf 248}\ \, = {\bf 120} + {\bf 128}\,, \nonumber \\
   & & {\bf 3875} = {\bf 135} + {\bf 1820}+ {\bf 1920} \quad .
   \end{eqnarray}
The real longest weights are in the ${\bf 120}$, ${\bf 135}$ and ${\bf 1820}$, which correspond  again to two antisymmetric, two symmetric and four antisymmetric vector indices of   $\text{SO}(4,12)$. To summarise, if one decomposes the relevant representations under the subgroups in eq.~\eqref{FEEEunderSOD=3}, one finds that the branes are exactly in the same representations as those of the $\text{SO}(4,n)$ theory, and thus their number is always the same, i.e.~24 0-branes and 8+16 1-branes, because of the light-cone rules.

The same analysis can be repeated in four and five dimensions. In four dimensions the relevant maximal embeddings are
given by
  \begin{eqnarray}
  & & \text{Sp}(6,\mathbb{R}) \supset \text{SO}(2,3) \times \text{SU}(1,1)\,, \nonumber \\
  & & \text{SU}(3,3) \supset \text{SO}(2,4) \times \text{SU}(1,1)\times \text{U}(1)\,,  \nonumber \\
  & & \text{SO}^*(12) \supset \text{SO}(2,6) \times \text{SU}(1,1)\times \text{SU}(2)\,, \nonumber \\
  & & \text{E}_{7(-25)} \supset \text{SO}(2,10) \times \text{SU}(1,1) \quad . \label{FEEEunderSOD=4}
  \end{eqnarray}
The maximal subgroups are  always of the form  $\text{SO}(2,n-2) \times \text{SU}(1,1)$ up to compact factors.
 Similarly, in five dimensions one has the following maximal embeddings:
  \begin{eqnarray}
  & & \text{SL}(3,\mathbb{R}) \supset \text{SO}(1,2) \times \mathbb{R}^+\,, \nonumber \\
  & & \text{SL}(3,\mathbb{C}) \supset \text{SO}(1,3) \times \mathbb{R}^+ \times \text{U}(1)\,,  \nonumber \\
  & & \text{SU}^*(6) \supset \text{SO}(1,5) \times \mathbb{R}^+  \times \text{SU}(2)\,, \nonumber \\
  & & \text{E}_{6(-26)} \supset \text{SO}(1,9) \times \mathbb{R}^+   \quad . \label{FEEEunderSOD=5}
  \end{eqnarray}
 All maximal subgroups are given by $\text{SO}(1,n-3) \times \mathbb{R}^+ $ up to the same compact factors. One can show that decomposing the relevant representations of Tables \ref{F4chain}-\ref{E8chain} one finds that the real longest weights are always in the representations of Table \ref{branesofSOtheories}, which explains why the brane structure of all these theories is universal.

The last three cases in Table \ref{upliftchain} are special because none of them can be uplifted to six dimensions. While the three-dimensional global symmetry groups of the cases analysed above have all four non-compact Cartan generators, $\text{G}_{2(2)}$ and $\text{SU}(n,2)$ have two non-compact Cartan generators, and the theories can be uplifted at most to five and four dimensions respectively. The global symmetry group $\text{USp}(2n,2)$ has only one non-compact Cartan
generator and the corresponding theory only exists in three dimensions. One can repeat the analysis for all these cases. In the case of the $\text{G}_{2(2)}$ chain of theories, the only fields associated to  roots of the Kac-Moody algebra with maximum positive squared length are a 1-form in the ${\bf 14}$ (adjoint) of $\text{G}_{2(2)}$ in the three-dimensional theory and a 1-form in the ${\bf 4}$ of $\text{SU}(1,1)$ in the four-dimensional theory. The number of branes are
  \begin{eqnarray}
   & & D=3: \quad 6  \ \ {\rm 0-branes}\,, \nonumber \\
     & & D=4: \quad 2  \ \ {\rm 0-branes} \quad .
     \end{eqnarray}
In the case of the $\text{SU}(n,2)$ chain of theories, the only representations with real highest weight are
\begin{eqnarray}
& D=3: \quad  &{\rm 1-forms}: \ (1\ 0 \ 0 \ ...\ 0 \ 0\ 1) \ \rightarrow \ 4 \quad {\rm
0-branes}\,, \nonumber \\
&   &{\rm 2-forms}: \ (0\ 1 \ 0 \ ...\ 0 \ 1\ 0) \ \rightarrow \ 4 \quad {\rm
1-branes}\,, \nonumber \\
& D=4: \quad  &{\rm 2-forms}: \ (1\ 0 \ 0 \ ...\ 0 \ 0\ 1) \ \rightarrow \ 2 \quad {\rm
1-branes} \quad ,
\end{eqnarray}
where we have denoted the representations in terms of the Dynkin indices of the highest weight.
Finally, in the $\text{USp}(2n,2)$ theory in three dimensions only the highest weight of the 2-form, with Dynkin labels $(0 \ 2 \ 0\ ...\ 0\ 0)$, is a real weight, and this results into 2 1-branes.

\subsection{Inclusion of hypermultiplets}

While the six-dimensional classification of branes that we obtained in subsection 4.1 included the hypermultiplet sector, the analysis in any dimension of subsection 4.2 involved theories that are oxidations of symmetric three-dimensional theories, which do not include hypermultiplets in dimensions higher than three. This is due to the fact that the
world-volume of the branes we are considering can be at most four-dimensional. This implies that in six dimensions the 1/2-BPS branes have at least codimension two, but in five dimensions they can also have  codimension one and there are no constraints on the codimension in dimensions four and three. Therefore, the knowledge of the propagating degrees of freedom of the theory allows us to determine the full brane spectrum  in six dimensions only, while in lower dimensions one needs the knowledge of the $(D-1)$- and $D$-forms in the supersymmetric multiplets, that do not carry on-shell degrees of freedom but are associated to domain walls and space-filling branes, respectively. The Kac-Moody analysis gives precisely this information for the theories considered in subsection 4.2. Actually, we have seen that these theories only have 0-branes and 1-branes, which means that this analysis is {\it a posteriori} only crucial  for the domain walls in three dimensions.

We now wish to determine the branes in the hyper-sector in dimensions less than six for the special class
of theories with eight supercharges that result as truncations of the theories with sixteen supercharges. We consider here theories corresponding to  the first chain of coset manifolds in Table \ref{upliftchain} in the vector  multiplet sector and with hyperscalars parametrising the first manifold in eq.~\eqref{quatsymmspaces}. This means that we take the symmetry of the three-dimensional theory to be  $\text{SO}(4,n)\times \text{SO}(4,m)$, corresponding to the product of two symmetric quaternionic manifolds. We consider the theory in higher dimensions with the first simple factor uplifted to vector multiplets and the second factor unchanged. We will show that
the brane structure in any dimensions can be obtained by requiring that this theory is a truncation of the theory with sixteen supercharges whose three-dimensional symmetry is $\text{SO}(8,n+m)$.
Given the universality in the brane counting found in the previous subsection, we expect the results of this subsection to apply also to the cases in which the vector multiplets and the hypermultiplets are of the $\text{F}_{4(4)}$, $\text{E}_{6(2)}$, $\text{E}_{7(-5)}$ and $\text{E}_{8(-24)}$ type.

We start our discussion with the three-dimensional case. We consider a theory whose scalars parametrise the product of two quaternionic spaces,
\begin{equation}
\dfrac{\text{SO}(4,n)}{\text{SO}(4)\times \text{SO}(n)}\times \dfrac{\text{SO}(4,m)}{\text{SO}(4)\times \text{SO}(m)}
\quad ,
\end{equation}
with $m,n\geq 4$. Following the results of the previous subsection we can say in all generality that together with the fields in eq.~\eqref{threedimSOSimplie} with vector indices in the first orthogonal group, there will be an equivalent set of fields with indices in the second group. Denoting these indices with $M,N,...$, this leads, on top of the branes of the previous subsection, to the same number of branes for the second hypermultiplet sector,
  \begin{eqnarray}
  & & A_{1, M_1 M_2} \ \ \ \, \rightarrow  \ 24 \ \quad  {\rm 0-branes}\, \nonumber \\
  & & A_{2, M N} \ \ \ \ \ \, \rightarrow  \ \ \, 8 \ \quad {\rm 1-branes}\, \nonumber \\
    & & A_{2, M_1 ...M_4} \ \ \rightarrow  \ 16 \ \quad {\rm 1-branes} \quad .
    \end{eqnarray}

We still have to determine the fields with indices of both groups that give rise to 1/2-BPS branes. We determine such fields by requiring that this theory is a truncation of the half-maximal theory with symmetry $\text{SO}(8,n+m)$. Denoting with $\hat{A},\hat{B},...$ the indices of $\text{SO}(8,n+m)$, the fields that are associated to branes in the half-maximal theory are $A_{1, \hat{A}_1 \hat{A}_2}$, $A_{2, \hat{A} \hat{B}}$, $A_{2, \hat{A}_1 ...\hat{A}_4}$ and $A_{3, \hat{A} \hat{B}_1 ...\hat{B}_5}$, in analogy with eq.~\eqref{threedimSOSimplie}. The truncation projects $A_{1, \hat{A}_1 \hat{A}_2}$ to $A_{1, A_1 A_2}$ and $A_{1, M_1 M_2}$. The remaining fields all follow by requiring that the gauge algebra, {\it i.e.}~the algebra of
gauge transformations of all the potentials in the theory, is a consistent algebra. Associating to each field an operator with the dual index structure, this requirement stems from imposing the consistency of the algebra of these operators\,\footnote{It is precisely this correspondence between fields and operators that in the case of a single orthogonal group leads to the very-extended Kac-Moody algebra associated to the theory.}.
Denoting by $R_{1, \hat{A}_1 \hat{A}_2}$ the operator associated to the field $A_{1, \hat{A}_1 \hat{A}_2}$, we obtain the commutator
\begin{equation}
[ R_{1, \hat{A}_1 \hat{A}_2} , R_{1, \hat{B}_1 \hat{B}_2}] = R_{2, \hat{A}_1 \hat{A}_2 \hat{B}_1 \hat{B}_2 }  + R_{2, [\hat{A}_1  [ \hat{B}_1 } \eta_{\hat{B}_2 ] \hat{A}_2 ]} \quad .
\end{equation}
This commutator implies that the first operator on the right-hand side is completely antisymmetric and the second is symmetric, exactly as the corresponding 2-form fields. As far as this commutator is concerned, the consistently truncated algebra is
  \begin{eqnarray}
  & & [ R_{1, A_1 A_2} , R_{1, B_1 B_2} ] = R_{2, A_1 A_2 B_1 B_2} + R_{2, [ A_1 [B_1} \eta_{ B_2 ]  A_2 ]} \nonumber \\
  & & [ R_{1, M_1 M_2} , R_{1, N_1 N_2} ] = R_{2, M_1 M_2 N_1 N_2} + R_{2, [ M_1 [N_1} \eta_{ N_2 ]  M_2 ]} \nonumber \\
  & & [ R_{1, A_1 A_2} , R_{1, M_1 M_2} ] = R_{2, A_1 A_2 M_1 M_2}  \quad .
  \end{eqnarray}
This implies that, apart from the 2-form fields that we have already introduced, associated to the operators on the right-hand side of the first two equations, the third equation leads to the field $A_{2, A_1 A_2 M_1 M_2}$ with indices on both groups. Using the light-cone rule this gives $24 \times 24 = 576$ 1-branes. Similarly, one can show using algebraic arguments that the only fields that are associated to branes and result from the truncation of the field $A_{3, \hat{A} \hat{B}_1 ... \hat{B}_5}$ are $A_{3, A B_1 B_2 B_3 M_1 M_2}$ and $A_{3, M N_1 N_2 N_3 A_1 A_2}$. Using the light-cone rule, each of these two fields gives ${4 \choose 3} \times 2^3 \times 3 \times 24 = 2304 $ 2-branes.

We next consider the four-dimensional theory with symmetry $\text{SO}(2, n-2) \times \text{SU}(1,1) \times \text{SO}(4,m)$, with the first two groups associated to the vector multiplets and the third to the hypermultiplets.
 We require that this theory results from the truncation of the half-maximal theory with symmetry $\text{SO}(6, n+m-2) \times \text{SU}(1,1)$. The fields associated to branes with indices only in the vector-multiplet sector are contained in eq.~\eqref{fourdimSOSimplie}.
The lowest-rank field with internal indices in the hypermultiplet sector is the 2-form $A_{2, M_1 M_2}$ dual to the hyper-scalars. Consistency of the truncation implies that there is a 3-form $A_{3, M_1 M_2 A a}$, leading to  $24 \times 4 \times 2 = 192$ 2-branes. There are also  4-forms $A_{4, M_1 M_2 A_1 A_2 ab}$ (giving 192 3-branes) and $A_{4, M N_1 N_2 N_3}$ and $A_{4, AB M_1 M_2 }$ (giving both 96 3-branes).

We now discuss the five-dimensional case. The symmetry is $\text{SO}(1,n-3) \times \mathbb{R}^+ \times \text{SO}(4,m)$, where the last factor is the symmetry of the quaternionic manifold of the hypermultiplets. We want to derive the branes of this theory considering it as a truncation of the half-maximal theory with  symmetry group $\text{SO}(5, n+m -3) \times \mathbb{R}^+$. The fields corresponding to branes in the half-maximal theory are given by
  \begin{eqnarray}
  & & \alpha=0 \ \ \,\qquad A_{1,\hat{A}} \quad A_2\,, \nonumber \\
& & \alpha=-2 \qquad A_{1} \quad A_{2,\hat{A}} \quad A_{3, \hat{A}_1 \hat{A}_2} \quad A_{4, \hat{A}_1 \hat{A}_2 \hat{A}_3} \quad A_{5 , \hat{A}_1 ...\hat{A}_4}\,, \nonumber \\
  & & \alpha=-4 \qquad A_{4,\hat{A}_1 \hat{A}_2} \quad A_{5, \hat{A}, \hat{B}_1 \hat{B}_2} \quad .
  \end{eqnarray}
The analysis of the branes of the truncated theory, as far as the vector multiplets are concerned, was performed in the previous subsection. The lowest-rank form fields in the hyper-sector are the 3-forms $A_{3, M_1 M_2}$, with $\alpha=-2$. Such fields are dual to the hyper-scalars and lead to 24 defect branes. Imposing consistency of the truncation implies that the only higher-rank fields that are associated to branes are the 4-forms $A_{4, A M_1 M_2}$ with $\alpha=-2$ and $A_{4, M_1 M_2}$ with $\alpha=-4$. The first field gives 48 3-branes, while the second one corresponds to 24 3-branes.

\begin{table}[t]
\begin{center}
\begin{tabular}{|c|c|c|c|}
\hline \rule[-1mm]{0mm}{6mm} dim. & type of brane & field & $\#$ of branes \\
\hline
\hline  \rule[-1mm]{0mm}{6mm}
$D=3$ & 0-brane & $A_{1, M_1 M_2}$ & 24\\
\cline{2-4} \rule[-1mm]{0mm}{6mm} & 1-brane & $A_{2, MN}$ & $8$\\
\cline{3-4} \rule[-1mm]{0mm}{6mm} &  & $A_{2, M_1... M_4}$ & $16$\\
\cline{3-4} \rule[-1mm]{0mm}{6mm} &  & $A_{2, A_1 A_2 M_1 M_2}$ & 576\\
\cline{2-4} \rule[-1mm]{0mm}{6mm} & 2-brane & $A_{3,M N_1 N_2 N_3 A_1 A_2}$ & 2304\\
\cline{3-4} \rule[-1mm]{0mm}{6mm} &  & $A_{3,A B_1 B_2 B_3 M_1 M_2}$ & 2304\\
\hline \hline \rule[-1mm]{0mm}{6mm}
$D=4$ & 1-brane & $A_{2,M_1 M_2}$ &  24\\
\cline{2-4} \rule[-1mm]{0mm}{6mm} & 2-brane & $A_{3, M_1 M_2 A a}$ & 192\\
\cline{2-4} \rule[-1mm]{0mm}{6mm} & 3-brane & $A_{4, M_1 M_2 A_1 A_2 ab}$ & 192\\
\cline{3-4} \rule[-1mm]{0mm}{6mm} &  & $A_{4, M N_1 N_2  N_3}$ & 96\\
\cline{3-4} \rule[-1mm]{0mm}{6mm} &  & $A_{4, AB M_1 M_2 }$ & 96\\
\hline \hline \rule[-1mm]{0mm}{6mm}
$D=5$ & 2-brane  & $ A_{3, M_1 M_2}$ $(\alpha=-2)$ & 24\\
\cline{2-4} \rule[-1mm]{0mm}{6mm} & 3-brane & $A_{4, A M_1 M_2}$  $(\alpha=-2)$& 48\\
\cline{3-4} \rule[-1mm]{0mm}{6mm} &  & $A_{4, M_1 M_2}$  $(\alpha=-4)$& 24\\
\hline \hline \rule[-1mm]{0mm}{6mm}
$D=6$ & 3-brane &  $A_{4, M_1 M_2}$ $(\alpha=-2)$& 24\\
\hline
\end{tabular}
\end{center}
  \caption{\sl The branes that are added to those in Table \ref{branesofSOtheories} when hypermultiplets are included.
   \label{branesinhypersector}}
\end{table}

The six-dimensional case was already discussed in subsection 4.1. The only relevant fields in the hyper-sector are the 4-forms $A_{4, M_1 M_2}$ leading to 24 3-branes. Considering in particular the case in which the global symmetry is $\text{SO}(1,1)\times \text{SO}(n-4) \times \text{SO}(4,m)$ (corresponding to the first case in Table \ref{upliftchain} as far as vector and tensor multiplets are concerned) the $\text{SO}(1,1)$ dilaton scaling of this 4-form field is $\alpha=-2$.

We have listed in Table \ref{branesinhypersector} the results derived in this subsection. In the next section we will show that considering these models as low-energy effective actions of the heterotic string compactified on ${\rm K3} \times T^p$, the number of branes derived in this section, for specific string dilaton scalings, can be derived from the branes of the six-dimensional heterotic string ({\it i.e.}~the heterotic string compactified on K3) by applying exactly the same wrapping rules that where derived previously for theories with more supersymmetry.

\section{Reduction of the heterotic string on K3 and wrapping rules}

In maximally supersymmetric theories, one can classify the branes according to how their tension $T$ scales with respect to the string coupling $g_S$ in terms of the non-positive integer number $\alpha$ defined as $T \sim g_S^\alpha$. This analysis was obtained in each dimension in \cite{Bergshoeff:2011zk,Bergshoeff:2011ee,Bergshoeff:2012ex} studying the properties of the representations of the T-duality group $\text{SO}(d,d)$ in $10-d$ dimensions. What this analysis reveals is that for $\alpha=0,-1,-2,-3$ the number of branes in a given dimension can be obtained from the branes in ten dimensions using different {\it wrapping rules}, that are specific for each value of $\alpha$ \cite{Bergshoeff:2011mh,Bergshoeff:2011ee,Bergshoeff:2012ex}.  In \cite{Bergshoeff:2012jb} it was then shown that the same applies to the heterotic theory compactified on a torus. In this case only even values of $\alpha$ are allowed, and the branes with $\alpha=0,-2$ are obtained from ten dimensions using the same wrapping rules that one obtains in the maximal case for these values of $\alpha$.

In this section we want to perform the same analysis for theories with eight supercharges. In particular, we consider the six-dimensional theory as the low-energy action of the heterotic string compactified on K3, and the lower-dimensional theories as its torus dimensional reductions. We will determine the number of branes for each value of $\alpha$ in any dimension and we will show that for $\alpha=0,-2$ the result can be obtained using the wrapping rules of the maximal and half-maximal case starting from the branes of the six-dimensional theory. Finally, we will show how also the six-dimensional branes can be obtained using the same wrapping rules on K3 cycles starting from the branes of the ten-dimensional heterotic theory.

The heterotic dilaton in theories with eight supercharges sits in a tensor multiplet in six dimensions and in a vector multiplet in five and four dimensions. At the perturbative level, one can only obtain six-dimensional models with a single tensor multiplet, while non-perturbatively one can also consider models with more tensor multiplets\,\footnote{This differs from the Type-I case, in which models with various tensor multiplets can be constructed at the perturbative level \cite{Bianchi:1990yu}.}. In any case, decomposing
$\text{SO}(1,n_T)\supset \text{SO}(1,1) \times \text{SO}(n_T-1 )$, where
 $\text{SO}(1,n_T)$ is the global symmetry of the theory with $n_T$ tensor multiplets, gives the scaling of the various fields with respect to the string dilaton, that  is the scalar parametrising $
\text{SO}(1,1)$. Moreover, the symmetry of the hypermultiplet sector is a T-duality symmetry because it does not affect the dilaton.
We can then consider the reduction to five dimensions and in particular the first chain of theories in Table \ref{upliftchain}. We already mentioned in the previous section that the string dilaton is the $\mathbb{R}^+$ dilaton and that the $\text{SO}(1,n-3)$ symmetry is a T-duality symmetry (together again with the symmetry of the hyper-sector). In four dimensions, the string dilaton is the dilaton of the $\text{SU}(1,1)/\text{U}(1)$ coset manifold, and finally in three dimensions one has to decompose $\text{SO}(4,n) \supset \text{SO}(1,1) \times \text{SO}(3,n-1)$, where again the dilaton is the $\text{SO}(1,1)$ scalar and $\text{SO}(3,n-1)$ is a T-duality symmetry. It follows from eqs.~\eqref{FEEEunderSOD=3}, \eqref{FEEEunderSOD=4} and \eqref{FEEEunderSOD=5} that the $\text{F}_{4(4)}$, $\text{E}_{6(2)}$, $\text{E}_{7(-5)}$ and $\text{E}_{8(-24)}$ chains of theories in Table \ref{upliftchain}
can be decomposed with respect to the groups in the $\text{SO}(4,n)$ chain giving exactly the same brane structure.  Therefore, the T-duality analysis of all these theories is the same as far as the branes are concerned.

We now count the branes in any dimension for the different values of $\alpha$. In six dimensions, there are two 1-branes, one with $\alpha=0$, which is the fundamental string, and one with $\alpha=-2$, which is its solitonic dual. The scalars in the hypermultiplet are in the perturbative sector, and thus the 24 3-branes which are magnetically charged under such scalars have $\alpha=-2$. In five dimensions, the number of branes for the various values of $\alpha$ can be read directly from Tables \ref{branesofSOtheories} and \ref{branesinhypersector}. In four dimensions, the fields transform under $\text{SU}(1,1) \simeq \text{SL}(2,\mathbb{R})$, and the value of $\alpha$ is determined by the relation \cite{Bergshoeff:2012jb}
\begin{equation}
 \alpha = n_1 - n_2 -p \quad ,
 \end{equation}
where $p$ is the rank of the form and $n_1$ and $n_2$ are the number of indices along the directions 1 and 2 of $\text{SL}(2,\mathbb{R})$ respectively. Reading the representations of the fields from Table
\ref{branesofSOtheories}, we get that $A_{1,Aa}$ gives 4 0-branes with $\alpha=0$ and 4 0-branes with $\alpha=-2$, $A_{2, ab}$ gives one 1-brane with $\alpha=0$ and one with $\alpha=-4$, while $A_{2, A_1 A_2}$ gives 4 1-branes with $\alpha=-2$. Similarly, from Table \ref{branesinhypersector} we get that $A_{2,M_1 M_2}$ gives 24 1-branes with $\alpha=-2$, $A_{3, M_1 M_2 A a}$ gives 96 2-branes with $\alpha=-2$ and 96 2-branes with $\alpha=-4$,  $A_{4, M_1 M_2 A_1 A_2 ab}$ gives 96 3-branes with $\alpha=-2$ and 96 3-branes with $\alpha=-6$, and finally both $A_{4, M N_1 N_2 N_3}$ and  $A_{4,ABM_1 M_2}$ give 96 3-branes with $\alpha=-4$. Finally, in three dimensions one considers $\text{SO}(4,n)\supset \text{SO}(1,1) \times \text{SO}(3,n-1)$, where again the heterotic dilaton parametrises $\text{SO}(1,1)$. Denoting with $+$ and $-$ the light-cone directions of $\text{SO}(1,1)$, and with $n_+$ and $n_-$ the number of $+$ and $-$ indices of the $p$-forms in Tables \ref{branesofSOtheories} and  \ref{branesinhypersector}, the value of $\alpha$ is given by
\begin{equation}
\alpha= 2(n_+ - n_- -p ) \quad .
\end{equation}
In Table \ref{branesofSOtheories}, the field $A_{1, A_1 A_2}$ gives 6 0-branes with $\alpha=0$, 12 0-branes with $\alpha=-2$ and 6 0-branes with $\alpha=-4$, $A_{2,AB}$ gives one 1-brane with $\alpha=0$, 6 with $\alpha=-4$ and one with $\alpha=-8$, while the field $A_{2,A_1 ...A_4}$ gives 8 1-branes with $\alpha=-2$ and 8 with $\alpha=-6$. Finally, we consider in Table \ref{branesinhypersector} only the fields giving branes with $\alpha=-2$, ignoring all the other branes with more negative $\alpha$. This is 24 0-branes from $A_{1,M_1 M_2}$, 144 1-branes from $A_{2, A_1 A_2 M_1 M_2}$ and 288  2-branes from $A_{3, A B_1 B_2 B_3 M_1 M_2}$.

\begin{table}[h]
\begin{center}
\begin{tabular}{|c||c|c|c|c|}
\hline \rule[-1mm]{0mm}{6mm} F-brane &6D&5D&4D&3D\\
\hline \hline \rule[-1mm]{0mm}{6mm} 0&&2&4&6\\
\hline \rule[-1mm]{0mm}{6mm} 1&1&1&1&1\\
\hline
\end{tabular}
\caption{\sl \footnotesize The number of fundamental ($\alpha=0$) branes in any dimension. The entries of the table in any dimension result from applying the fundamental wrapping rule (\ref{Fbranewrapping}) to the branes of one dimension above.
\label{Fwrapping}}
\end{center}
\end{table}

We summarise the result for $\alpha =0$ and $\alpha =-2$ in Tables \ref{Fwrapping} and \ref{Swrapping}. The reader can see that the number of $\alpha=0$ branes in a given dimension can be  derived from the number of branes in one dimension above using the wrapping rules
\begin{equation}
 \alpha=0 \  \left\{ \begin{array}{l}
{\rm wrapped}   \ \rightarrow\   \ {\rm doubled}\\
{\rm unwrapped} \  \rightarrow \  {\rm undoubled} \ .
 \end{array} \right. \label{Fbranewrapping}
 \end{equation}
This implies that their number in any dimension can be derived from the branes in six dimensions using the wrapping rules.
Similarly, the $\alpha=-2$ branes satisfy the wrapping rules
\begin{equation}
\alpha=-2 \  \left\{ \begin{array}{l}
{\rm wrapped}   \ \rightarrow\   {\rm undoubled}\\
{\rm unwrapped} \  \rightarrow  \ {\rm doubled} \ .
 \end{array} \right. \label{Sbranewrapping}
\end{equation}
From the branes in six dimensions and applying these wrapping rules one derives the numbers of all the $\alpha=-2$ branes in any dimension.

\begin{table}[h]
\begin{center}
\begin{tabular}{|c||c|c|c|c|}
\hline \rule[-1mm]{0mm}{6mm} S-brane &6D&5D&4D&3D\\
\hline \hline \rule[-1mm]{0mm}{6mm} 0&&1&4&12+24\\
\hline \rule[-1mm]{0mm}{6mm} 1&1&2&4+24&8+144\\
 \hline \rule[-1mm]{0mm}{6mm} 2&&24&96&288\\
 \hline \rule[-1mm]{0mm}{6mm} 3&24&48&96&\\
\hline
\end{tabular}
\caption{\sl \footnotesize The number of half-supersymmetric solitons ($\alpha=-2$ S-branes) in any dimensions. The number of branes in a given dimension are obtained applying the wrapping rule
rule (\ref{Sbranewrapping}) to the branes of one dimension above.
 \label{Swrapping}}
\end{center}
\end{table}

In \cite{Bergshoeff:2012jb,Bergshoeff:2013spa} the duality between the Type-IIA theory compactified on K3 and the heterotic theory on $T^4$ was used to determine how the wrapping rules can be applied in K3 compactifications. More specifically, one identifies a basis of six homology 2-cycles, and allowing the IIA branes to either wrap these 2-cycles, the whole of K3 or remain unwrapped, and using the same wrapping rules as derived for the torus reduction, one reproduces the branes of the heterotic side, in agreement with the duality. We can now apply this result to the heterotic theory compactified on K3, to see whether this reproduces the number of six-dimensional branes that we have derived in this paper. The only branes of the heterotic theory in ten dimensions are the fundamental string ($\alpha=0$) and its solitonic dual NS5-brane ($\alpha=-2$).
 The fundamental string cannot wrap, because there are no 1-cycles, and the wrapping rules imply that if it does not wrap, it does not double. Therefore one obtains a single $\alpha=0$ 1-brane in six dimensions, as in Table \ref{Fwrapping}.
 The NS5-brane gives the dual $\alpha=-2$ string when it wraps on the whole of K3, and the $\alpha=-2$ wrapping rules imply that it does not double, giving only one string as in Table \ref{Swrapping}.  The NS5-brane can also wrap on a 2-cycle to give a 3-brane. There are six possibilities, while the two directions in which this brane does not wrap give two factors 2 because of the doubling. This gives exactly 24 3-branes as in Table \ref{Swrapping}. The fully unwrapped NS5-brane is not allowed in six dimensions because of supersymmetry. A similar phenomenon occurs in the half-maximal case, where an additional halving takes place for the unwrapped NS5-branes, as discussed in \cite{Bergshoeff:2013spa}. In that case the halving was interpreted
 in  the orbifold limit $T^4/\mathbb{Z}_2$ of K3 as an additional action of $\mathbb{Z}_2$ on the charges, leading to a self-duality condition. In this case we expect that  $\mathbb{Z}_2$ projects out the charge completely.

\section{Central charges and degeneracies}

In this section we want to study the relation between the number of half-supersymmetric branes and the supersymmetry algebra in theories with 8 supercharges. We will only consider  the theories that can be uplifted to six dimensions and such that the symmetry group of the hyper-sector has four non-compact Cartan generators. These are indeed the theories whose brane structure is universal, as we have shown in section 4.
The R-symmetry of the theories with eight supercharges is  $\text{SU}(2)$ in six and five dimensions, $\text{U}(2)$ in four dimensions and  $\text{SU}(2) \times \text{SU}(2)$ in three dimensions.
In six dimensions,  the supercharge is a chiral spinor which is also a doublet of $\text{SU}(2)$ satisfying symplectic-Majorana conditions. In five dimensions the spinors satisfy the same symplectic-Majorana conditions. In four dimensions  the supercharges are doublets of Majorana spinors. Finally, in three dimensions the supercharges are Majorana spinors in the ${\bf (2,2)}$ representation. The resulting central charges in the supersymmetry algebra are summarised in Table \ref{centralcharges8}.

\begin{table}[h]
\begin{center}
\begin{tabular}{|c|c|c|c|c|c|}
\hline
$D$&$R$-symmetry&$n=0$&$n=1$&$n=2$&$n=3$\\[.1truecm]
\hline \rule[-1mm]{0mm}{6mm} 6&$\text{SU}(2)$ & & ${\bf 1}$& & ${\bf 3}^+$\\[.05truecm]
\hline \rule[-1mm]{0mm}{6mm} 5&$\text{SU}(2)$& ${\bf 1}$& ${\bf 1} $& {\bf 3}&\\[.05truecm]
\hline \rule[-1mm]{0mm}{6mm} 4&$\text{U}(2)$& ${\bf 1} + {\bf 1}$ & ${\bf 1} + {\bf 3}$ & ${\bf 3}^+ + {\bf {3}}^-$ &\\[.05truecm]
\hline \rule[-1mm]{0mm}{6mm} 3&$\text{SU}(2)\times\text{SU}(2)$& ${\bf (3,1) + (1,3)}$ & ${\bf (1,1) } + {\bf (3,3)}$ &&\\[.05truecm]
\hline
\end{tabular}
\end{center}
  \caption{\sl This table indicates the $R$-representations of the
  $n$-form central charges of $3\le D\le 6$ quarter-maximal supersymmetric theories. Momentum is included, corresponding to the always present $n=1$ singlet.
  If applicable, we have also indicated the space-time duality of
  the central charges with a superscript $\pm$. }
  \label{centralcharges8}
\end{table}

We want to relate the central charges in Table \ref{centralcharges8} and their duals (with the exception of the $n=0$ charge and the momentum operator, which cannot be dualised) to the half-supersymmetric branes
discussed in the previous sections.
The central charges determine the BPS-conditions of the corresponding branes, {\it i.e.} which supersymmetries are preserved on the brane. As we will see, the BPS conditions are degenerate, which means that to each central charge one associates different branes. This degeneracy implies that a bound state of these branes keeps preserving the same amount of supersymmetry.
The same degeneracy analysis was performed in \cite{Bergshoeff:2011zk,Bergshoeff:2011se,Bergshoeff:2012pm,Bergshoeff:2013sxa} for the maximal theories and in \cite{Bergshoeff:2012jb} for the half-maximal ones. In the maximal case, the branes with more than two transverse directions are in one-to-one correspondence with the central charges and thus there are no degeneracies, while all the defect branes have degeneracy 2 and the domain walls and space-filling  branes have even higher degeneracy. In the half-maximal case the degeneracy is twice the one of the maximal case, and in particular it is 2 for branes with more than two transverse directions and four for defect branes. As we will see in the following, in theories with eight supercharges the degeneracy is twice that of the half-maximal case and four times that of the maximal case. Indeed, we will find that the branes with more than two transverse directions have degeneracy 4 and the defect branes have degeneracy 8.
Below we discuss each dimension separately.

\vskip 0.5cm
{\bf 6D}:
As we have seen in section 4, in six dimensions there are two 1-branes and 24 3-branes. From Table
\ref{centralcharges8} we read that there is a singlet charge with $n=1$ (the momentum operator) and a self-dual $n=3$ charge in the ${\bf 3}$.
 The $n=1$ charge corresponds to the pp-wave, the KK-monopole and the two 1-branes. It thus has degeneracy four. The $n=3$ charge in the ${\bf 3}$,  instead,  corresponds to the 24 3-branes, giving a degeneracy 8. We note that in both cases the degeneracy is twice the degeneracy of the half-maximal case and four times the degeneracy of the maximal case. There are no other charges because we cannot dualise the momentum operator. This is consistent with the fact that we have no other branes.

 \vskip .5cm
{\bf 5D}: The branes with more than two transverse directions in five dimensions are three 0-branes and three 1-branes. There is a singlet $n=0$ central charge corresponding to the KK-monopole and the three 0-branes, while the singlet $n=1$ momentum operator corresponds to the pp-wave and the three 1-branes. In both cases we get degeneracy four. The $n=2$ charge is in the ${\bf 3}$, and it corresponds to the 24 2-branes so that the degeneracy is 8. This is again twice the degeneracy of the half-maximal case. The $n=2$ charge can be dualised to an $n=3$ charge. There are $48+24$ 3-branes, and this gives in total degeneracy 24. There are no other central charges because the momentum operator and the $n=0$ charges cannot be dualised.

\vskip .5cm
{\bf 4D}:  In four dimensions there are two singlet $n=0$ charges with different $\text{U}(1)$ weight. The four fundamental and four solitonic  0-branes are associated to each of the central charges, which thus have both degeneracy four. The singlet $n=1$ central charge corresponds not only to the pp-wave and the fundamental string, but also to the S-dual of the fundamental string (with $\alpha=-4$) and the four $\alpha=-2$ 1-branes in the vector-multiplet sector. We therefore have total degeneracy 7. Note that this case is special due to the fact that these defect branes have the same BPS condition as the pp-wave. A similar phenomenon occurs in the maximal and half-maximal case in four dimensions \cite{Bergshoeff:2011se,Bergshoeff:2012jb}. The other $n=1$ central charge is in the ${\bf 3}$ and it corresponds to the 24 defect branes in the hyper-sector, leading to a degeneracy 8 as usual. The $n=2$ central charges are associated to the domain walls. There are 6 charges and 192 branes, resulting in a degeneracy 32. Finally, the $n=1$ central
charge in the ${\bf 3}$ can be dualised to give an $n=3$ charge.  There are 96 3-branes with $\alpha=-2$ and with $\alpha=-6$, each with  degeneracy 32, and 192 3-branes with $\alpha=-4$, with degeneracy 64.

\vskip .5cm
{\bf 3D}: In three dimensions the 0-branes are defect branes. The $n=0$ charges are in the ${\bf (3,1)+(1,3)}$, and there are 24 0-branes in each hypermultiplet sector. This implies a total degeneracy 8 as usual. The $n=1$ charges are in the ${\bf (1,1)+(3,3)}$. The $8+16$ 1-branes in each hyper-sector are associated to the singlet, while the 576 1-branes in the mixed sector are associated to the charge in the ${\bf (3,3)}$. This latter charge can also be dualised to give an $n=2$ central charge. The total degeneracy of the $2 \times 2304$ space-filling branes is 512.

\vskip .5cm This finishes our discussion about the degeneracy of the central charges in theories with eight supercharges.

\section{Conclusions}

In this paper we have first given a group-theoretic characterisation of the 1/2-BPS branes in half-maximal theories.
In the maximal case these branes correspond to the components of the potentials associated to the longest weights of the representation of the global symmetry group \cite{Bergshoeff:2013sxa}. In this classification, the fact that the symmetry group is maximally non-compact plays a crucial rule.  In the half-maximal case, the supergravity theory in $10-d$ dimensions coupled to $d+n$ abelian vector multiplets  possesses a  global symmetry $\text{SO}(d,d+n)$, which is not maximally non-compact (for $n \neq 0,1$). The 1/2-supersymmetric branes correspond to the components of the representations of $\text{SO}(d,d+n)$  that satisfy the light-cone rules \cite{Bergshoeff:2012jb}. The Tits-Satake diagram associated to the given real form of the orthogonal group determines the reality properties of the roots and the weights of the corresponding algebra, and we have shown that the light-cone rules identify the components of each representation that correspond 
to 1/2-supersymmetric branes as the ones that are associated to the real longest weights.

We have  generalised this result to the supergravity theories with 8 supercharges. In particular, we have analysed theories with scalars parametrising coset manifolds. Considering for each simple factor of the global symmetry group  the corresponding Tits-Satake diagram, we have classified all the 1/2-BPS branes of these theories determining the number of real longest weights. We have first determined the number of 1/2-BPS branes of the theories whose coset manifolds are given in Table \ref{upliftchain}. These theories, that do not contain hypermultiplets in dimensions higher than three, are all the theories whose reduction to three dimensions gives scalars parametrising a symmetric manifold.  We have then considered the branes with scalars in the hypermultiplet sector.

What our classification shows is that the first five chains of theories in Table \ref{upliftchain} give all exactly the same number of branes in any dimension. The $\text{SO}(4,n)$ theories give the same number of branes for any $n \geq 4$ because changing $n$ does not lead to a different number of lightlike directions. This translates to the property that, for a given representation, the number of real longest weights does not depend on $n$, as long as $n \geq 4$.
For the   $\text{F}_{4(4)}$, $\text{E}_{6(2)}$, $\text{E}_{7(-5)}$ and $\text{E}_{8(-24)}$ chains, corresponding to the ``magic'' supergravities
(see the first reference in \cite{gunaydinsierratownsend}), exactly the same happens. The $\text{F}_{4(4)}$ chain is maximally non-compact, and thus one simply counts the longest weights of each representation to obtain the number of branes. The other three cases give exactly the same result as the $\text{F}_{4(4)}$ chain once the roots and weights are projected on their real part. In this sense, one can think of $\text{E}_{6(2)}$, $\text{E}_{7(-5)}$ and $\text{E}_{8(-24)}$ as being  to $\text{F}_{4(4)}$ exactly what $\text{SO}(4,n)$ with $n > 4$ is to $\text{SO}(4,4)$. We have also explained why all the magic supergravities give the same brane structure as the  $\text{SO}(4,n)$ theories. The final outcome is that the brane structure of all these theories is identical. In other words, there is a universal brane structure underlying the theories with 8 supercharges. What the theories just considered have in common is that they can all be uplifted to six dimensions. The last three chains of theories in Table \ref{upliftchain}, instead, cannot be uplifted to six dimensions and thus the number of branes that one gets is less than what supersymmetry allows.

An interesting spin-off of our analysis is that we are now able to give a full classification of all
`vector-branes', i.e.~half-supersymmetric branes whose worldvolume dynamics is determined by a single vector multiplet.
The worldvolume action for such a vector multiplet is given by a supersymmetric Born-Infeld theory.
 Vector branes have recently been studied in the context of constructing new supersymmetric invariants when studying the UV properties of perturbative supergravity \cite{Bergshoeff:2013pia}.
The most well-studied examples of vector branes are the Dirichlet branes (or D-branes) of IIA/IIB string theory whose worldvolume dynamics is governed by a supersymmetric Born-Infeld theory with 16 supercharges.  The tension of the D-branes scales with the inverse string coupling constant, i.e.~they have $\alpha=-1$, and this implies that fundamental strings can end on them. As a consequence, they can be described by imposing Dirichlet boundary conditions on the fundamental string. In the type-II theories, though, there are also additional vector branes that are more non-perturbative, i.e.~they have more-negative values of $\alpha$. In particular, in the IIB theory one has the NS5-brane, with $\alpha=-2$, the S-dual of the D7-brane, with $\alpha=-3$, and the S-dual of the D9-brane, with $\alpha=-4$.
In theories with sixteen supercharges there are no Dirichlet branes, i.e.~there are no branes that are defined by being the end-points of fundamental strings. Indeed, the Dirichlet branes of the Type-I theories have  a worldvolume dynamics that is described by a hypermultiplet and not by a vector multiplet. Nonetheless, there are vector branes with $\alpha \ne -1$.  In particular, in the heterotic theory compactified on a torus there are vector branes with  $\alpha=-4$. They correspond to branes on which non-perturbative solitonic strings end. The highest dimension in which these branes exist is six. More specifically, in six dimensions one finds that the $(1,1)$-supergravity theory has $V4$ and $V5$-branes whereas the chiral $(2,0)$-supergravity theory allows for $V3$-branes and $V5$-branes \cite{Bergshoeff:2012jb}. Moving to the theories with eight supercharges studied in this paper, one can derive the worldvolume content from  the Wess-Zumino terms that arise from the fields in Tables \ref{branesofSOtheories} and \ref{branesinhypersector}. We find that the only vector branes that are present are the V3-branes that couple to the potentials $A_{4,MN_1N_2N_3}$ in four dimensions. Such branes have $\alpha=-4$ and there are 96 of them. Like the vector branes in six dimensions they correspond to branes on which solitonic strings can end. 
To summarise, we report in Table \ref{tableEric} the highest dimension in which vector branes appear, together with their value of $\alpha$ and the number of supercharges preserved on the world-volume.

\begin{table}[t!]
\begin{center}
\begin{tabular}{|c|c|c|c|c|}
\hline \rule[-1mm]{0mm}{1mm} \# supercharges & highest dimensions & type & $\alpha$\\[.2truecm]
\hline \hline \rule[-1mm]{0mm}{1mm}
16&D=10&D$p$-branes&--1\\[.2truecm]
&& IIB NS5-brane&--2\\[.2truecm]
&& S-dual D7-brane&--3\\[.2truecm]
&& S-dual D9-brane&--4\\[.2truecm]
\hline
8&D=6&$(V3), V4, V5$&--4\\[.2truecm]
\hline
4&D=4&$V3$-brane&--4\\[.1truecm]
\hline
\end{tabular}
\caption{ \label{tableEric} \sl
This table summarises the vector branes ($Vp$-branes) whose worldvolume dynamics is governed by a supersymmetric Born-Infeld theory with 16, 8 and 4 supercharges. Only the highest dimension is given. The V3-brane between bracket belongs to the
six-dimensional chiral supergravity
theory for which one cannot define $\alpha$.
}
\end{center}
\end{table}

An important outcome of the brane classification performed in this work is that the same wrapping rules we derived in the
maximal case and verified in the half-maximal case also work for string theories with 8 supercharges. This shows the
universal nature of these wrapping rules. They precisely tell us how to relate branes in different dimensions. In some sense the wrapping rules tell us how the different branes `see' the underlying stringy geometry. We expect that this particular approach of describing the stringy geometry should be equivalent and complementary to other approaches in
the literature, such as the doubled geometry \cite{Hull1}, double field theory \cite{Aldazabal:2013sca} and  the exotic brane description of U-folds \cite{deBoer1}. Hopefully, our work will help in clarifying these relations and will be a useful step in unraveling the mysteries of the proper geometry underlying string theory.

\section*{Acknowledgements}
E.B.~wishes to thank the University of Rome ``La
Sapienza'' and F.R.~wishes to thank the University of Groningen for hospitality at different stages of this work.


\vskip 1.5cm





\begin{thebibliography}{99}

\bibitem{Ferrara:1997ci}
  S.~Ferrara and J.~M.~Maldacena,
  ``Branes, central charges and U duality invariant BPS conditions,''
  Class.\ Quant.\ Grav.\  {\bf 15} (1998) 749
  [hep-th/9706097].

\bibitem{Ferrara:1997uz}
  S.~Ferrara and M.~Gunaydin,
  ``Orbits of exceptional groups, duality and BPS states in string theory,''
  Int.\ J.\ Mod.\ Phys.\ A {\bf 13} (1998) 2075
  [hep-th/9708025].

\bibitem{Lu:1997bg}
  H.~Lu, C.~N.~Pope and K.~S.~Stelle,
  ``Multiplet structures of BPS solitons,''
  Class.\ Quant.\ Grav.\  {\bf 15} (1998) 537
  [hep-th/9708109].

\bibitem{Greene:1989ya}
  B.~R.~Greene, A.~D.~Shapere, C.~Vafa and S.~-T.~Yau,
  ``Stringy Cosmic Strings and Noncompact Calabi-Yau Manifolds,''
  Nucl.\ Phys.\ B {\bf 337} (1990) 1.

\bibitem{Gibbons:1995vg}
  G.~W.~Gibbons, M.~B.~Green and M.~J.~Perry,
  ``Instantons and seven-branes in type IIB superstring theory,''
  Phys.\ Lett.\ B {\bf 370} (1996) 37
  [hep-th/9511080].

\bibitem{Romans}
  L.~J.~Romans,
  ``Massive N=2a Supergravity in Ten-Dimensions,''
  Phys.\ Lett.\ B {\bf 169} (1986) 374.

\bibitem{Polchinski:1995df}
  J.~Polchinski and E.~Witten,
  ``Evidence for heterotic - type I string duality,''
  Nucl.\ Phys.\ B {\bf 460} (1996) 525
  [hep-th/9510169].

\bibitem{Polchinski:1995mt}
  J.~Polchinski,
  ``Dirichlet Branes and Ramond-Ramond charges,''
  Phys.\ Rev.\ Lett.\  {\bf 75} (1995) 4724
  [hep-th/9510017].

\bibitem{Angelantonj:2002ct}
  See, {\it e.g.}, C.~Angelantonj and A.~Sagnotti,
  ``Open strings,''
  Phys.\ Rept.\  {\bf 371} (2002) 1
   [Erratum-ibid.\  {\bf 376} (2003) 339]
  [hep-th/0204089].

\bibitem{Bergshoeff:2005ac}
  E.~A.~Bergshoeff, M.~de Roo, S.~F.~Kerstan and F.~Riccioni,
  ``IIB supergravity revisited,''
  JHEP {\bf 0508} (2005) 098
  [hep-th/0506013].

\bibitem{Bergshoeff:2006qw}
  E.~A.~Bergshoeff, M.~de Roo, S.~F.~Kerstan, T.~Ort\'\i n and F.~Riccioni,
  ``IIA ten-forms and the gauge algebras of maximal supergravity theories,''
  JHEP {\bf 0607} (2006) 018
  [hep-th/0602280].

\bibitem{Bergshoeff:2010mv}
  E.~A.~Bergshoeff, J.~Hartong, P.~S.~Howe, T.~Ort\'\i n and F.~Riccioni,
  ``IIA/IIB Supergravity and Ten-forms,''
  JHEP {\bf 1005} (2010) 061
  [arXiv:1004.1348 [hep-th]].

\bibitem{Riccioni:2007au}
  F.~Riccioni and P.~C.~West,
  ``The E(11) origin of all maximal supergravities,''
  JHEP {\bf 0707} (2007) 063
  [arXiv:0705.0752 [hep-th]].

\bibitem{Bergshoeff:2007qi}
  E.~A.~Bergshoeff, I.~De Baetselier and T.~A.~Nutma,
  ``E(11) and the embedding tensor,''
  JHEP {\bf 0709} (2007) 047
  [arXiv:0705.1304 [hep-th]].

\bibitem{West:2001as}
  P.~C.~West,
  ``E(11) and M theory,''
  Class.\ Quant.\ Grav.\  {\bf 18} (2001) 4443
  [hep-th/0104081].

\bibitem{deWit:2008ta}
  B.~de Wit, H.~Nicolai and H.~Samtleben,
  ``Gauged Supergravities, Tensor Hierarchies, and M-Theory,''
  JHEP {\bf 0802} (2008) 044
  [arXiv:0801.1294 [hep-th]].

\bibitem{Nicolai:2000sc}
  H.~Nicolai and H.~Samtleben,
  ``Maximal gauged supergravity in three-dimensions,''
  Phys.\ Rev.\ Lett.\  {\bf 86} (2001) 1686
  [hep-th/0010076];
  B.~de Wit, H.~Samtleben and M.~Trigiante,
  ``On Lagrangians and gaugings of maximal supergravities,''
  Nucl.\ Phys.\ B {\bf 655} (2003) 93
  [hep-th/0212239].




\bibitem{Bergshoeff:2011qk}
  E.~A.~Bergshoeff and F.~Riccioni,
  ``The D-brane U-scan,''
  arXiv:1109.1725 [hep-th].

\bibitem{Bergshoeff:2012ex}
  E.~A.~Bergshoeff, A.~Marrani and F.~Riccioni,
  ``Brane orbits,''
  Nucl.\ Phys.\ B {\bf 861} (2012) 104
  [arXiv:1201.5819 [hep-th]].

\bibitem{Kleinschmidt:2011vu}
  A.~Kleinschmidt,
  ``Counting supersymmetric branes,''
  JHEP {\bf 1110} (2011) 144
  [arXiv:1109.2025 [hep-th]].

\bibitem{Bergshoeff:2013sxa}
  E.~A.~Bergshoeff, F.~Riccioni and L.~Romano,
  ``Branes, Weights and Central Charges,''
  JHEP {\bf 1306} (2013) 019
  [arXiv:1303.0221 [hep-th]].

\bibitem{Bergshoeff:2011zk}
  E.~A.~Bergshoeff and F.~Riccioni,
  ``String Solitons and T-duality,''
  JHEP {\bf 1105} (2011) 131
  [arXiv:1102.0934 [hep-th]].

\bibitem{Bergshoeff:2011ee}
  E.~A.~Bergshoeff and F.~Riccioni,
  ``Branes and wrapping rules,''
  Phys.\ Lett.\ B {\bf 704} (2011) 367
  [arXiv:1108.5067 [hep-th]].

\bibitem{Bergshoeff:2012jb}
  E.~A.~Bergshoeff and F.~Riccioni,
  ``Heterotic wrapping rules,''
  JHEP {\bf 1301} (2013) 005
  [arXiv:1210.1422 [hep-th]].

\bibitem{Borsten:2011ai}
  L.~Borsten, M.~J.~Duff, S.~Ferrara, A.~Marrani and W.~Rubens,
  ``Small Orbits,''
  Phys.\ Rev.\ D {\bf 85} (2012) 086002
  [arXiv:1108.0424 [hep-th]].

\bibitem{Bergshoeff:2011mh}
  E.~A.~Bergshoeff and F.~Riccioni,
  ``Dual doubled geometry,''
  Phys.\ Lett.\ B {\bf 702} (2011) 281
  [arXiv:1106.0212 [hep-th]].

\bibitem{Bergshoeff:2013spa}
  E.~A.~Bergshoeff, C.~Condeescu, G.~Pradisi and F.~Riccioni,
  ``Heterotic-Type II duality and wrapping rules,''
  JHEP {\bf 1312} (2013) 057
  [arXiv:1311.3578 [hep-th], arXiv:1311.3578].

\bibitem{Bergshoeff:2011se}
  E.~A.~Bergshoeff, T.~Ort\'\i n and F.~Riccioni,
  ``Defect Branes,''
  Nucl.\ Phys.\ B {\bf 856} (2012) 210
  [arXiv:1109.4484 [hep-th]].

\bibitem{grouptheory}
   S. Araki, ``On root systems and an infinitesimal classification of irreducible symmetric spaces,'' Journal of Mathematics, Osaka City University, Vol. 13, No. 1 (1962); S. Helgason, ``Differential geometry, Lie groups and symmetric spaces,'' New York, Academic Press (1978) (Pure and applied mathematics, 80);   M.~Henneaux, D.~Persson and P.~Spindel,
  ``Spacelike Singularities and Hidden Symmetries of Gravity,''
  Living Rev.\ Rel.\  {\bf 11} (2008) 1
  [arXiv:0710.1818 [hep-th]].

\bibitem{Romans6D}
  L.~J.~Romans,
  ``Selfduality for Interacting Fields: Covariant Field Equations for Six-dimensional Chiral Supergravities,''
  Nucl.\ Phys.\ B {\bf 276} (1986) 71.

\bibitem{gunaydinsierratownsend}
  M.~Gunaydin, G.~Sierra and P.~K.~Townsend,
  ``Exceptional Supergravity Theories And The Magic Square,''
  Phys.\ Lett.\  B {\bf 133} (1983) 72;
  ``The Geometry Of N=2 Maxwell-Einstein Supergravity And Jordan Algebras,''
  Nucl.\ Phys.\  B {\bf 242} (1984) 244;
  ``More On D = 5 Maxwell-Einstein Supergravity: Symmetric Spaces And Kinks,''
  Class.\ Quant.\ Grav.\  {\bf 3} (1986) 763.


\bibitem{Kleinschmidt:2008jj}
  A.~Kleinschmidt and D.~Roest,
  ``Extended Symmetries in Supergravity: The Semi-simple Case,''  JHEP {\bf 0807} (2008) 035  [arXiv:0805.2573 [hep-th]].  

\bibitem{Riccioni:2008jz}
  F.~Riccioni, A.~Van Proeyen and P.~C.~West,
  ``Real forms of very extended Kac-Moody algebras and theories with eight supersymmetries,''
  JHEP {\bf 0805} (2008) 079
  [arXiv:0801.2763 [hep-th]].


\bibitem{Keurentjes:2002xc}
  A.~Keurentjes,
  ``The Group theory of oxidation,''
  Nucl.\ Phys.\ B {\bf 658} (2003) 303
  [hep-th/0210178];
    ``The Group theory of oxidation 2: Cosets of nonsplit groups,''
  Nucl.\ Phys.\ B {\bf 658} (2003) 348
  [hep-th/0212024].

\bibitem{Schnakenburg:2004vd}
  I.~Schnakenburg and P.~C.~West,
  ``Kac-Moody symmetries of ten-dimensional nonmaximal supergravity theories,''
  JHEP {\bf 0405} (2004) 019
  [hep-th/0401196].

\bibitem{Bergshoeff:2007vb}
  E.~A.~Bergshoeff, J.~Gomis, T.~A.~Nutma and D.~Roest,
  ``Kac-Moody Spectrum of (Half-)Maximal Supergravities,''
  JHEP {\bf 0802} (2008) 069
  [arXiv:0711.2035 [hep-th]].

\bibitem{Bergshoeff:2010xc}
  E.~A.~Bergshoeff and F.~Riccioni,
  ``D-Brane Wess-Zumino Terms and U-Duality,''
  JHEP {\bf 1011} (2010) 139
  [arXiv:1009.4657 [hep-th]].

\bibitem{Bianchi:1990yu}
  M.~Bianchi and A.~Sagnotti,
  ``On the systematics of open string theories,''
  Phys.\ Lett.\ B {\bf 247} (1990) 517;
  ``Twist symmetry and open string Wilson lines,''
  Nucl.\ Phys.\ B {\bf 361} (1991) 519.

\bibitem{Bergshoeff:2012pm}
  E.~A.~Bergshoeff, A.~Kleinschmidt and F.~Riccioni,
  ``Supersymmetric Domain Walls,''
  Phys.\ Rev.\ D {\bf 86} (2012) 085043
  [arXiv:1206.5697 [hep-th]].



\bibitem{Bergshoeff:2013pia}
  E.~Bergshoeff, F.~Coomans, R.~Kallosh, C.~S.~Shahbazi and A.~Van Proeyen,
  ``Dirac-Born-Infeld-Volkov-Akulov and Deformation of Supersymmetry,''  JHEP {\bf 1308} (2013) 100  [arXiv:1303.5662 [hep-th]].  


\bibitem{Hull1}
  C.~M.~Hull,
  ``A Geometry for non-geometric string backgrounds,''  JHEP {\bf 0510} (2005) 065  [hep-th/0406102]; {\sl ibidem}, 
  ``Doubled Geometry and T-Folds,''  JHEP {\bf 0707} (2007) 080  [hep-th/0605149];
  C.~M.~Hull and R.~A.~Reid-Edwards,
  ``Gauge symmetry, T-duality and doubled geometry,''  JHEP {\bf 0808} (2008) 043  [arXiv:0711.4818 [hep-th]].  


\bibitem{Aldazabal:2013sca}
For a recent review and further references, see  G.~Aldazabal, D.~Marques and C.~Nunez,
  ``Double Field Theory: A Pedagogical Review,''  Class.\ Quant.\ Grav.\  {\bf 30} (2013) 163001  [arXiv:1305.1907 [hep-th]].  



\bibitem{deBoer1}
  J.~de Boer and M.~Shigemori,
  ``Exotic branes and non-geometric backgrounds,''  Phys.\ Rev.\ Lett.\  {\bf 104} (2010) 251603  [arXiv:1004.2521 [hep-th]]; {\sl ibidem},
  ``Exotic Branes in String Theory,''  Phys.\ Rept.\  {\bf 532} (2013) 65  [arXiv:1209.6056 [hep-th]].  












\end{thebibliography}
\end{document}